\documentclass[12pt]{article}
\usepackage{amssymb,amsmath,amsthm,graphicx,ulem}

\pdfoutput=1

\usepackage{graphicx,subfigure}
\usepackage{epsfig}
\usepackage{amsmath}
\usepackage{amsfonts}
\usepackage{amssymb}
\usepackage[usenames]{color}
\usepackage[a4paper,left=2.cm,right=2.cm,top=2.5cm,bottom=2.5cm]{geometry}

\newcommand{\beq}{\begin{equation}}
\newcommand{\eeq}{\end{equation}}
\newcommand{\be}{\begin{equation}}
\newcommand{\ee}{\end{equation}}
\newcommand{\beqa}{\begin{eqnarray}}
\newcommand{\eeqa}{\end{eqnarray}}
\newcommand{\beqar}{\begin{eqnarray*}}
\newcommand{\eeqar}{\end{eqnarray*}}
\newcommand{\bea}{\begin{eqnarray}}
\newcommand{\eea}{\end{eqnarray}}

\newcommand{\eg}{{\it e.g.}\ }
\newcommand{\ie}{{\it i.e.}\ }

\numberwithin{equation}{section}

\begin{document}

\allowdisplaybreaks

\normalem

\title{Black holes with only one Killing field}

\vskip1cm
\author{\'Oscar J. C. Dias${}^{\,a}$, Gary T. Horowitz${}^{\,b}$,
Jorge E. Santos${}^{\,b}$\\ \\ ${}^{\,a}$ DAMTP, Centre for Mathematical Sciences, University of Cambridge, \\
 Wilberforce Road, Cambridge CB3 0WA, UK \\ ${}^{\,b}$ Department of Physics, UCSB, Santa Barbara, CA 93106, USA \\ \\
 \small{  O.Dias@damtp.cam.ac.uk, gary@physics.ucsb.edu, jss55@physics.ucsb.edu}}
 
 \date{}

\maketitle 

\begin{abstract}
\noindent We present the first examples of black holes with only one Killing field. The solutions describe five dimensional AdS black holes with scalar hair. The black holes are neither stationary nor axisymmetric, but are invariant under a single Killing field which is tangent to the null generators of the horizon. Some of these solutions can be viewed as putting black holes into rotating boson stars. Others are related to the endpoint of a superradiant instability. For given mass and angular momentum (within a certain range) several black hole solutions exist.
\end{abstract}

\newpage


\tableofcontents

\section{Introduction}

All known black hole solutions have at least two Killing fields. This is not just due to the fact that finding analytic solutions of Einstein's equation usually requires the assumption of symmetries, but also follows from  theorems which say that stationary black holes must be axisymmetric \cite{Hawking:1971vc,Hollands:2006rj,Moncrief:2008mr}. However, it has been proposed that black holes in anti-de Sitter (AdS) space might exist with just a single Killing field \cite{Kunduri:2006qa}. The argument is based on superradiance. Recall that a mode $e^{-i\omega t + im\psi}$ of a scalar field can increase its amplitude by scattering off a rotating black hole with angular velocity $\Omega_H$ satisfying $\omega < m\Omega_H$. In asymptotically AdS spacetimes, this leads to an instability since the outgoing wave is reflected back onto the black hole and scatters again further increasing its amplitude. This process decreases $\Omega_H$ and eventually results in a black hole with ``lumpy" scalar hair rotating around it. Such a black hole would  not be stationary or axisymmetric, but invariant under just a single Killing field which co-rotates with the scalar hair.

The main goal of this paper is to numerically construct such black holes and study their properties\footnote{Another way to construct asymptotically AdS black holes with only one Killing field is to modify the boundary conditions at infinity so that the boundary metric is not conformal to a round sphere cross time \cite{Anderson:2002xb}. In this case, the spatial symmetries are broken by the boundary conditions. We are interested only in black holes with the standard asymptotic behavior.}. It will be convenient to work in five dimensions. This is because a clever ansatz was recently found for five dimensional scalar fields in which the coupled Einstein scalar field equations reduce to ordinary differential equations even though the scalar fields are invariant only under one Killing field \cite{Hartmann:2010pm}.  In \cite{Hartmann:2010pm} a  potential was added for the scalar fields and asymptotically flat solutions without black holes were found. These were interpreted as rotating boson stars. We will set the potential to zero (so the scalars are massless) and add a negative cosmological constant. We will see that rotating boson stars  still exist in this case. The first example of a rotating boson star in AdS was given in \cite{Astefanesei:2003rw}, where the authors constructed a cohomogeneity-1 solution by working in three dimensions.

Of more interest, we show that one can add small spherical black holes in the center of these boson stars without destroying the solution. There are results in the literature stating that one cannot add black holes inside boson stars \cite{Pena:1997cy, Astefanesei:2003qy}. These results apply to static black holes and follow essentially because the boson star has $e^{-i\omega t}$ time dependence and $t\to \infty$ at the horizon of a black hole. The scalar field thus oscillates infinitely often near the horizon and cannot be smoothly continued inside. We will see that if you put a rotating black hole inside a rotating boson star, this obstruction goes away.  One way to see this is to note that a mode $e^{-i\omega t + im\psi}$ is invariant under $K = \partial/\partial t + (\omega/m) \partial/\partial \psi$. If the black hole has angular velocity $\Omega_H  = \omega/m$, the Killing field $K$ is the null generator of  the horizon, and the scalar field is essentially stationary in a co-rotating frame near the horizon.

The special ansatz for the scalar fields corresponds to an $m=1$ mode. For this mode we determine which Myers-Perry black holes are unstable to superradiance and find that they all have radii much less than the AdS radius. For the same energy $E$ and angular momentum $J$, we find black holes with scalar hair which represent the endpoint of the superradiant instability for this mode.
In the full theory with no restriction on the scalar fields, these hairy black holes are still subject to superradiant instabilities associated with higher $m$ modes. In the final section we discuss the endpoint of these black holes in the full theory.

Similar to the asymptotically flat case,  rotating boson stars in AdS have quite an intricate structure. As one increases the central density,   $E$ and $J$  execute   damped oscillations around  critical values $E_c$ and $J_c$. Thus, not only is there a maximum $E$ and $J$, but there are many different boson stars with energy and angular momentum near $E_c$ and $J_c$. For each of these boson stars, one can add  black holes of various sizes. The area of the black hole can range from zero up to a maximum. This leads to a high level of black hole nonuniqueness. There can be several different hairy black holes with the same energy and angular momentum. We will see that as one increases the size of the hairy black hole, the solution approaches a singular, extremal configuration.

\section{Model}

\subsection{Equations of motion\label{sec:EOM}}

We consider five-dimensional  Einstein gravity with a negative cosmological constant  minimally coupled to two complex
massless scalar fields $\vec{\Pi}$ , \ie with action (we set Newton's constant $G\equiv 1$)
\begin{equation} \label{eq:action5d} 
S=\frac{1}{16\pi}
\int_{\mathcal{M}}d^5 x\,\sqrt{-g}\left[R+\frac{12}{\ell^2}-2\left|\nabla\vec{\Pi}\right|^2\right].
\end{equation}
We will look for boson star and black hole solutions of this theory whose gravitational and scalar fields obey the {\it ansatz}:
\begin{eqnarray}
&& \mathrm{d}s^2 = -f\,g\,\mathrm{d}t^2+\frac{\mathrm{d}r^2}{f}+r^2\left[h\left(\mathrm{d}\psi+\frac{\cos\theta}{2}\mathrm{d}\phi-\Omega \mathrm{d}t\right)^2+\frac{1}{4}\left(\mathrm{d}\theta^2+\sin^2\theta \mathrm{d}\phi^2\right)\right],
\label{eq:metricansatz} 
\\
&& \vec{\Pi} = \Pi\, e^{-i\,\omega\,t+i \,\psi}
\left[
\begin{array}{c}
\sin\left(\frac{\theta}{2}\right) \, e^{-i\frac{\phi}{2}}
\\
\\
\cos\left(\frac{\theta}{2}\right) \, e^{i\frac{\phi}{2}}
\end{array}
\right].
\label{eq:scalaransatz}
\end{eqnarray}
Here, $f,g,h,\Omega$ and $\Pi$ are real functions only of the radial coordinate $r$, \ie the  ansatz is cohomogeneity-1.
Surfaces of constant time $t$ and radial $r$ coordinates have the geometry of a homogeneously squashed
$S^{3}$, written as an $S^1$ fibred over $S^2\equiv CP^1$. The fibre is
parameterized by the coordinate $\psi$, which has period $2\pi$, while $\theta,\phi$ are the standard coordinates on $S^2$. Since $\phi$ also has period $2\pi$, it might appear that $\vec \Pi$ is not single valued, but the Hopf fibration requires that  $\psi\to \psi + \pi$  when $\phi \to \phi + 2\pi$. Thus $\vec \Pi$ is indeed single valued on spacetime.

The motivation for the particular  ansatz  \eqref{eq:metricansatz} for $g_{ab}$ will become clear later in this section. 
This metric  has five linearly independent 
Killing vector fields, namely $\partial_t$, $\partial_\psi$ and the three rotations of $S^2$. However, it is easy to check that the only linear combination which leaves $\vec{\Pi}$ invariant is  
\begin{equation} \label{eq:TheKV} 
 K=\partial_t +\omega \partial_\psi \,.
 \end{equation}
Since a symmetry of the solution must leave both the metric and matter fields invariant, any solution with nonzero scalar fields will have only one Killing field. Of course, one does not usually expect there to be solutions in which the matter fields have much less symmetry than the metric. However the ansatz \eqref{eq:scalaransatz} is special\footnote{It was first considered in \cite{Hartmann:2010pm}.} in that its  stress tensor
\be
T_{ab}=\left(\partial _a\vec{\Pi }^*\partial _b\vec{\Pi }+\partial _a\vec{\Pi }\partial _b\vec{\Pi }^*\right)-g_{\text{ab}}\left(\partial _c\vec{\Pi }\partial ^c\vec{\Pi }^*\right)
\ee
has the same symmetry as the metric \eqref{eq:metricansatz}.  One way to see this is to think of the two complex components of $\vec{\Pi}$ as coordinates on $C^2$. The fact that $|\vec{\Pi}|^2$ is only a function of radius means that at each $(r,t)$, $\vec{\Pi}$ maps the squashed 3-sphere in spacetime into a round 3-sphere. The first term in the stress tensor is just the pull-back of the highly symmetric metric on this sphere and the second term is proportional to the spacetime metric. 

We thus expect our ansatz to have nontrivial solutions, and indeed the field equations $G_{ab}-6\ell^{-2}g_{ab}=T_{ab}$ and $\nabla^2\vec{\Pi} = 0$, lead to a consistent set of ODE's:
\begin{subequations}
\begin{align}
&f''-\frac{6 f'}{f r} \left(\frac{r f'}{6}-\frac{f}{6}+\Xi\right)+\frac{4 h'}{r}+\frac{8}{r^2}+\frac{8}{\ell ^2}+\frac{8 \Pi\,\Pi '}{r}+\frac{4 \Pi ^2 (\omega -\Omega )^2}{f g}-\frac{8\Xi^2}{f r^2}-\frac{4(1+h) \Pi^2}{h r^2}=0\,,
\label{eq:EOMf}
\\
&g''-g' \left(\frac{4\Xi}{f r}+\frac{g'}{g}-\frac{1}{r}\right)-4 g \left[\frac{(\Xi r \sqrt{h})'}{f r^2 \sqrt{h}}+\frac{h^2-\Pi ^2}{f h r^2}+\frac{6}{f \ell ^2}\right]-\frac{8 \Pi ^2 (\omega -\Omega )^2}{f^2}-\frac{h r^2 {\Omega'}^2}{f}=0\,,
\label{eq:EOMg}
\\
&h''+\frac{h'}{r}-\frac{2 h'}{f r} \left(\Xi+\frac{f r h'}{2h}\right)+\frac{h^2 r^2 {\Omega'}^2}{f g}+\frac{4 (1-h)(\Pi ^2+2 h)}{f r^2}=0\,,
\label{eq:EOMh}
\\
&\Omega''+\Omega ' \left(\frac{f'}{f}+\frac{2\Xi}{f r}+\frac{2 h'}{h}+\frac{7}{r}\right)+\frac{4\,\Pi ^2 (\omega -\Omega )}{f h\,r^2}=0\,,
\label{eq:EOMO}
\\
&\Pi ''-\frac{2}{f r} \left(\Xi-\frac{f}{2}\right)\Pi ' +\frac{\Pi  (\omega -\Omega )^2}{f^2 g}-\frac{(1+2 h) \Pi }{f h r^2}=0\,,
\label{eq:EOMP}
\\
&C_1\equiv\frac{(r^4hgf^2)'}{r^3f}+4 g h \Xi=0\,,
\label{eq:EOMC1}
\\
&\frac{C_2}{r^2\ell^2f^2h^2g}\equiv\frac{\Pi^2(\Omega-\omega)^2}{f^2g}+\frac{h'(f h)'}{4 h^2 f}+\frac{(r^4h)'\Xi}{r^5 h f}+\frac{3 (f h r^2)'}{2h r^3 f}+{\Pi'}^2+\frac{2}{f \ell^2}+\frac{h^2-\Pi^2}{r^2hf}-\frac{r^2h{\Omega'}^2}{4fg}=0\,,
\label{eq:EOMC2}
\end{align}
\label{eqs:EOM}
\end{subequations}
where $\Xi \equiv h+\Pi ^2-2 r^2/\ell^2-2$, and $'$ denotes differentiation with respect to $r$. The last two equations can be regarded as constraint equations, 
which under the flow of the remaining five, obey:
\begin{subequations}
\begin{align}
&C_1' = \left(\frac{3 f'}{f}+\frac{4 \Xi }{f r}+\frac{2 h'}{h}+\frac{5}{r}\right)C_1\,,
\\
&C_2' = \left[\frac{\Xi(r^4h)'}{r^4}+\frac{h^2-\Pi^2}{r}+\frac{r h'(f h)'}{4 h}+\frac{3(f h r^2)'}{2 r^2}+\frac{2 r h}{\ell^2}+r f h {\Pi'}^2\right]C_1+\frac{r^4}{h}\left(\frac{h}{r^4}\right)'C_2\,.
\end{align}
\label{eqs:preserve}
\end{subequations}

If we had considered a single complex scalar field, we could still assume a Fourier 
decomposition along the time $t$ and azimuthal coordinates $\psi,\phi$ of $S^3$. This guarantees that $T_{ab}$ is independent of $t,\psi,\phi$. However the stress tensor would still  have contributions depending on the polar coordinate $\theta$. When plugged into the Einstein equation, such a $T_{ab}$  necessarily sources 
a gravitational field that is cohomogeneity-2, \ie that has a radial and polar dependence. The simple ansatz  \eqref{eq:metricansatz} would then not be appropriate and finding a hairy solution would 
require solving a coupled system of PDEs; a much more difficult task. 
Note also that a more general ansatz for the doublet scalar field would consider a more general  Fourier decomposition $e^{i m \psi}$, \ie with an arbitrary integer azimuthal quantum number $m$  not necessarily restricted to the case $m=1$ that we consider. Although we certainly do not rule out the possibility of   finding a  cohomogeneity-1 ansatz for the gravitational field also when $m>1$, this is not a trivial task and we will not attempt it here. Moreover, as we will prove latter, the superradiant instability of Myers-Perry$-$AdS black holes is stronger for the $m=1$ mode than for higher $m$ modes. So the
hairy (black hole) solutions with $m=1$ that we will construct are the first to form from a generic perturbation.

\subsection{Boundary conditions for boson stars and black holes\label{sec:BCs}}

To discuss the boundary conditions of the solutions, we first notice that the ansatz \eqref{eq:metricansatz} and \eqref{eq:scalaransatz} still has a residual gauge freedoom,  
that leaves the gravitational field \eqref{eq:metricansatz} invariant:
\begin{equation}\label{eq:residual}
 \psi \to \psi +\lambda  t\,, \qquad \Omega\to \Omega +\lambda\,,\qquad \omega\to \omega-\lambda \,,
\end{equation}
for an arbitrary constant $\lambda$.
Unless otherwise stated\footnote{In section \ref{sec:grj} we will construct numerically the hairy solutions. We will find that in certain regions of the parameter space the numerical convergence is better if we work in the unphysical gauge where we choose $\lambda$ to be such that $\omega=0$  (which implies $\Omega(\infty)\neq 0$). When we find it necessary to do the computations in this gauge, we will however present our final results in the physical gauge where $\Omega(\infty)=0$ (and thus $\omega\neq 0$). Note also that the boundary conditions in the gauge $\omega=0$ can be easily read from the boundary conditions for the gauge $\Omega(\infty)=0$ once \eqref{eq:residual} is applied.}, we use this gauge freedom to require that $\Omega\rightarrow 0$ as $r\rightarrow \infty$. We are thus in a frame where the solution does not rotate at infinity, 
\ie the conformal boundary of the hairy solutions is an Einstein universe $R\times S^3$. This also guarantees that, in the black hole case,
the angular velocity measured at the horizon, $\Omega_H$, is the appropriate quantity for a thermodynamic analysis \cite{Hawking:1998kw,Hawking:1999dp,Gibbons:2004ai}.

The asymptotic boundary conditions are the same for the boson star and black hole. More concretely, we require that our solutions $f,g,h,\Omega$  asymptote to global AdS
with the next-to-leading order terms in the expansion (determined by  constants $C_f,\,C_h,\,C_\Omega$) being responsible for the generation of mass and 
 angular momentum. The boundary condition for $\Pi$ is determined by the requirement of normalizability. This demands that the massless scalar field decays as $r^{-4}$. 
In the above non-rotating frame, the asymptotic boundary conditions are then: 
\begin{equation}
\label{eq:asympBC}
\begin{aligned}
& f{\bigl |}_{r\to\infty }=  \frac{r^2}{\ell ^2}+1+\frac{C_f\,\ell ^2}{r^2}+\mathcal{O}\left( r^{-3}\right)\,,\quad  
 g{\bigl |}_{r\to\infty }=  1-\frac{C_h\,\ell ^4}{r^4}+\mathcal{O}\left( r^{-5}\right)\,, \\
& h{\bigl |}_{r\to\infty }=  1+\frac{C_h\,\ell ^4}{r^4}+\mathcal{O}\left( r^{-5}\right)\,, \quad \Omega{\bigl |}_{r\to\infty }=  \frac{C_{\Omega }\,\ell ^4}{r^4}+\mathcal{O}\left( r^{-5}\right)\,,\quad  \Pi{\bigl |}_{r\to\infty }= \frac{\epsilon \,\ell^4}{r^4}+\mathcal{O}\left(r^{-5}\right)\,.\\
\end{aligned}
\end{equation}
 Note that $\epsilon$ is a quantity that will be frequently used henceforth and is a dimensionless asymptotic measure of the amplitude of the scalar field (condensate). 
 
The second set of boundary conditions naturally depends on whether we are looking for boson star or black hole solutions. Regular boson stars are smooth horizonless solutions (with harmonic time dependence), 
\ie its boundary conditions at the origin are:
\begin{equation}
\label{eq:BSoriginBC}
 f{\bigl |}_{r\to 0 }= 1 +\mathcal{O}\left( r^2 \right)\,,\quad g{\bigl |}_{r\to 0 }=\mathcal{O}\left( 1 \right)\,,\quad  h{\bigl |}_{r\to 0}= 1+ \mathcal{O}\left( r^2 \right)\,, \quad 
\Omega{\bigl |}_{r\to 0}=  \mathcal{O}\left( 1 \right)\,,\quad \Pi{\bigl |}_{r\to 0 }= \mathcal{O}\left( r \right)\,.
\end{equation}
Note that regularity of $\vec{\Pi}$ requires $\Pi(r)$ to vanish linearly at the origin. 

For the black hole, the inner boundary is at its horizon where the function $f(r)$ must vanish. We take this condition as our definition for the
location of the black hole horizon $r_+$  (the largest root of $f$). The other functions must be regular at this hypersurface:
\begin{equation}
\label{eq:BHhorizonBC}
f{\bigl |}_{r\to r_+ }=\mathcal{O}\left( r-r_+ \right)\,,\quad  g{\bigl |}_{r\to  r_+ }=   \mathcal{O}\left( 1 \right)\,,\quad  h{\bigl |}_{r\to r_+}=   \mathcal{O}\left( 1 \right)\,,\quad 
 \Omega{\bigl |}_{r\to  r_+}=  \mathcal{O}\left( 1 \right)\,,\quad  \Pi{\bigl |}_{r\to  r_+ }= \mathcal{O}\left( 1 \right)\,.
\end{equation}
Regularity of $\vec \Pi$ on the horizon imposes a further restriction. Multiplying \eqref{eq:EOMP} by $f^2$ and evaluating on the horizon shows that the frequency $\omega$ of the scalar field must equal the  angular velocity of the black hole $\Omega_H$, \ie 
\begin{equation}
\label{eq:oO}
\omega=\Omega_H\equiv \Omega(r_+)  \qquad (\hbox{black hole case})\,.
\end{equation}

Recall that the only Killing field of our solution is $ K=\partial_t +\omega \partial_\psi$ \eqref{eq:TheKV}. In general, this has norm $|K|^2=-f g+r^2 h (\omega - \Omega)^2$. On the horizon, \eqref{eq:BHhorizonBC} and \eqref{eq:oO} imply $K$ is null, so the event horizon is also a Killing horizon associated with the only Killing field of the solution.  $K$ is always timelike just outside the event horizon  and in a neighboorhood of the origin $r=0$ for the boson star. The asymptotic boundary conditions \eqref{eq:asympBC}  imply that 
$|K|\rightarrow r^2\left( \omega^2 - 1/\ell^2\right)$ as $r\to\infty$. 
Therefore the Killing  field of the hairy solution will be asymptotically timelike, null or spacelike depending on whether $\omega \ell<1$ ,  $\omega \ell=1$ or  $\omega \ell>1$, respectively. The solutions we find will all have $\omega \ell>1$.
Thus, the solutions are not globally stationary. There is an effective ergoregion at large radius. It turns out that the scalar field is actually concentrated in this region.

\subsection{Asymptotic conserved quantities and thermodynamics \label{sec:charges}}

Even though  the vectors $\partial_t$ and $\partial_\psi$ are not Killing vectors of the full hairy solution, 
they are asymptotic Killing vectors (since the scalar field vanishes asymptotically).  We can therefore define conserved quantities 
associated with these asymptotic symmetries.  
To compute the conserved quantities,  we use the Astekhar-Das formalism \cite{Ashtekar:1999jx}, which  we now briefly review.

Start by introducing a conformal metric $\widetilde{g}$ related to the physical metric $g$ through $\widetilde{g}_{ab}=\Lambda^2 g_{ab}$.
In our case,  the physical metric is given by \eqref{eq:metricansatz} and the appropriate conformal factor is $\Lambda=1/r$. The conformal boundary surface $\Sigma_{\Lambda=0}$ is defined by
$\Lambda=0$ and has normal vector $n_a=\partial_a \Lambda$. Let  $\widetilde{C}_{abcd}$ be the Weyl tensor associated to the conformal metric. Define  
the tensors $\widetilde{K}_{abcd} = \Lambda^{3-d} \widetilde{C}_{abcd} {\bigl |}_{\Lambda=0}$ and $\widetilde{\varepsilon}_{ab} = \ell^2 \widetilde{K}_{abcd} n^b n^d$. On the conformal boundary $\Sigma_{\Lambda=0}$ consider now a timelike surface with $t=$constant and with normal $t_a=\partial_a t$. 
This defines a $(d-2)$-dimensional hypersurface with induced metric $h$ that we call $\Sigma$. 
Take now an asymptotic conformal Killing vector $\xi$. The conserved charge $Q_{\xi}$ associated to this  symmetry is  \cite{Ashtekar:1999jx} 
\begin{equation}
\label{eq:ChargesDef}
Q_{\xi}=\pm\frac{d-3}{8 \pi}\int _{\Sigma}\,\widetilde{\varepsilon}_{ab} \,\xi ^a t^b \, \mathrm{d}\Sigma 
\,,
\end{equation}
where $\pm$ for a timelike/spacelike conformal Killing vector, respectively. 

For solutions described by \eqref{eq:metricansatz} and \eqref{eq:scalaransatz}, we can take $\xi=\left( \partial_t,\,\partial_\psi \right) $ and $d=5$.
The associated conserved quantities are the energy, $E=Q_{\partial_t}$, and the angular momentum, $J=Q_{\partial_\psi}$, given by   
\begin{equation}
\label{eq:Charges}
E=\frac{ \pi \, \ell ^2}{8 }\left(4 C_h-3 C_f\right)\,,\qquad J=\frac{\pi \, \ell^3\, C_{\Omega }}{2}\,,
\end{equation}
where the constants $C_f,\, C_h,\,  C_{\Omega}$ can be read from the asymptotic decay of the gravitational fields in the boundary conditions \eqref{eq:asympBC}.

Boson stars have no horizon and therefore they are zero entropy objects. They do not have an intrinsically defined temperature.
If the solution is a black hole, we have just seen that the Killing vector $K=\partial_t+\omega \partial_\psi$ becomes null at the event horizon $r=r_+$. This is therefore a Killing horizon and its temperature is given by the surface gravity divided by $2\pi$, $T_H=  \sqrt{-(\nabla K)^2{\bigl |}_{r_+}}/\left( 2\sqrt{2}\,\pi \right) $. The black hole entropy is, as usual, the area of the event horizon divided by $4$. This gives
\begin{equation}
\label{eq:TSdef}
T_H=  \frac{f'\sqrt{g}}{4 \pi }{\biggl |}_{r_+}\,,\qquad S=\frac{1}{4}\,A_{d-2}\,r_+^{d-2} \sqrt{h(r_+)}\,,
\end{equation}
 where $A_{d-2}$ is the area of a unit ($d-2$)-sphere.
 
  Any family of  solutions to  Einstein's equation with a Killing field satisfies an analog of  the first law of thermodynamics. This follows from a Hamiltonian derivation of the first law \cite{Wald:1993ki}. When the Killing is given by \eqref{eq:TheKV}, this takes the form:

\begin{eqnarray}
&& \hbox{Boson star:}\qquad \mathrm{d}E=\omega \mathrm{d}J\,,
\label{eq:BS1stlaw} 
\\
&&   \hbox{Black hole:}\qquad  \mathrm{d}E=\omega \mathrm{d}J+T_H\mathrm{d}S\,, \qquad \hbox{with}\quad \omega\equiv \Omega_H\,.
\label{eq:BH1stlaw} 
\end{eqnarray}

\subsection{Myers-Perry$-$AdS solutions\label{sec:MP}}

A very special black hole solution of \eqref{eqs:EOM} is the one where the scalar field vanishes. It describes a rotating black hole where the angular momenta
along the two possible rotation planes in $d=5$ is the same. This solution is the $d=5$ case  of ``Myers-Perry$-$AdS'' (MP-AdS) black holes  \cite{Hawking:1998kw}, with equal angular momenta and with cohomogeneity-1 that  exist in odd dimensions $d=2 N+3$ \cite{Gibbons:2004uw,Gibbons:2004js}. Although we are mainly interested in the $d=5$ solutions, 
we find it useful to discuss briefly the higher-dimensional counterpart of these black holes. Therefore in this subsection, we present the results for an arbitrary odd dimension. 
The equal angular momenta MP-AdS black holes have a gravitational field given by
\begin{equation} \label{eq:MPads} 
  \mathrm{d}s^2 = -f\,g\,\mathrm{d}t^2+\frac{\mathrm{d}r^2}{f}+r^2\left[h\left(\mathrm{d}\psi+A_a  \mathrm{d}x^a-\Omega \mathrm{d}t\right)^2+\hat{g}_{ab}  \mathrm{d}x^a  \mathrm{d}x^b\right],
\end{equation}
 where
\begin{equation}\label{eq:MPadsAUX} 
 f=1+\frac{r^2}{\ell ^2}-\frac{r_M^{2N}}{r^{2N}}\left(1-\frac{a^2}{\ell^2}\right)+\frac{r_M^{2N} a^2}{r^{2(N+1)}}\,,
 \qquad h=1+\frac{r_M^{2N} a^2}{r^{2(N+1)} }\,, \qquad g=\frac{1}{h}\,, \qquad \Omega =\frac{r_M^{2N} a}{r^{2(N+1) }h}\,,  
\end{equation}
and $\hat{g}_{a b}$ is the Fubini-Study metric on $\mathbb{CP}^{N}$ with
Ricci tensor  $\hat{R}_{ab} =2(N+1) \hat{g}_{ab}\,$, and $A = A_a
dx^a\,$ is related to the K\"ahler form $J$ by $dA=2J$. 

Asymptotically, the solution approaches AdS space. The event horizon is located at $r=r_+$ (the largest real root of
$f$) and has a null tangent vector $\partial_t +\Omega_H \partial_\psi$ where the angular velocity is:
\begin{equation} \label{angvel}
  \Omega_H = \frac{r_M^{2N} a}{r_+^{2N+2}+r_M^{2N}a^2}
 \leq \Omega_H^{\rm ext}\,\qquad \hbox{where}\quad \Omega_H^{\rm ext}=\frac{1}{\ell}\sqrt{1+\frac{N \ell^2}{(1+N)r_+^2}} \,. 
\end{equation}
The solution saturating the bound in the angular velocity
corresponds to an extreme black hole with a regular, but degenerate,
horizon. Note that this upper bound in $\Omega_H\ell$ is always
greater than one and tends to the unit value in the limit of large
$r_+/\ell$.

It is convenient to parameterize the solution in terms of $(r_+,\Omega_H)$
 instead of $(r_M,a)$ through the relations,
\begin{equation}
\label{eq:MPrMa}
 r_M^{2N}=\frac{r_+^{2(N+1)} \left(r_+^2+\ell^2\right)}{r_+^2 \ell^2-a^2 \left(r_+^2+ \ell^2\right)}\,,
 \qquad a=\frac{r_+^2 \ell ^2 \Omega_H}{r_+^2+\ell^2}\,.
\end{equation}
The temperature and the entropy as defined in \eqref{eq:TSdef} are given by
\begin{equation}
\label{eq:MPts}
 T_H = \sqrt{\frac{r_+^2+\ell^2}{\ell^2-r_+^2 \left( \Omega_H^2\ell^2-1\right)}}
 \frac{N \ell^2-(N+1) r_+^2 \left(\Omega_H^2\ell^2-1\right)}{2 \pi  r_+
 \ell^2}\,,\quad S=\frac{A_{2N+1}}{4}r_+^{2 N+1} \sqrt{\frac{r_+^2+\ell ^2}{\ell ^2+r_+^2 \left(1-\ell ^2 \Omega _H^2\right)}}\,,
\end{equation}
and the temperature vanishes for the extreme configuration. The mass $E$ and angular
momentum $J$, defined in \eqref{eq:ChargesDef}, 
are~\cite{Gibbons:2004ai}
\begin{equation} \label{eq:MPEJ} E =
\frac{A_{2N+1}}{8\pi}r_M^{2N}\left [N+\frac{1}{2} + \frac{r_+^4
\ell^2 \Omega_H^2}{2(r_+^2+\ell^2)^2} \right]\,, \qquad J  =
\frac{A_{2N+1}}{8\pi} \frac{(N+1) r_M^{2N} r_+^2 \ell ^2
\Omega_H}{r_+^2+\ell^2}\,. 
\end{equation}

The  $N=1$ case describes the $d=5$ black hole that we are most interested in. In this case \eqref{eq:MPads}-\eqref{eq:MPadsAUX}
reduce to  \eqref{eq:metricansatz}-\eqref{eq:scalaransatz}  with $\mathbb{CP}^1=S^2$, $2 A_a  \mathrm{d}x^a=\cos\theta \mathrm{d}\phi$,  and $\Pi(r)=0$.
The line element \eqref{eq:MPads} for the equal angular momenta MP-AdS solution and the fact that it is superradiant unstable 
(thus indicating a possible bifurcation to a new branch of hairy black holes) motivated our choice of the gravitational ansatz \eqref{eq:metricansatz} for the hairy solutions.
Henceforth, unless otherwise stated, when we refer to the  equal angular momenta Myers-Perry$-$AdS black hole we will assume that $d=5$. 

\section{Perturbative results\label{sec:Perturbative}}

In this section we construct  rotating  boson stars and hairy black holes that have a single Killing field \eqref{eq:TheKV}. 
This is a perturbative construction, \ie valid only for small energy and angular momentum.   

\subsection{Rotating boson stars\label{sec:PertBstar}}

Boson stars are smooth horizonless geometries with harmonic time dependence. There is a one-parameter family of solutions, which in the perturbative regime can be parametrized by $\epsilon$. Recall that this  is a measure of the amplitude of the doublet scalar field at infinity  \eqref{eq:asympBC}. We can construct perturbatively the fields of these rotating stars through a power expansion in the condensate amplitude $\epsilon$ around global AdS,
\begin{equation}
\label{eqP:BSexpandDef}
 F(r,\epsilon)=\sum _{j=0}^n  F_{2j}(r)\,\epsilon^{2j}\,, \qquad \Pi(r,\epsilon)=\sum _{j=0}^n  \Pi_{2j+1}(r)\,\epsilon^{2j+1}\,, \qquad \omega(\epsilon)= \sum _{j=0}^{n}\omega_{2j}\,\epsilon^{2j}\,,
\end{equation}
where we use $F$ to represent schematically one of the gravitational fields in \eqref{eq:metricansatz}. That is, each gravitational field $F=\{f,g,h,\Omega\}$ has a power expansion in even powers of the condensate $\epsilon$, with expansion coefficients  $ F_{2j}(r)\equiv\{  f_{2j}(r), g_{2j}(r), h_{2j}(r), \Omega_{2j}(r)\}$ at order $\epsilon^{2j}$. The doublet scalar field has an expansion in odd powers of $\epsilon$.
Note that we expand not only the gravitational and scalar fields but also the frequency. The reason being that at linear order the frequency is constrained to be one of the normal modes in AdS but, when non-linear effects are included, this frequency receives corrections. When $\epsilon=0$, \eqref{eqP:BSexpandDef} describes global AdS spacetime.
Using a standard perturbation theory strategy, we plug the expansions  \eqref{eqP:BSexpandDef} in the equations of motion \eqref{eqs:EOM} to solve for the coefficients $f_{2j}, g_{2j}, h_{2j}, \Omega_{2j}, \Pi_{2j+1}$ and $\omega_{2j}$. At any order in the expansion, the fields must decay asymptotically according to the  boundary conditions \eqref{eq:asympBC}, and the solutions must be regular at the origin as dictated by the boundary conditions \eqref{eq:BSoriginBC}.

The leading order contribution in the expansion, $n=0$, describes the linear perturbation problem where we introduce a non-trivial massless doublet scalar in the AdS background, but this condensate does not back-react on the gravitational field. The most general massless doublet scalar field in AdS that decays with the power law given in \eqref{eq:asympBC} and satisfies  \eqref{eq:EOMP} is given by a sum of normal modes that are labelled by $k$ and given by ($_2F_1$ is the hypergeometric function),
\begin{equation}
\label{eqP:BSlinearPi}
 \Pi(r)=\frac{\epsilon \ell^4 r}{\left(r^2+\ell^2\right)^{5/2}}\, _2F_1\left[\frac{5-\omega \ell}{2},\frac{5+\omega \ell}{2},3,\frac{\ell^2}{r^2+\ell^2}\right],\qquad \omega \ell=5+2k\,,\quad (k=0,1,2,\cdots)\,, 
\end{equation}
where the quantization condition on the allowed frequencies $\omega$ follows from requiring the regularity condition \eqref{eq:BSoriginBC} at the origin. This selects the discrete frequency spectrum of stationary scalar normal modes that can fit into the AdS box. The non-negative integer $k$ describes the several possible radial modes or overtones. 

 A non-linear boson star with Killing vector \eqref{eq:TheKV} is sourced by just one of the modes described in \eqref{eqP:BSlinearPi}. In this paper, we will focus on just the lowest ($k=0$) normal mode,  since this is the one that describes the ground state boson star with the lowest mass to angular momentum ratio.
For $k=0$, \eqref{eqP:BSlinearPi} simplifies.  In the nomenclature of \eqref{eqP:BSexpandDef}, the full solution that satisfies the equations of motion \eqref{eqs:EOM} up to order $\mathcal{O}(\epsilon)$ is then,
\begin{equation}
\label{eqP:BSn1Expansion}
 f_0=1+\frac{r^2}{\ell^2}\,,\qquad g_0=1\,,\qquad h_0=1\,,\qquad \Omega_0=0\,,\qquad \Pi_1=\frac{\ell^4 r}{\left(r^2+\ell^2\right)^{5/2}}\,,\qquad \omega_0=\frac{5}{\ell}\,.
\end{equation}

 If we keep moving in the expansion ladder to the non-linear level, the scalar field back-reacts in the metric (and vice-versa), with the lower order contributions sourcing the higher order fields. At each order $n$ we have to solve the coupled system of equations \eqref{eqs:EOM} up to order $\mathcal{O}\left(\epsilon^{n}\right)$. This can be done analytically. Typically, these are second order ODE's so the most general solution is a  linear combination of two solutions for each field. The two sets of boundary conditions   \eqref{eq:asympBC} and  \eqref{eq:BSoriginBC} fix the two coefficients of the linear combinations that yield the physical solution. At order $\epsilon^2$ and actually at any even order $\epsilon^n$, the equation for the scalar field \eqref{eq:EOMP} is trivially satisfied and the gravitational field equations  \eqref{eq:EOMf}-\eqref{eq:EOMO} are the ones whose solution encodes the non-trivial information on the back-reaction of the scalar field on the gravitational field. On the other hand, the equations at any odd order $\epsilon^n$ are used to find the corrections to the scalar field. This construction can be carried with relative ease up to arbitrary higher orders in $\epsilon$ with the aid of {\it Mathematica}. The expressions for the  fields become  increasingly cumbersome. For our purposes it is enough to present the expansion up to order  $\mathcal{O}\left(\epsilon^{5}\right)$, which is explicitly written in equations \eqref{BS:functions} of Appendix  \ref{sec:AppExpanSoliton}.\footnote{We should however note that when we compare our perturbative results of this section with the numerical results we will be using an expansion that encodes information up to order $\mathcal{O}\left(\epsilon^{7}\right)$.}
The companion expansion for the frequency is given by 
\begin{equation}
\label{eqP:omegaBstar}
\omega  \ell  = 5-\frac{15}{28 }\,\epsilon ^2-\frac{22456447}{35562240 }\,\epsilon^4 +\mathcal{O}\left( \epsilon ^6\right) \,.
\end{equation}

Having at our disposal the field expansion \eqref{BS:functions} for $\{f,g,h,\Omega\}$ we can now easily study their asymptotic fall-off and identify the relevant constants $C_f, C_h, C_\Omega$ 
introduced in \eqref{eq:asympBC}. When plugged into \eqref{eq:Charges} we find that the mass $E$ and angular momentum $J$ of the  boson star are given, respectively, by
\begin{equation}
\label{eqP:ChargesBstar}
E=\ell^2\,\frac{\pi }{4}\left[\frac{5}{6}\,\epsilon^2+\frac{77951}{127008}\,\epsilon^4+\mathcal{O}\left( \epsilon ^6\right) \right], \qquad J=\ell^3\,\frac{\pi }{2}\left[\frac{1}{12}\,\epsilon^2+\frac{83621}{1270080}\,\epsilon^4+\mathcal{O}\left( \epsilon ^6\right) \right].
\end{equation}
Perturbative rotating boson stars are then a one-parameter family of solutions parametrized by the scalar amplitude. In particular,  for small $E,J$, their mass is uniquely determined as a function of their angular momentum. (We will see in the next section that this does not remain true for larger $E,J$.)

\subsection{Hairy black holes\label{sec:PertBHs}}

Hairy black holes are a two-parameter family of solutions. In the perturbative regime, they can be parametrized  by the asymptotic scalar amplitude $\epsilon$ and, in addition, by the horizon radius $r_+$. Consequently, to construct these solutions we need to do a double expansion in powers of $\epsilon$ and also in powers of $r_+$. Plugging this double expansion into the equations of motion we immediately find that, contrary to what happens in the boson star case, we cannot solve  analytically the coupled system of perturbative equations everywhere. This occurs already at the linear level and is not a surprise since we already encounter a similar situation in the familiar studies of, \eg the Klein-Gordon equation in the  Kerr black hole. The traditional way to proceed is to use a {\it matched asymptotic expansion} procedure that applies quite well to the problem at hand, both at linear and non-linear level. In short, the main idea behind this approach is the following. We divide the exterior spacetime of our black hole into two regions; a near-region where $r_+\leq r\ll \ell$ and a far-region where $r\gg r_+$. In each of these regions, some of the terms in the equations of motion make a sub-dominant contribution that is effectively discarded in a lower order expansion analysis. This yields a coupled system of perturbative equations that have an analytical solution in the spacetime region where the approximation is valid. The next important step is to restrict our attention to small black holes\footnote{A coordinate invariant definition of ``small" black hole requires the black hole to have small energy and angular momentum. In the end of our analysis we will find that this definition is equivalent to have  black holes with small $r_+/\ell$ and $\epsilon$.}  that have $r_+/\ell \ll 1$ (and $\epsilon\ll 1$). In this regime, the far and near regions have an overlaping zone, $r_+ \ll r\ll \ell$, where the far and near region solutions are simultaneously valid. In this matching region, we can then match/relate the set of independent parameters that are generated in each of the two regions. Our matched asymptotic analysis of the rotating hairy black holes is similar in spirit to the analysis done in \cite{Basu:2010uz} to construct static charged hairy black holes. 

We are ready to apply the  matched asymptotic expansion to the construction of our hairy black holes. The first observation is that the expansion of the scalar frequency $\omega$ must be independent of the region we look at since it defines  the Killing vector \eqref{eq:TheKV} of the full solution. Its double expansion is given by
\begin{equation}
\label{eqP:BHexpandDefw}
 \omega(\epsilon,r_+)= \sum _{j=0}^{n}\omega_{2j}(r_+)\,\epsilon^{2j},\qquad \omega_{2j}(r_+)=\sum _{i=0}^p \omega_{2j,2i}\,\left( \frac{r_+}{\ell}\right)^{2i}; \qquad 
 \Omega_H\equiv \omega\,;
\end{equation}
\ie at each order of the $\epsilon$-expansion we do a second expansion in $r_+/\ell$. The  last relation in \eqref{eqP:BHexpandDefw} guarantees that the only Killing vector of the hairy black hole  is also the null generator of its Killing horizon, and it is also required by regularity of the scalar field at the horizon; recall discussion associated with \eqref{eq:oO}.

The procedure now proceeds with the aforementioned split of the spacetime into the far and near regions.
Describing the full details of this double expansion and matching procedure is in practice unfeasible due to the large expressions involved and intermediate steps required before matching/fixing the constants of the problem. In  the next three subsections we will describe in detail the formalism and all the ingredients required to do the far-region, near-region and matching analysis. However we will not do the explicit analysis except in a small number of cases just to illustrate  the computational task. The explicit final results for the expansions of all the fields up to an appropriate order is then presented in Appendix \eqref{sec:AppExpanBH}. These expansions contain all the information we need to compute the energy, angular momentum and  thermodynamic quantities in subsection \ref{sec:FinalPertBHs}. 

Like the boson star, we will focus on only the ground state hairy black holes that have the lowest radial excitation of the scalar field. They form a two-parameter family that corresponds to the lowest possible energy to angular momentum ratio. The $k^{\rm th}$-excited hairy black hole family of solutions could also  be constructed following similar steps. 

\subsubsection{Far-region expansion \label{sec:FRPertBHs}}

The far-region is defined by $r\gg r_+$. In this region, a small hairy black hole is a small perturbation in $r_+/\ell$ and $\epsilon$  around global AdS. 

To construct the hairy black holes, in addition to the expansion \eqref{eqP:BHexpandDefw}, we  also do a similar double expansion in the gravitational fields that we denote collectively as $F^{out}$, \ie $F^{out}=\{f^{out},g^{out},h^{out},\Omega^{out}\}$, and in the doublet scalar $\Pi^{out}$:
\begin{eqnarray}
\label{eqP:BHexpandDefFR}
&& \hspace{-0.5cm} F^{out}(r,\epsilon,r_+)=\sum _{j=0}^n  F_{2j}^{out}(r,r_+)\,\epsilon^{2j}, \qquad  F_{2j}^{out}(r,r_+)=\sum _{i=0}^p  F_{2j,2i}^{out}(r)\,\left( \frac{r_+}{\ell}\right)^{2i}; \nonumber\\
&& \hspace{-0.5cm}  \Pi^{out}(r,\epsilon,r_+)=\sum _{j=0}^n  \Pi^{out}_{2j+1}(r,r_+)\,\epsilon^{2j+1}, \qquad  \Pi_{2j+1}^{out}(r,r_+)=\sum _{i=0}^p  \Pi_{2j+1,2i}^{out}(r)\,\left( \frac{r_+}{\ell}\right)^{2i}.
\end{eqnarray}
We use the superscript ``out" to emphasize that these expansions are valid only in the far-region. We want to find the gravitational and scalar fields that obey the equations of motion \eqref{eqs:EOM} and the asymptotic boundary conditions  \eqref{eq:asympBC}. The boundary conditions \eqref{eq:BHhorizonBC} at the horizon are not imposed since the far-region does not extend all the way down to this boundary of the full spacetime.

Start by considering the leading order expansion in $\epsilon$, \ie $n=0$. Like in the boson star, this order describes the linear perturbation problem where we introduce a non-trivial massless doublet scalar in the spacetime background but this condensate does not back-react in the gravitational field. The difference with the boson star case is that now the background spacetime is not just global AdS but the ($d=5$) MP-AdS black hole described in subsection \ref{sec:MP}. 
Since we know this solution exactly, the $n=0$ coefficients  $F_{0,2i}^{out}$ are already known up to any desired order $\mathcal{O}(r_+^p/\ell^p)$. All we need to do is to insert the expansion \eqref{eqP:BHexpandDefw} for $\omega_0$ into \eqref{eq:MPrMa}, that defines the mass-radius $r_M$ and rotation parameter $a$ of the MP-AdS black hole in terms of $r_+$ and $\Omega_H\equiv\omega_0$, and replace the resultant expressions into the gravitational fields  \eqref{eq:MPadsAUX}. A simple series expansion in $r_+/\ell$ gives the desired coefficients $F_{0,2i}^{out}(r)$ in terms of the constants $\omega_{0,2i}$ (the latter are left undetermined at this stage). By construction,  the associated $F^{out}(r,0,r_+)$ satisfies trivially the gravitational field equations  \eqref{eq:EOMf}-\eqref{eq:EOMO} and the asymptotic boundary conditions \eqref{eq:asympBC} up to order $\mathcal{O}(\epsilon,r_+^p/\ell^p)$. 
To find the coefficients $\Pi_{1,2i}^{out}$ we now have to solve the scalar equation \eqref{eq:EOMP} up to order  $\mathcal{O}(\epsilon,r_+^p/\ell^p)$ subject to the asymptotic boundary condition in  \eqref{eq:asympBC} for $\Pi$.  

To illustrate the computational task let us discuss the details of the determination of the first two contributions $\Pi^{out}_{1,0}(r)$ and $\Pi^{out}_{1,2}(r)$. 
The lowest order in the horizon radius expansion has no dependence on $r_+$, \ie it encodes no information about the existence of an horizon. Effectively we are determining the lowest normal mode in global AdS and the desired coefficients are precisely given by  \eqref{eqP:BSn1Expansion} after the relabelling $\{ F_0,\Pi_0,\omega_0\} \to \{F_{0,0}^{out},\omega_{0,0},\Pi^{out}_{1,0}\}$.   The next-to-leading order is the first to encode a $r_+$ dependence. We already described how we can get  $F_{0,2}^{out}$. All we need to find is $ \Pi^{out}_{1,2}$. The requirement that the scalar equation is solved up to $\mathcal{O}(\epsilon^1,r_+^2/\ell^2)$ and the field decays as  \eqref{eq:asympBC} demands that
\begin{eqnarray}
\label{eqP:BHpi12out}
&& \Pi^{out}_{1,2}=\frac{\ell^4 }{24 r^3 \left(r^2+\ell ^2\right)^{7/2}}{\biggl(}- \ell ^6(\ell \omega_{0,2}+3)-3 r^2 \ell ^4 (3 \ell  \omega_{0,2}+11)-2r^4 \ell ^2 (4 \ell  \omega_{0,2}+45) \nonumber\\
&& \hspace{1.3cm}
+6 r^4 \left(r^2+\ell ^2\right) (2 \ell \omega_{0,2}-15)\, {\rm log}\left(\frac{r^2}{r^2+\ell ^2}\right){\biggr )}.
\end{eqnarray}
Note that the asymptotic boundary condition was used to fix two integration constants that would give the more general but unphysical solution. At this stage the constant  $\omega_{0,2}$ is left undetermined. We will fix it in subsection \ref{sec:MatchPertBHs}.
We leave a discussion of the linear analysis of the higher order terms in $r_+/\ell$, as well as the discussion of the non-linear terms $\mathcal{O}(\epsilon^n)$ ($n\geq 2$) to subsection   \ref{sec:MatchPertBHs}.

At this point we still need to justify why the far-region analysis is valid only in the coordinate range  $r\gg r_+$. In subsection \ref{sec:MatchPertBHs} we will take the small radius expansion of $\Pi^{out}$ and we will find that it is singular as $r\to 0$, diverging with a power of $\frac{r_+}{r}$. This shows that the far-region analysis certainly breaks down at $r\sim r_+$. Therefore we must take its regime of validity to be $r\gg r_+$.

\subsubsection{Near-region expansion \label{sec:NRPertBHs}}

The near-region is defined by the radial range $r_+\leq r\ll \ell$. In this region, the rotational field $\Omega\sim \mathcal{O}(1)$ of the  MP-AdS/hairy black holes is negligible compared with the mass scale $\ell/r_+\gg 1$ set by the horizon radius. The best way to see this is to introduce rescaled  time and radial coordinates, $\tau=t \,\ell/r_+$ and $z=r \,\ell/r_+$, such that the near-region is described by $z\simeq \ell$. In these rescaled coordinates the $d=5$ MP-AdS black hole \eqref{eq:MPads}-\eqref{eq:MPadsAUX}   (to which we want to add a rotating condensate) is written as
\begin{eqnarray} \label{eq:MPadsNR} 
 &&\hspace{-1cm} \mathrm{d}s^2 =\frac{r_+^2}{\ell^2}\left[    -f\,g\,\mathrm{d}\tau^2+\frac{\mathrm{d}z^2}{f}+z^2\left[h\left(\mathrm{d}\psi+\frac{\cos\theta}{2}\mathrm{d}\phi-\frac{r_+}{\ell}\,\Omega \mathrm{d}\tau\right)^2+\frac{1}{4}\left(\mathrm{d}\theta^2+\sin^2\theta \mathrm{d}\phi^2\right)\right]  \right]; \nonumber \\
&& f=\frac{\left(z^2-\ell ^2\right) }{z^4\ell ^4}\left(\left(z^2+\ell ^2\right) \left(r_+^2 z^2+\ell ^4\right)+\frac{\ell ^6 \left(r_+^2+\ell ^2\right)}{r_+^2\left(\ell ^2 \Omega _H^2-1\right)-\ell ^2}\right), \qquad g=\frac{1}{h}\,,\nonumber \\
&&
h=1-\frac{r_+^2 \ell ^6 \Omega _H^2}{z^4 \left(r_+^2\left(\ell ^2 \Omega _H^2-1\right)-\ell ^2\right)}, \qquad
\Omega=\frac{\Omega _H\ell ^4\left(r_+^2+\ell ^2\right)}{r_+^2 \ell ^6 \Omega _H^2-z^4 \left(r_+^2\left(\ell ^2 \Omega _H^2-1\right)-\ell ^2\right)}\,.
\end{eqnarray}
This coordinate transformation is illuminating because the rotation field of the conformal metric,  $g_{\tau\psi}\propto \frac{r_+}{\ell}\,\Omega$,  has an explicit factor of $r_+/\ell\ll 1$. In the near-region the rotation field is therefore weak and it is to be seen as a small perturbation around the static configuration. It makes also explicit that an expansion in the small rotation field is is an expansion in a power series of $r_+/\ell$. These statements should continue to hold when a small condensate is added to the system, \ie when we do an additional  power series expansion in $\epsilon$. 

These arguments justify that the expansion \eqref{eqP:BHexpandDefw} is to be accompanied by a similar double expansion in the gravitational fields $F^{in}=\{f^{in},g^{in},h^{in},\Omega^{in}\}$, and in the doublet scalar $\Pi^{in}$:
\begin{eqnarray}
\label{eqP:BHexpandDefNR}
&& \hspace{-0.5cm} F^{in}(z,\epsilon,r_+)=\sum _{j=0}^n  F_{2j}^{in}(z,r_+)\,\epsilon^{2j}, \qquad  F_{2j}^{in}(z,r_+)=\sum _{i=0}^N  F_{2j,2i}^{in}(z)\,\left( \frac{r_+}{\ell}\right)^{2i}; \\
&& \hspace{-0.5cm}  \Pi^{in}(z,\epsilon,r_+)=\sum _{j=0}^n  \Pi^{out}_{2j+1}(z,r_+)\,\epsilon^{2j+1}, \qquad  \Pi_{2j+1}^{in}(z,r_+)=\sum _{i=0}^N  \Pi_{2j+1,2i+1}^{out}(z)\,\left( \frac{r_+}{\ell}\right)^{2i+1};\nonumber
\end{eqnarray}
The superscript ``in" indicates that these expansions are valid only in the near-region. The near-region fields must obey the equations of motion \eqref{eqs:EOM} and the horizon boundary conditions  \eqref{eq:BHhorizonBC}. The asymptotic region $z\to \infty$ is not part of the near-region so we do not impose the asymptotic  boundary conditions \eqref{eq:asympBC}.

The strategy for the next steps of the near-region analysis is now similar to the far-region one. At leading order in the $\epsilon$-expansion, we again have the linear problem where we want to find the scalar field that propagates in the MP-AdS background without back-reacting with it.  The gravitational coefficients   $F_{0,2i}^{in}$ are those of the MP-AdS background \eqref{eq:MPadsNR}: we just replace the frequency expansion  \eqref{eqP:BHexpandDefw} into \eqref{eq:MPadsNR} and take a series expansion in $r_+/\ell$ up to the desired order $\mathcal{O}(r_+^p/\ell^p)$. The non-trivial task is to determine the scalar coefficients $\Pi_{1,2i+1}^{in}$ which satisfy \eqref{eq:EOMP} up to order  $\mathcal{O}(\epsilon,r_+^p/\ell^p)$ and the last boundary condition in \eqref{eq:BHhorizonBC}.  

As an  example of the procedure,  we find here the leading order contribution in the $r_+/\ell$ expansion of the linear scalar field.  Equation  \eqref{eq:EOMP} is satisfied up to order  $\mathcal{O}(\epsilon,r_+^2/\ell^2)$ by a linear combination, $K^{in}_{11} P_{1/2}\left(2 z^2/\ell^2-1\right)+ C^{in}_{11} Q_{1/2}\left(2 z^2/\ell^2-1\right)$, of a Legendre polynomial of the first kind $P$ and of a  Legendre polynomial of the second kind $Q$. The latter has a logarithmic divergence at the horizon $z=\ell$. The boundary condition  \eqref{eq:BHhorizonBC} therefore requires that we eliminate this contribution by killing its amplitude:  $C^{in}_{11} =0$. The physical solution is then
\begin{equation}
\label{eqP:BHpi11in}
 \Pi^{in}_{1,1}=K^{in}_{11} P_{1/2}\left(2 z^2/\ell^2-1\right),
\end{equation}
where the integration constant $K^{in}_{11}$  will be fixed only by the matching analysis of  subsection \ref{sec:MatchPertBHs}. This is a good moment to note the curious fact that the far-region expansion of the scalar field is an expansion in even powers of  $r_+/\ell$, see \eqref{eqP:BHexpandDefFR}, while in the near-region $\Pi^{in}$ has instead an expansion in odd powers of $r_+/\ell$, see \eqref{eqP:BHexpandDefNR}. The reason why the $\Pi^{in}$ expansion starts at $r_+/\ell$ instead of at $r_+^0/\ell^0$ is that  $\Pi$  must have a smooth limit when we send the horizon radius to zero, in order for perturbation theory to be consistent. One has $\Pi^{in}_{1,1}{\bigl |}_{r_+\to 0}= \frac{4 K^{in}_{11} r}{\pi  r_+}$ and thus 
$\Pi^{in}=\epsilon \left[\Pi^{in}_{1,1}\, r_+/\ell + \mathcal{O}(r_+^2/\ell^2)\right]$ is regular when $r_+\to 0$ as required\footnote{To be more clear, $\Pi^{in}=\epsilon \left[\Pi^{in}_{1,0}+\Pi^{in}_{1,1}\, r_+/\ell + \mathcal{O}(r_+^2/\ell^2)\right]$ with $\Pi^{in}_{1,0}=K^{in}_{10} P_{1/2}\left(2 z^2/\ell^2-1\right)$ would satisfy both the field equations and the horizon boundary condition. But it would not have a smooth limit when $r_+\to 0$ and thus we have to set $K^{in}_{10}=0$. The  expansion then starts with the linear term in $r_+/\ell$.}. Ultimately, this occurs because, as we will find in the end of our analysis, the boson start describes the $r_+\to 0$ limit of the hairy black hole and in the former case one has $\Pi|_{r\to 0}= 0$; see \eqref{BS:functions} or \eqref{eq:BSoriginBC}.  

The discussion of the linear analysis of the higher order terms in $r_+/\ell$, and the discussion of the non-linear terms $\mathcal{O}(\epsilon^n)$ ($n\geq 2$) is left to subsection   \ref{sec:MatchPertBHs}.
A question we still have to address is why the near-region analysis is valid only for  $r\ll \ell$. In subsection \ref{sec:MatchPertBHs} we will take the large radius expansion of $\Pi^{in}$ and we will conclude that it diverges asymptotically as $\frac{r}{\ell}$. Consequently,  the near-region analysis breaks down at $r\sim \ell$, and we must take its regime of validity to be  $r\ll \ell$. 

\subsubsection{Matching region \label{sec:MatchPertBHs}}

Small black holes have $r_+\ll \ell$ and the far and near region solutions intersect in the region  $r_+ \ll r\ll \ell$ where they are  simultaneously valid. Therefore, this overlapping region provides a set of matching conditions that fix the set of integration constants and frequencies $\omega_{2j,2i}$. After this process the solution is fully determined everywhere. The matching procedure should be done iteratively order by order in the $r_+/\ell$ expansion for a given order of the $\epsilon$-expansion. That is, at a given order $\mathcal{O}(\epsilon^k r_+^q/\ell^q)$ we determine the associated unknown constants and only then we move to the next order  $\mathcal{O}(\epsilon^k r_+^{q+2}/\ell^{q+2})$ to do its matching analysis. The set of matching conditions are generated by matching the large radius behavior of the near-region solution with the small radius expansion of the far-region solution. 
To get the former,  we take the large-$z$  Taylor expansion of \eqref{eqP:BHexpandDefNR} and then we return to the original radial coordinate $r=z \,r_+/\ell$. 
To get the small radius expansion of the far-region we do the small-$r$ Taylor series of \eqref{eqP:BHexpandDefFR}. 
The two golden rules for the matching procedure are the standard ones: at each order we should match only terms that will not receive a further contribution at next order of the expansion; and the divergent terms in the large radius near-region expansion (that go as powers of $r/\ell$) and the divergent terms in the small radius far-region expansion (that go as powers of $r_+/r$) must be eliminated or at most exactly matched in the other region.

Once more this matching procedure is better explained with an example. Consider the linear problem, \ie the order $\mathcal{O}(\epsilon)$. Here, we want to explicitly determine the scalar field up to order $\mathcal{O}(\epsilon\, r_+^2/\ell^2)$. Take first the order $\mathcal{O}(\epsilon \,r_+/\ell)$. Insert $\Pi^{in}_{1,1}$ as given by \eqref{eqP:BHpi11in} into \eqref{eqP:BHexpandDefNR}; do a Taylor expansion in $z$ and replace $z=r \,\ell/r_+$. This yields the asymptotic near-region expansion that we call $\Pi^{in}_{\rm large}$.  To get the small-$r$ far-region expansion $\Pi^{out}_{\rm small}$,  do the Taylor expansion of  $\Pi=\epsilon \Pi^{out}_{1,0}$. Matching these two expansions fixes the integration constant $ K^{in}_{11}$ introduced in  \eqref{eqP:BHpi11in}: 
\begin{equation}
\left \{
\begin{array}{c}
\Pi^{in}_{\rm large}= \epsilon\,\frac{4 K^{in}_{11}}{\pi}\,\frac{r}{\ell}+ \mathcal{O}\left(\frac{r_+^2}{\ell\, r}\right)\,,
\\
\\
\hspace{-0.9cm}\Pi^{out}_{\rm small}= \epsilon\,\frac{r}{\ell}+ \mathcal{O}\left( \frac{r^3}{\ell^3}\right)\,,
\end{array}
\right . \qquad \rightarrow  \quad  \Pi^{in}_{\rm large}= \Pi^{out}_{\rm small} \qquad \Rightarrow \quad  K^{in}_{11}=\frac{\pi}{4}\,.
\label{eqP:matchPi11}
\end{equation}
 Take now the order $\mathcal{O}(\epsilon\, r_+^2/\ell^2)$.  There is no coefficient $\Pi^{in}_{1,2}$ so the near-region contribution to $\mathcal{O}(r_+^2)$ comes exclusively from  $\Pi^{in}_{1,1}$: we repeat the steps used to get \eqref{eqP:matchPi11} but this time we keep also the $r_+^2$ contribution in $\Pi^{in}_{\rm large}$ that has to be matched  by the small-$r$ far-region expansion. The latter  is generated by the Taylor expansion of  $\Pi=\epsilon \left( \Pi^{out}_{1,0}+\Pi^{out}_{1,2}\,r_+^2/\ell^2\right)$. Matching the two expansions determines  $ \omega _{0,2}$ introduced in  \eqref{eqP:BHpi12out}: 
\begin{equation}
\hspace{-0.1cm}\left \{
\begin{array}{c}
\hspace{-4.7cm}\Pi^{in}_{\rm large}= \epsilon\, \left(\frac{r}{\ell }-\frac{1}{4}\frac{r_+^2 }{\ell  r}\right)+ \mathcal{O}\left(\frac{r_+^4}{\ell r^3}\right)\,,
\\
\\
\Pi^{out}_{\rm small}= \epsilon\,\left(  \frac{r}{\ell } -\frac{\left(45+11 \ell  \omega _{0,2}\right)}{48}\frac{r_+^2}{\ell r}
 -\frac{\left(3+\ell  \omega _{0,2}\right)}{24 }\frac{r_+^2 \ell }{r^3}\right) + \mathcal{O}\left( \frac{r}{\ell}\frac{r_+^2}{\ell^2},\frac{r^3}{\ell^3} \right)\,,
\end{array}
\right .  \hspace{-0.5cm} \rightarrow  \Pi^{in}_{\rm large}= \Pi^{out}_{\rm small}  \Rightarrow  \omega _{0,2}=-\frac{3}{\ell}\,,
\label{eqP:matchPi12}
\end{equation}
that we can now also insert in the expressions for the gravitational coefficients $F_{0,2i}^{in}$ and  $F_{0,2i}^{out}$ fixing them totally.
We could now proceed and do the analysis at order $\mathcal{O}(\epsilon\, r_+^3)$ and  $\mathcal{O}(\epsilon\, r_+^4)$. The final results, after the matching, are presented in Appendix \eqref{sec:AppExpanBH}.

Having done the analysis at linear order $\mathcal{O}(\epsilon)$, we must next consider the non-linear ($n\geq 2$) contribution  $\mathcal{O}(\epsilon^n,\, r_+^p/\ell^p)$. Indeed, only at these non-linear orders the  scalar field back-reacts in the metric giving the necessary information about the gravitational field of the hairy black holes. At each order $(n,p)$ we have to solve the coupled system of equations \eqref{eqs:EOM} at order $\mathcal{O}\left(\epsilon^{n},r_+^p/\ell^p\right)$. At  any even order $\epsilon^n$, the equation for the scalar field \eqref{eq:EOMP} is trivially satisfied and the gravitational field equations  \eqref{eq:EOMf}-\eqref{eq:EOMO} are the ones whose solution encodes the non-trivial information on the back-reaction of the scalar field on the gravitational field. On the other hand, the equations at any odd order $\epsilon^n$ are used to find the corrections to the scalar field. The analysis again starts with the far and near region construction of the solutions that are then matched in the overlaping region. The process is done order by order and it is similar to the few cases we illustrated at the linear level. The analysis is however cumbersome for several reasons: the  lower order contributions source the higher order fields; the expressions are quite often large; and there are many fields involved. For this reason we do not give the technical details here. The explicit final results  for the expansions of all the fields up to combined order $(n+p)=4$ is then presented in Appendix \eqref{sec:AppExpanBH}. The near-region solutions with the matching analysis already done is presented in Equation \eqref{NR:functions}, while the far-region fields are written in Equation \eqref{FR:functions}, again after having fixed the integration constants of the problem by the matching procedure. The companion expansion for the frequency will be given below in equation \eqref{eqP:omegaBH}. 
In the presentation of these results,  we take  $\epsilon\ll 1$ and $r_+\ll \ell$ and we assume that $\mathcal{O}(\epsilon)\sim\mathcal{O}(r_+/\ell)$. The latter assumption implies that terms with the same $(n+p)$ contribute equally to the perturbative expansion, \ie  $\mathcal{O}\left(\epsilon^{0},r_+^4/\ell^4\right)\sim \mathcal{O}\left(\epsilon^{2},r_+^2/\ell^2\right)\sim \mathcal{O}\left(\epsilon^{4},r_+^0/\ell^0\right)$. We have done the consistent perturbative expansion analysis up to $(n+p)=4$, which is sufficient for our purposes.  Extending the perturbative analysis to higher orders is certainly possible, although increasingly cumbersome.

\subsubsection{Energy, angular momentum, and thermodynamic quantities \label{sec:FinalPertBHs}}

The matched asymptotic expansion analysis of the previous subsections constructed perturbatively the scalar and gravitational fields of the small  rotating hairy black holes. With this information we can now compute their energy, angular momentum, and thermodynamic quantities and study some of their main properties.

The angular velocity of the black hole, which is also the frequency of the doublet scalar field, was determined side by side together with the field expansions in the matched asymptotic expansion construction; see {\it eg} \eqref{eqP:matchPi12}. It is given by, 
\begin{eqnarray}
\label{eqP:omegaBH}
&&\hspace{-1cm} \Omega_H \ell \equiv \omega  \ell =\left[5-3\frac{r_+^2}{\ell ^2}-\left(\frac{959}{16}-\frac{3}{2}\, {\rm log}\left(\frac{r_+}{4\ell }\right)\right)\frac{r_+^4}{\ell ^4}+\mathcal{O}\left( \frac{r_+^6}{\ell^6} \right)  \right]+\epsilon ^2\left[ -\frac{15}{28}-\frac{4469}{840}\frac{r_+^2}{\ell ^2}+\mathcal{O}\left(\frac{r_+^4}{\ell^4}\right)  \right] \nonumber \\        
&& \hspace{1.4cm} -\epsilon ^4\,\frac{22456447}{35562240}\left[ 1+\mathcal{O}\left( \frac{r_+^2}{\ell^2} \right) \right]  +\mathcal{O}\left( \epsilon ^6\right).
\end{eqnarray}
The perturbative near-region solution for  $\{f,g,h\}$  is written in Equation \eqref{NR:functions} of  Appendix \eqref{sec:AppExpanBH}. It then follows from \eqref{eq:TSdef} that the temperature and entropy of the rotating hairy black holes are, respectively, given by
\begin{eqnarray}
\label{eqP:TSbh}
&&\hspace{-1cm} T_H=\frac{1}{\pi r_+}{\biggl [} \left(\frac{1}{2 }-\frac{71}{4 }\frac{r_+^2}{\ell ^2}-\frac{2665}{16 }\frac{r_+^4}{\ell ^4} +O\left(\frac{r_+^6}{\ell ^6}\right)\right)
  -\epsilon ^2{\biggl (}\frac{1}{6 }+\frac{r_+^2}{\ell ^2}\frac{1}{2016 }{\bigl (}2197+63 \pi ^2-441\, {\rm log}(2)\nonumber \\ 
&& \hspace{0.8cm} -336\, {\rm log}\left(\frac{r_+}{\ell }\right){\bigr )}+O\left(\frac{r_+^4}{\ell ^4}\right) {\biggr )}  -\epsilon ^4\left(\frac{101341}{508032 }+O\left(\frac{r_+^2}{\ell ^2}\right)\right)+\mathcal{O}\left( \epsilon ^6\right)  {\biggr ]},    \\ 
 &&\hspace{-1cm} S=r_+^3\,\frac{\pi ^2}{2}\left[\left(1+\frac{25}{2}\frac{r_+^2}{\ell ^2}+\frac{1655}{8}\frac{r_+^4}{\ell ^4}+\mathcal{O}\left(\frac{r_+^6}{\ell ^6}\right)\right)+\epsilon ^2\left(\frac{265}{84}\frac{r_+^2}{\ell ^2}+\mathcal{O}\left(\frac{r_+^4}{\ell ^4}\right)\right) +\epsilon ^4\mathcal{O}\left(\frac{r_+^2}{\ell ^2}\right)+\mathcal{O}\left(\epsilon ^6\right)\right]. \nonumber
\end{eqnarray}
Given the perturbative far-region solution \eqref{FR:functions} for $\{f,g,h,\Omega\}$ we can read their asymptotic fall-off and identify the relevant constants $C_f, C_h, C_\Omega$ 
introduced in \eqref{eq:asympBC}. Plugging these into \eqref{eq:Charges} we find that the energy $E$ and angular momentum $J$ of the  rotating hairy black hole are given, respectively, by
\begin{eqnarray}
\label{eqP:ChargesBH}
&& \hspace{-0.7cm} E=\ell ^2\,\frac{\pi }{4}\left[\left(\frac{3}{2}\frac{r_+^2}{\ell ^2}+39\frac{r_+^4}{\ell ^4}+ \mathcal{O}\left(\frac{r_+^6}{\ell^6}\right) \right)+\epsilon ^2\left(\frac{5}{6}+\frac{191}{48}\frac{r_+^2}{\ell ^2} + \mathcal{O}\left(\frac{r_+^4}{\ell^4}\right)  \right)+\epsilon ^4 \left(  \frac{77951}{127008}+ \mathcal{O}\left(\frac{r_+^2}{\ell^2}\right) \right) \right]\nonumber \\ 
&& \hspace{1cm} +\mathcal{O}\left(\epsilon ^6\right),  \nonumber \\
&& \hspace{-0.7cm}  J=\ell ^3\,\frac{\pi }{2}\left[\left( 5\frac{r_+^4}{\ell ^4} + \mathcal{O}\left(\frac{r_+^6}{\ell^6}\right)  \right)+\epsilon ^2\frac{1}{12}\left(1+\frac{129}{24}\frac{r_+^2}{\ell ^2} + \mathcal{O}\left(\frac{r_+^4}{\ell^4}\right)  \right)+\epsilon ^4 \left( \frac{83621}{1270080} + \mathcal{O}\left(\frac{r_+^2}{\ell^2}\right) \right)\right]\nonumber \\ 
&& \hspace{1cm} +\mathcal{O}\left(\epsilon ^6\right). 
\end{eqnarray}
It can be explicitly checked that relations \eqref{eqP:omegaBH}-\eqref{eqP:ChargesBH} obey the first law of thermodynamics \eqref{eq:BH1stlaw} up to fourth order.

To discuss these results, start by noticing that the perturbative analysis constructs small black holes. Indeed  $r_+/\ell \ll 1$ and  $\epsilon \ll 1$ implies that $E/\ell^2\ll 1$ and $J/\ell^3 \ll 1$. One of the main properties of the hairy black holes is that they reduce to the boson stars constructed in section \ref{sec:PertBstar} when the horizon radius is sent to zero. {\it \ A priori} there is no reason why this had to be the case. The boson stars may have been  analogous to marginally bound states which would disappear when a small black hole is added. However, we indeed find that solutions exist and the scalar field frequency \eqref{eqP:omegaBH} and the conserved quantities \eqref{eqP:ChargesBH} reduce to the frequency  \eqref{eqP:omegaBstar} and energy and angular momentum \eqref{eqP:ChargesBstar} of the rotating boson stars in the limit $r_+\to 0$. Naturally, the black hole entropy also vanishes in this limit, while its temperature diverges (a similar situation $-$  see \eqref{eq:MPts} $-$ occurs when  we send $r_+ \to 0$ in the MP-AdS black hole). 
Another special curve is the one-parameter sub-family of hairy black holes with $\epsilon=0$. In this limit the scalar condensate vanishes: this curve describes the merger of the hairy black holes with the MP-AdS black holes. This merger line describes non-extreme MP-AdS black holes that are at the onset of the superradiant instability, as will be discussed later in section \ref{subsec:merger}.

In sum, in a phase diagram of $E/\ell^2$ {\it vs} $J/\ell^3$ the two parameter family of hairy black holes is bounded below by the curve describing the one-parameter family of boson stars, and bounded above by the merger curve with the MP-AdS black holes. We will postpone further discussion of the properties of the system to section \ref{sec:grj} where we will construct numerically the most general family of rotating boson stars and hairy black holes whose energy and angular momentum are not restricted to be (very) small. We will find that the phase diagram develops a very interesting structure as the energy and angular momentum of the  boson stars and black holes grow to larger values. 
 
\section{\label{sec:grj}General  results}

The perturbative construction of the small  rotating boson stars and  hairy black holes of the previous section has the advantage of being an analytic analysis. However, as we will find latter in this section,
there are interesting properties of the system that can only be revealed by studying solutions of (\ref{eqs:EOM}) in regimes where the energy and angular momentum are not small, \emph{i.e.} where the analytical results are not valid. To probe this region of the parameter space we thus need to perform a frontal numerical approach in solving the equations of motion (\ref{eqs:EOM}). This will be the aim of this section. As a check of our numerical work, for small energy and angular momentum we will find agreement with the analytical results found in section \ref{sec:Perturbative}, for both the  boson stars and  black holes. 

Before proceeding, we recall that the metric ansatz (\ref{eq:metricansatz}) has the residual gauge freedom \eqref{eq:residual}. We will use this freedom to optimize the numerical construction of our  solutions. In this computational task, we either assume $\lambda$ to be such that $\Omega(\infty)=0$ (the physical gauge, as explained in subsection \ref{sec:BCs}) or $\omega = 0$. Note that the latter necessarily implies that the boundary metric is rotating, which corresponds to a rather unphysical situation. Any calculations developed in this gauge, should (and will) therefore be recast in a gauge where $\Omega(\infty)=0$, which in turn implies $\omega \neq 0$. It turns out that these two gauges  are not equivalent numerically, in the sense that convergence is achieved at different rates in different regions of the parameter space whether we use one or the other.

In constructing both the rotating boson stars and the rotating hairy black holes we use a numerical method which relies on a standard relaxation procedure, combined with spectral methods defined on a Chebyshev grid. In order to solve the coupled system of differential equations (\ref{eqs:EOM}), we first discretize Eqs.~(\ref{eq:EOMf}-\ref{eq:EOMP}) and use the constraint equations, Eqs.~(\ref{eq:EOMC1}-\ref{eq:EOMC2}), to gather information about the appropriate boundary conditions at asymptotic infinity and either horizon (black holes) or origin (boson stars). Once the information about the constraints is imposed on the boundary conditions \eqref{eq:asympBC},  \eqref{eq:BSoriginBC} and \eqref{eq:BHhorizonBC}, the flow of the equations of motion into the bulk ensures that the constraints are satisfied everywhere, see Eqs.~(\ref{eqs:preserve}).

A brief but incomplete  outline of this section follows. We start by presenting the growth rates of the superradiant instability for several values of the azimuthal quantum number $m$. Not surprisingly, we will find that the onset of this instability signals, in a phase diagram, a merger or bifurcation line that connects the MP-AdS black holes with the rotating hairy black holes (one merger line for each $m$). This conclusion is firmly established once we compare the $m=1$ threshold instability curve with the hairy black hole curve we obtain independently when we send the condensate to zero in our hairy black hole solutions. We find that the most unstable mode is the one with $m=1$. As discussed previously, the cohomogeneity-1  ansatz \eqref{eq:metricansatz}, \eqref{eq:scalaransatz} is appropriate to study the non-linear solutions that bifurcate from the $m=1$ merger line. We thus use this  ansatz to construct the rotating boson stars and rotating hairy black holes, this time numerically and for any values of the energy and angular momentum. 
This section will culminate with the presentation of a full phase diagram of solutions where we gather the information on the $m=1$  hairy solutions (both numerical and perturbative results) and on the MP-AdS black hole. The hairy solutions we present here are associated with the scalar superradiant instability. However, MP-AdS black holes are also superradiant unstable against gravitational perturbations. We will also study the growth rate of this gravitational instability (the onset of the instability was first studied in \cite{Kunduri:2006qa}), and find that the properties of this instability are similar to those of the scalar instability.

\subsection{\label{subsec:lemj}Linear expectations}
Here, we intend to provide three distinct properties about rotating hairy black holes solutions that branch off the MP-AdS black holes. First, we determine the \emph{locus} in the moduli space of solutions where the hairy black holes merge with the MP-AdS family. Next, we investigate the growth rate of the superradiant instability associated with the scalar field $\vec{\Pi}$, in the background of a MP-AdS black hole. Finally, 
we will find that the properties of this scalar instability are similar to those of the purely gravitational superradiant instability. 

The equation we want to solve is that of a massless scalar field doublet, $\vec{\Pi}$, in the background of the MP-AdS solution, described in section \ref{sec:MP},
\begin{equation}
\nabla^2 \vec{\Pi} = 0.
\label{eq:BOXL}
\end{equation}
Because we would like to compare the growth rates associated with different values of the azimuthal quantum number $m$, we need to generalize (\ref{eq:scalaransatz}). Even though we did not manage to extend the inclusion of $m\geq 2$ modes in the construction of the non-linear solutions,  at the linear level this generalization is easily achieved:
\begin{equation}
\vec{\Pi} = \Pi(r)\, e^{-i\,\omega\,t+i\,m \,\psi}
\left[
\begin{array}{c}
\sin\left(\frac{\theta}{2}\right)^{|m|} \, e^{-i\frac{m\,\phi}{2}}
\\
\\
\cos\left(\frac{\theta}{2}\right)^{|m|} \, e^{i\frac{m\,\phi}{2}}
\end{array}
\right].
\label{eq:EMPLA}
\end{equation}
Inserting this expression in ({\ref{eq:BOXL}}), we obtain the following ordinary linear differential equation\footnote{This expression matches expression (33) in \cite{Kunduri:2006qa}, for $\mu = k = 0$, $N = \sigma = 1$ and $\Psi = r^{3/2}\sqrt{h} \Pi$.}:
\begin{equation}
\frac{(r^3f\,\Pi')'}{r^3}+\left[\frac{h(\omega-m\,\Omega)^2}{f}-\frac{2 |m|}{r^2}-\frac{m^2}{r^2h}\right]\Pi=0.
\label{eq:EMPL}
\end{equation}

In order to solve this equation we need to discuss first the  boundary conditions to be imposed. Naturally, we are interested in ingoing boundary conditions on the future event horizon, 
which means that close to the horizon $r=r_+$, $\Pi(r)$  has the behavior:
\begin{equation}
\Pi(r){\bigl |}_{r\to r_+} \approx  {\rm Exp} \left[-\frac{i\,\omega_{\star}\,r_+}{2}\frac{\ell ^2}{\ell ^2+2 r_+^2 \left(1-\ell ^2 \Omega _H^2\right)}\sqrt{1-\frac{r_+^2 \ell ^2 \Omega _H^2}{r_+^2+\ell ^2}}\log\left(1-\frac{r_+^2}{r^2}\right)\right] \left[a_0+\mathcal{O}\left(1-\frac{r_+^2}{r^2}\right)\right],
\end{equation}
where  $a_0$ is an arbitrary constant, and we defined
\begin{equation}
\omega_{\star} \equiv \omega-m \Omega_H\,.
\end{equation}
On the other hand, at infinity there is one normalizable mode only, for which $\Pi$ decays as
\begin{equation}
\Pi(r){\bigl |}_{r\to \infty} \approx \mathcal{O}(r^{-4})\,,
\end{equation}
like the non-linear solution.

\subsubsection{\label{subsec:merger}Merger}

As we mentioned above, the rotating hairy black holes will branch-off the MP-AdS solution for specific values of the MP-AdS angular velocity $\Omega_H $ and horizon size $r_+$. The strategy to determine this zero-mode is simple. We know that $\omega = m \Omega_H$ ($\omega_{\star} = 0$) for all hairy black holes: see discussion associated with \eqref{eq:oO} for the particular case $m=1$; the result for $m\neq 1$ follows from a similar reasoning. In particular, this statment is valid for the merger line (where the condensate vanishes), so we can choose values for both $m$ and $r_+/\ell$ and ask what is the value of $\Omega_H \ell$ for which the solution exists. Unfortunately, we did not manage to write (\ref{eq:EMPL}) as an eigenvalue problem in $\Omega_H \ell$, so we resort to a shooting procedure using a standard fourth order Runge-Kutta method. This can be easily implemented in the following way: 1) we choose a pair $(r_+/\ell,m)$ and determine the solution near the horizon and in the asymptotic infinity region using the Frobenius method; 2) then we integrate from both boundaries to a point in the middle of the integration domain where we demand the two solutions and their derivative to agree; 3) this, in turn, will only occur for a specific value of $\Omega_H\ell$, which we determine using Newton's method.

\begin{figure}[t]
\centerline{
\includegraphics[width=0.45\textwidth]{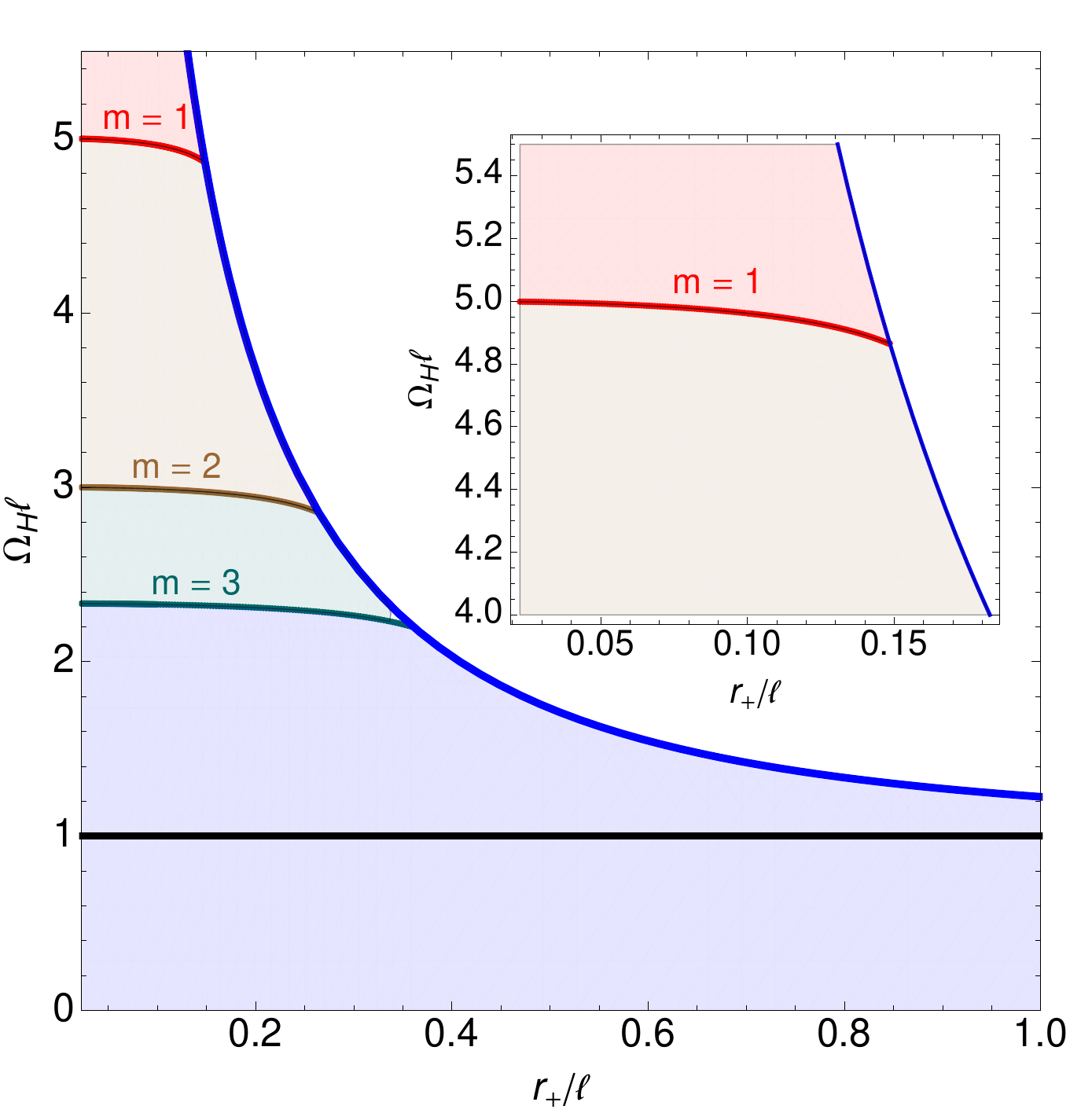}
}
\caption{Plot of angular velocity versus  horizon size. MP-AdS black holes exist in the shaded regions and its boundary corresponds to extremal black holes. Curves of $\Omega_H(r_+,m)$ corresponding to the merger lines are displayed for different values of $m$. Above each line, black holes are unstable to perturbations with that $m$. Below $\Omega_H \ell =1$, black holes are stable. The inset plot zooms in the region where the $m = 1$ merger exists.}
\label{fig:merger}
\end{figure}

Each value of $m$ labels a different merger line. We are primarily interested in the $m=1$ mode because we will find that these are the most unstable, and the associated black holes are those we construct at the full non-linear level. This is the first step to determine the full phase-space of solutions, since it provides a rough estimate of which energy and angular momentum scales we need to sweep. The results are illustrated in Fig.~\ref{fig:merger}. We see that $\Omega_H \ell$ starts being five, for very small black holes, in agreement with the analytical results for the radial mode $k=0$ of the previous section, and slowly decreases to $\Omega_H \ell \approx 4.8652$, for extremal MP-AdS. These results are in very good agreement with the analysis performed in \cite{Kunduri:2006qa}. The $m = 1$ merger line, here plotted in the $(r_+/\ell,\Omega_H\ell)$ plane, can equivalently be expressed in terms of the energy and angular momentum of the MP-AdS solution using Eqs.~(\ref{eq:MPEJ}). This is the representation we will use in sections \ref{subsec:rhbhj} and \ref{subsec:cpdj}.

The other two merger lines in Fig.~\ref{fig:merger} correspond to the merger lines of the $m =2$ and $m = 3$ modes. They share most of the characteristics of the $m = 1$ mode, except that the critical angular velocity above which small black holes are unstable changes. For  $m = 2$ this happens around $\Omega_H \ell = 3$ while for  $m = 3$, $\Omega_H \ell = 7/3$, in agreement with \cite{Kunduri:2006qa}. Also, the higher the value of $m$, the larger the unstable MP-AdS black holes can be, because the unstable region extends all the way up to extremality. The limiting value of $\Omega_H$ as $r_+ \to 0$ is directly related to the modes of a massless scalar field in $AdS_5$. The lowest radial mode with angular dependence  $m$ (\ie having eigenvalue $m(m+2)$ for the Laplacian on $S^3$) has frequency $\omega \ell= 4 + m$ \cite{Burgess:1984ti}. It is thus invariant under the Killing vector $\partial_t + (\omega/m) \partial_\psi$. If we want to add a small black hole, it must have angular velocity $\Omega_H \ell= \omega\ell/m = 1 + 4/m$. This reproduces the values $5,3,7/3$ that we find for $m=1,2,3$. The point is simply that the mode of the scalar field in $AdS_5$ is the marginal mode we are looking for when the black hole is very small.

When $\Omega_H \ell \le 1$ the MP AdS black holes do not have a superradiant instability for any $m$. This is because the Killing field $\partial_t + \Omega_H \partial_\psi$ remains timelike everywhere outside the horizon. Using this symmetry one can show that there is no superradiance for any field satisfying the dominant energy condition \cite{Hawking:1999dp}.

For values of the angular velocity above the merger lines, we expect the MP-AdS solution to be unstable. This is precisely what we are going to confirm  in the next subsection.

\subsubsection{\label{subsec:growth}Growth rates}

At this point, we know  exactly where the rotating hairy black holes branch-off the MP-AdS solution. Now, we want to quantify how unstable the MP-AdS black holes are. For this we need to provide the growth rates of the superradiant instability which afflicts the MP-AdS black holes. Namely, we need to go further in solving (\ref{eq:EMPL}) and consider solutions for which $\omega \neq m\Omega_H$. In order to work with analytic functions only, we define the auxilary function $q$ as
\begin{equation}
\Pi(r) = {\rm Exp}\left[-\frac{i\,\omega_{\star}\,r_+}{2}\frac{\ell ^2}{\ell ^2+2 r_+^2 \left(1-\ell ^2 \Omega _H^2\right)}\sqrt{1-\frac{r_+^2 \ell ^2 \Omega _H^2}{r_+^2+\ell ^2}}\log\left(1-\frac{r_+^2}{r^2}\right)\right] \left(\frac{\ell}{r}\right)^4q\left(1-\frac{r_+^2}{r^2}\right),
\end{equation}
where we have implicitly performed the change of variables $y = 1-r_+^2/r^2$. Note that $y$ is a compact variable taking values in the unit interval. By construction, $q(y)$ is an analytic function both at the future event horizon and at asymptotic infinity. Substituting this expression into (\ref{eq:EMPL}), gives the following quadratic Sturm-Liouville problem
\begin{equation}
\mathcal{L}_0 q-\omega_{\star} \mathcal{L}_1 q-\omega_{\star}^2 \mathcal{L}_2 q=0,
\label{eq:eigen}
\end{equation}
where each $\mathcal{L}_i$ is a second order differential operator in $y$, independent of $\omega_{\star}$. The procedure is now clear: we first transform $q(y)$ into a vector, say $\vec{q}$, where each of its entries is given by $q(y)$ evaluated at the Chebyshev points $y_j$. In this discretization scheme, the operators $\mathcal{L}_i$ are matrices, say $L_i$, that act on $\vec{q}$. Consequently, equation (\ref{eq:eigen}) reduces to a generalized eigenvalue problem in $\omega_{\star} $,
\begin{equation}
\left(\left[
\begin{array}{cc}
-L_1 & L_0
\\
\mathbb{I} & 0
\end{array}
\right]-\omega_{\star}
\left[
\begin{array}{cc}
L_2 & 0
\\
0 & \mathbb{I}
\end{array}
\right] \right)
\left[
\begin{array}{c}
\omega_{\star}\, \vec{q}
\\
\vec{q}
\end{array}
\right]=0,
\label{eq:eigenvaluesquadratic}
\end{equation}
where $\mathbb{I}$ is the identity matrix. This linear system can be easily solved by the in-built \emph{Mathematica} routine \verb Eigensystem . In order to generate the growth rates, we can choose values for $r_+/\ell$, $\Omega_H \ell$ and $m$ and then compute $\omega_{\star}$ using (\ref{eq:eigenvaluesquadratic}). Note also that there are no reality conditions on $\omega_{\star}$. Therefore, shooting methods would certainly face serious challenges in solving (\ref{eq:EMPL}) for generic values of $r_+/\ell$, $\Omega_H \ell$ and $m$, because $\omega_{\star}$ is generically complex.

\begin{figure}[t]
\centerline{
\includegraphics[width=0.9\textwidth]{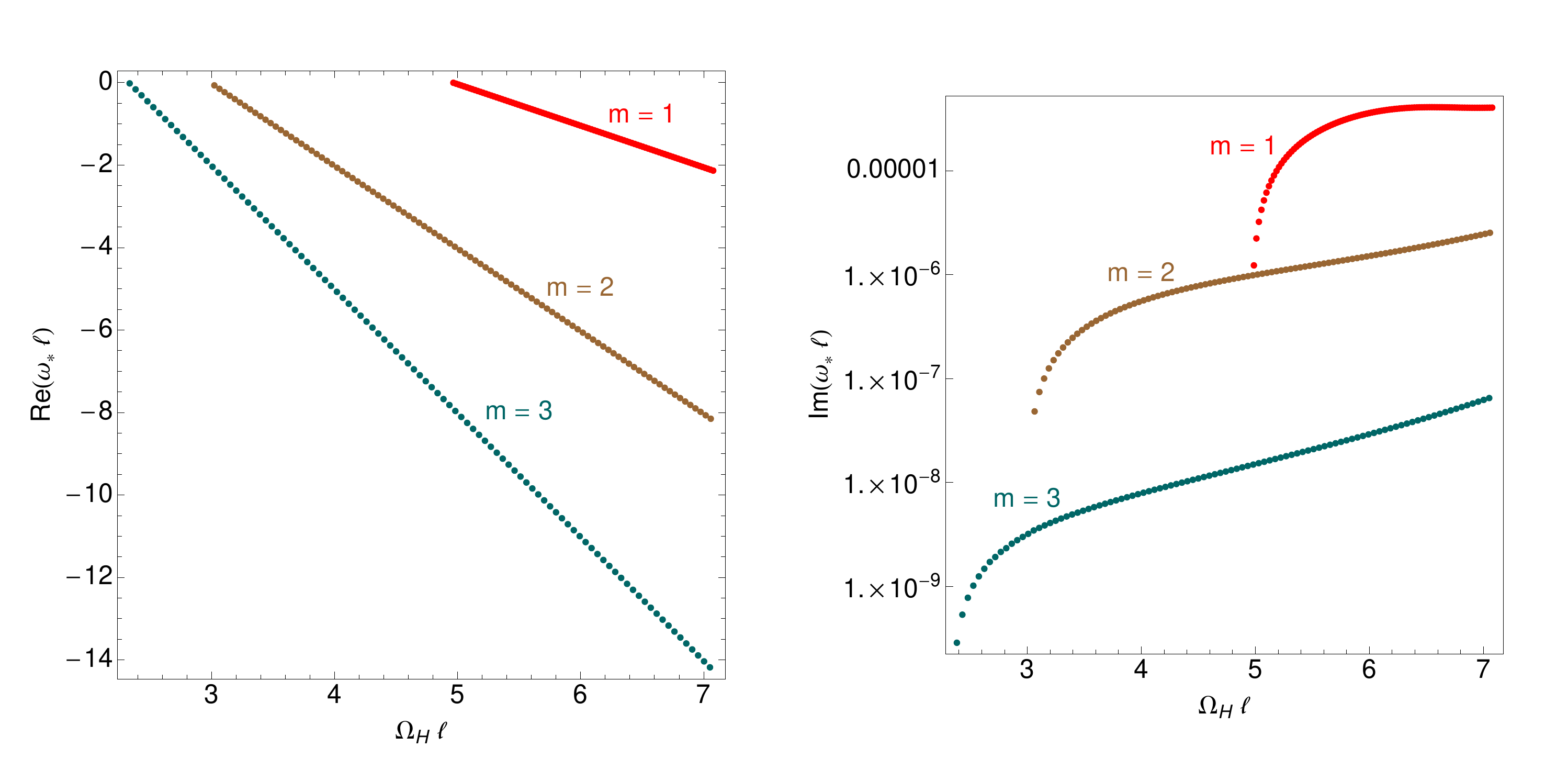}
}
\caption{The real (left panel) and imaginary (right panel) parts of the perturbation frequency as a function of angular velocity, for a constant value of the MP-AdS horizon size, $r_+/\ell = 0.1$. Different curves correspond to different values of $m$.  Note that the plot on the right  uses a logarithmic scale.}
\label{fig:growth}
\end{figure}

The results we obtained were generated along a line of constant MP-AdS horizon size, $r_+/\ell = 0.1$, and are illustrated in Fig.~\ref{fig:growth}. As a  check, we note that the value of $\Omega_H \ell$ for which $\mathrm{Im}(\omega \ell)=0$ is precisely the zero-mode described in the Merger subsection \ref{subsec:merger}. The concordance between the onset critical angular velocity obtained using the shooting method of the previous subsection and the spectral methods of this subsection is excellent; in particular, they agree with each other within a $0.001\%$ error.

The first striking property about the growth rates in Fig.~\ref{fig:growth} is that they are all small, leading to a secular type instability. This is reminiscent of what happens in the case of a massive scalar field in the background of the asymptotically flat Kerr black hole \cite{Detweiler:1980uk}-\cite{Dolan:2007mj}. Also, as $m$ increases, the growth rates become increasingly smaller. In fact, each unit of $m$ seems to change the growth rate by two orders of magnitude. These results indicate that the $m = 1$ mode is the most unstable mode, when coexisting with the remaining ones, and that modes with higher $m$ will become increasingly less unstable, in the sense that their maximum growth rates become increasingly smaller. The fact that the real part of $\omega_\star$ is negative just means that $0<Re(\omega)<m \Omega_H$, which is a well know property of the superradiant instability \cite{bardeen}-\cite{press}.

\subsubsection{Linear gravitational superradiance}

Before proceeding to the numerical construction of the rotating boson stars, we would like to compare the scalar instability described in the previous subsections, with the superradiant instability in a purely gravitational setup. It is well know that Kerr-AdS black holes, and their natural extension to higher dimensions \cite{Gibbons:2004uw}, exhibit the superradiance instability \cite{Kunduri:2006qa,Hawking:1999dp}, \cite{Cardoso:2004hs}-\cite{Uchikata:2009zz}. In this subsection, we will argue that much can be understood about this instability by inspecting our (more tractable) cohomegeneity-1 system. In particular, we will show that the purely gravitational instability has a spectrum which is very similar to the one in Fig.~\ref{fig:growth}.

To study this problem, we would need to perturb the five-dimensional MP-AdS line element described in section \ref{sec:MP}. This, however, turns out to lead to a complicated set of coupled ordinary differential equations \cite{Dias:2010eu}, which will be studied elsewhere. Here we take a simpler route which suffices for our purposes. The reason why it is so challenging to perturb the five-dimensional MP-AdS metric is related with the fact that no tensor harmonics exist on the $\mathbb{CP}_1 \equiv S^2$ base space, which in turn implies that we have to deal with either vector-type or scalar-type perturbations.  In order to avoid this conundrum, we change gears and consider the seven-dimensional counterpart of this geometry, which was studied in \cite{Kunduri:2006qa}. The $d=7$ equal angular momenta MP-AdS is still cohomogeneity-1 (see subsection \ref{sec:MP}) and its base space is $\mathbb{CP}_2$, where tensor harmonics do exist. This is still a reasonable comparison because we do not expect the physics involved here to depend on the number of dimensions. The equation for the tensor-mode was derived in \cite{Kunduri:2006qa}, and can be written as:
\begin{subequations}
\begin{equation}
-\frac{f}{\sqrt{h}}\left(\frac{f}{\sqrt{h}}\Psi'\right)'+V\Psi = 0,
\end{equation}
where
\begin{equation}
V(r)=\frac{f}{r^{5/2}h^{3/4}}\left[\frac{f}{\sqrt{h}}(h^{1/4}r^{5/2})'\right]'-(\omega - m \Omega)^2+\frac{f}{r^2h}\left[\ell_1(\ell_1+4)-m^2\left(1-\frac{1}{h}\right)\right],
\label{eq:poten7D}
\end{equation}
\end{subequations}
and $f$ and $g$ are given by (\ref{eq:MPadsAUX}) with $N = 2$. Here, $\Psi$ is the master variable and $\ell_1$ is a positive integer that can take certain values that depend on $m$. We will choose the values $\ell_1 = m = 3$, which were studied in \cite{Kunduri:2006qa}. Also, for the sake of concreteness, we will only consider a line of constant MP-AdS horizon size, namely $r_+/\ell = 0.1$. The authors of \cite{Kunduri:2006qa} studied the merger line associated with the gravitational superradiance. Here, we will go further and compute also the instability growth rate.
\begin{figure}[t]
\centerline{
\includegraphics[width=0.8\textwidth]{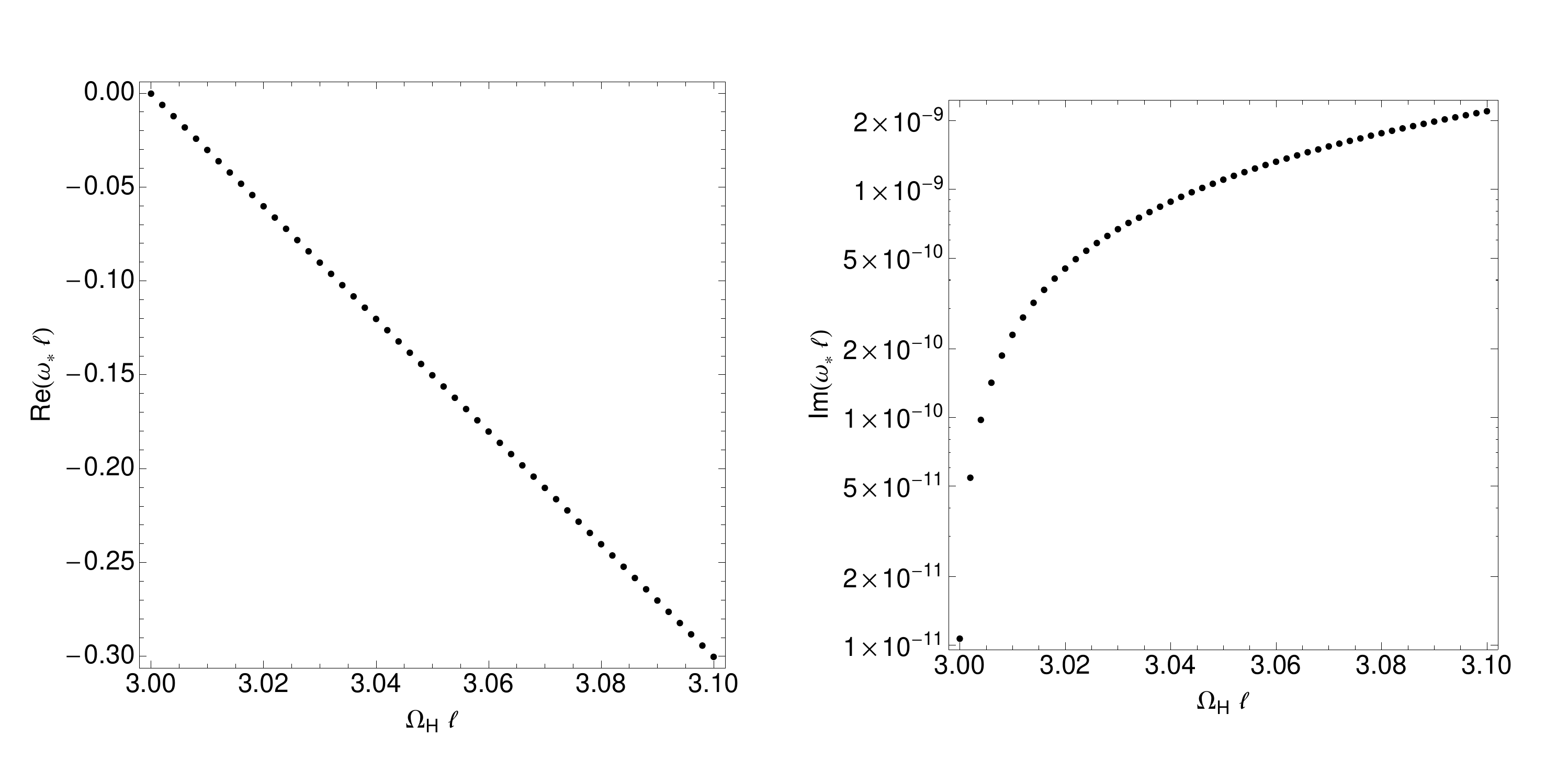}
}
\caption{The real (left panel) and imaginary (right panel) parts of the perturbation frequency as a function of angular velocity,  for the purely gravitational superradiance instability, at fixed MP-AdS horizon size $r_+/\ell = 0.1$. The mode shown has $\ell_1 = m = 3$.}
\label{fig:growthgravitational}
\end{figure}

We present the results in Fig.~\ref{fig:growthgravitational}, where we can see that the purely gravitational setup shares most of the features of the scalar instability described in the previous subsections. There is, however, a quantitative difference, namely, the instability growth rate seems to be even smaller than the one associated with the doublet complex scalar field, at least in the tensor-type sector of the perturbations. It is not clear whether this difference arises because we are comparing a five-dimensional result to a seven-dimensional result or because tensor-type perturbations are generically smaller. Another important difference, which we will address next, is that for the complex scalar field doublet we can find the endpoint of the instability in the $m = 1$ sector.  The gravitational case seems to require solving PDE's which are not of the standard elliptic type (since the Killing field becomes spacelike at large radius). A full time-dependent evolution may be needed for any type of gravitational superradiant perturbation.

\subsection{\label{subsec:rbsj}Rotating boson stars}
We now turn to the numerical construction of boson star solutions to (2.6). 
Recall that rotating boson stars form a one-parameter family of solutions. In the perturbative construction of section  \ref{sec:PertBstar}, the scalar amplitude $\epsilon$ was an appropriate quantity to parametrize the solution. This is no longer the case when we extend to larger values of the energy and angular momentum, since $\epsilon$ no longer defines uniquely the solution. A better choice is the energy density at the center of the boson star.  For the ansatz (\ref{eq:scalaransatz}), the energy momentum density at the centre of the boson star turns out to be proportional to $\Pi'(0)^2$, and as such we will parametrize our solutions by $\Pi'(0)$. 

Before proceeding, a few words about the numerical construction of these solutions are in order. Since we are using relaxation methods on a Chebyshev grid we need to guarantee that the functions we solve for are analytic in the domain of integration. In order to achieve this, we define the following auxiliary functions $\{q_f, q_h, q_\Omega, q_g, q_\Pi\}$:
\begin{gather}
f(r) = \frac{r^2}{\ell^2}+1+\frac{\ell^2\,q_f(r)}{r^2+\ell^2},\qquad h(r) = 1+\frac{\ell^4\, q_h(r)}{(r^2+\ell^2)^2},\qquad \Omega = \ell^{-1}q_\Omega(r) \nonumber \\
g(r) = 1+\frac{\ell^4\, q_g(r)}{(r^2+\ell^2)^2},\qquad \Pi(r) = \frac{r \ell^4\, q_{\Pi}(r)}{(r^2+\ell^2)^{5/2}}.
\label{eqs:spectralfs}
\end{gather}
For numerical purposes it is also easier to work in a gauge where the line element (\ref{eq:metricansatz}) is rotating at infinity, and for which $\omega = 0$. In the end we will change gauge, and set $\Omega(\infty) = 0$, thus generating a non-vanishing value for $\omega$. Finally, in using spectral methods it is convenient to compactify the integration domain, which can be readily achieved by introducing the new radial variable,
\begin{equation}
y = \frac{r^2}{r^2+\ell^2},
\end{equation}
whith $y\in(0,1)$. At $y=0$, regularity fixes the boundary conditions  \eqref{eq:BSoriginBC}, and at $y = 1$ we demand the spacetime to be asymptotically global AdS, \ie that it obeys \eqref{eq:asympBC}. These boundary conditions imply the non-trivial relations:
\begin{subequations}
\begin{gather}
q_f(0) =0,\qquad q_h(0) = 0\,,\qquad q_g'(0)-2 q_g(0)-q_h'(0)[1+q_g(0)]-\frac{2}{3}[1+q_g(0)]q_0^2=0\,,\nonumber \\
q_\Omega'(0) - \frac{q_0^2}{3}q_\Omega(0)=0,\qquad q_\Pi(0) - q_0 = 0\,,
\end{gather}
and
\begin{gather}
q_f'(1)+q_f(1)+q_g(1) =0, \qquad q_h(1)+q_g(1) = 0,\qquad q_g'(1)+2 q_g(1) = 0\,, \nonumber \\
q_\Omega'(1)=0,\qquad q_\Pi'(1) +\frac{1}{12}[25-q_\Omega(1)^2] q_\Pi(1)=0\,,
\end{gather}
\label{eqs:bcs}
\end{subequations} 
 where $q_0 = \ell\,\Pi'(0)$ should be regarded as a number that we vary as we move along the boson star family of solutions.

\begin{figure}
\centering
\subfigure[$E/\ell^2$ as a function of $\ell\, \Pi'(0)$.]{
\includegraphics[width = 0.4 \textwidth]{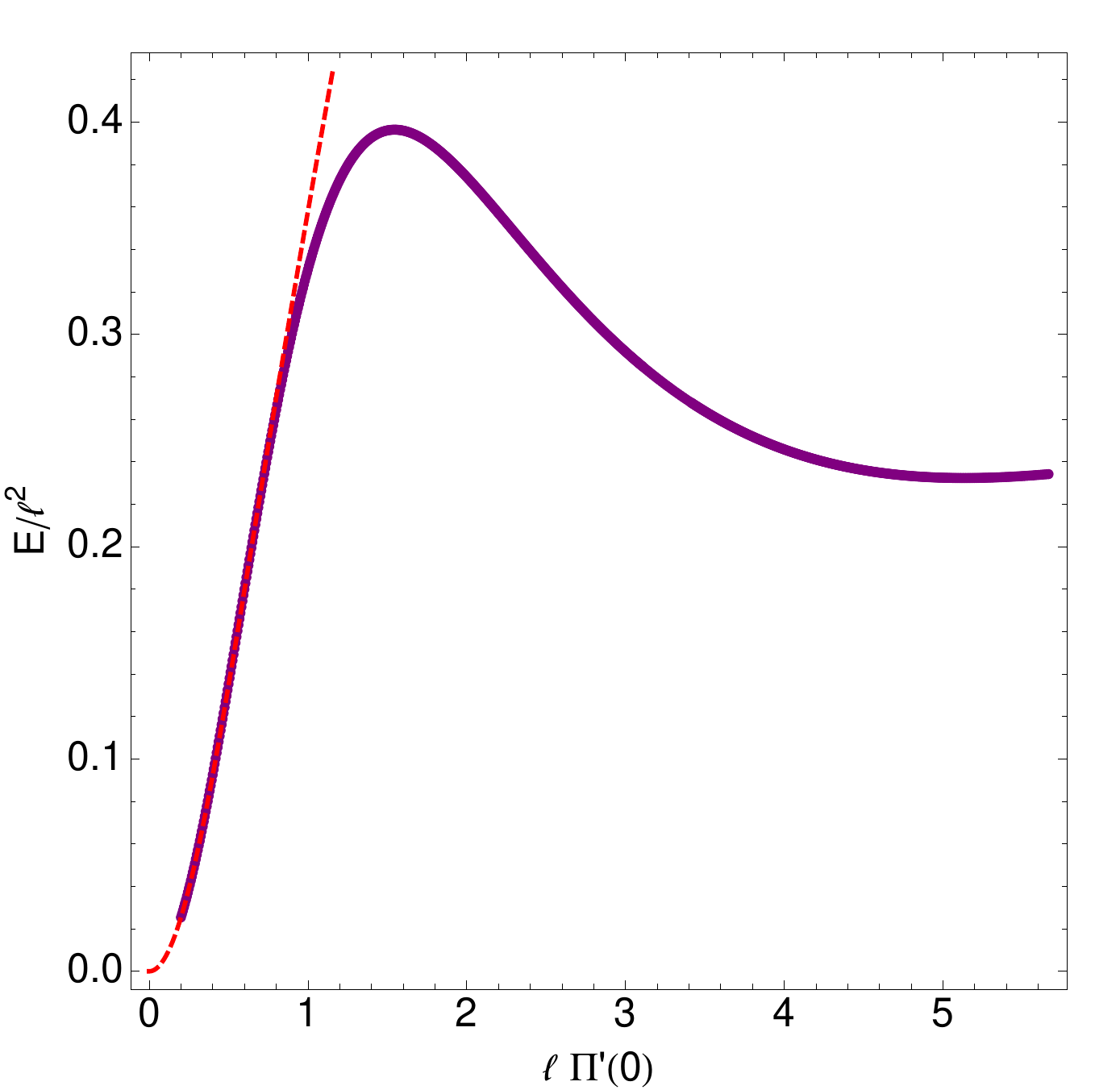}
\label{fig:plotssolitonsa}
}\hspace{1.5cm}
\subfigure[$J/\ell^3$ as a function of $\ell\, \Pi'(0)$.]{
\includegraphics[width = 0.4 \textwidth]{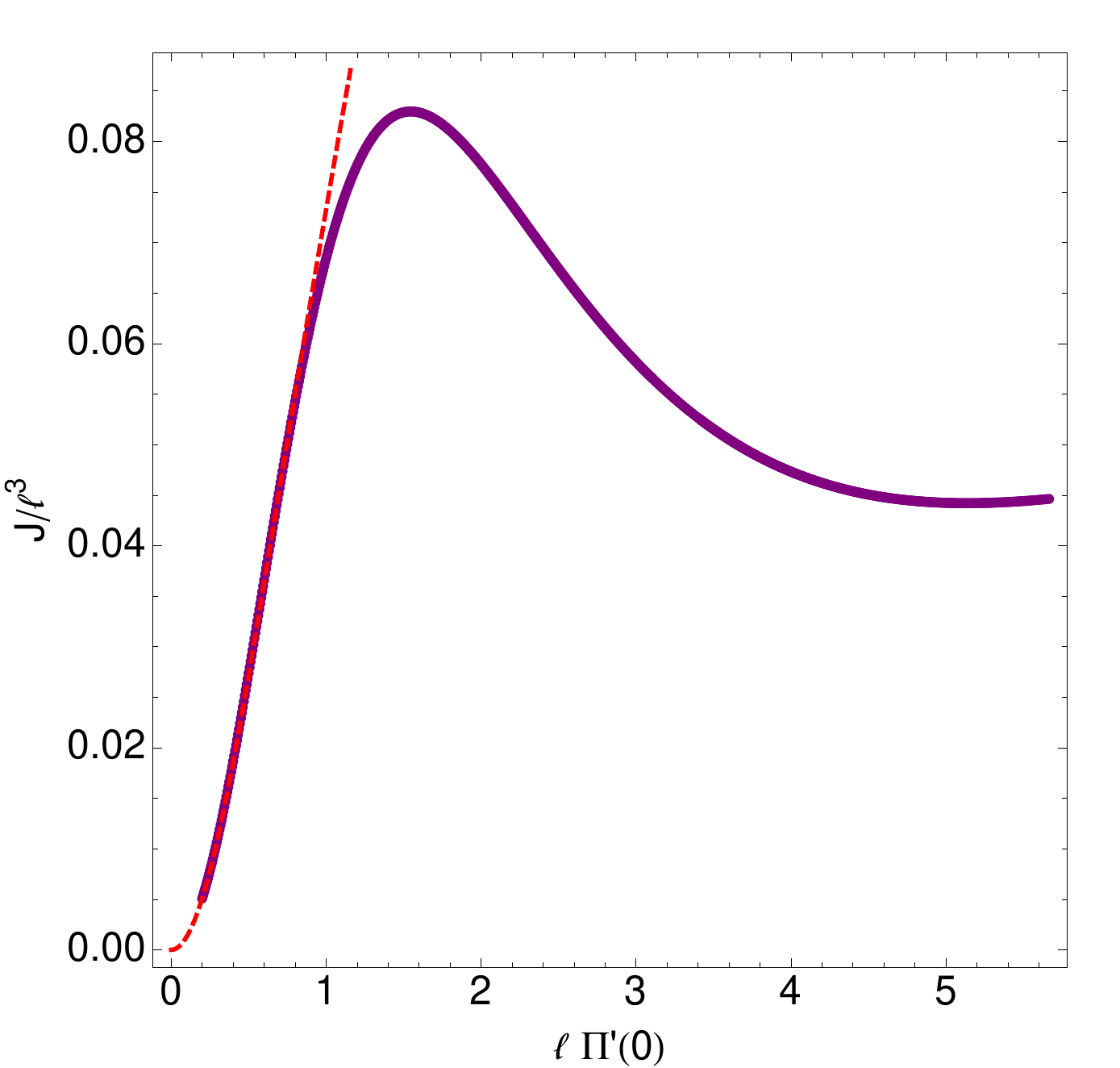}
\label{fig:plotssolitonsb}
}
\\
\subfigure[$\omega \,\ell$ as a function of $\ell\, \Pi'(0)$.]{
\includegraphics[width = 0.4 \textwidth]{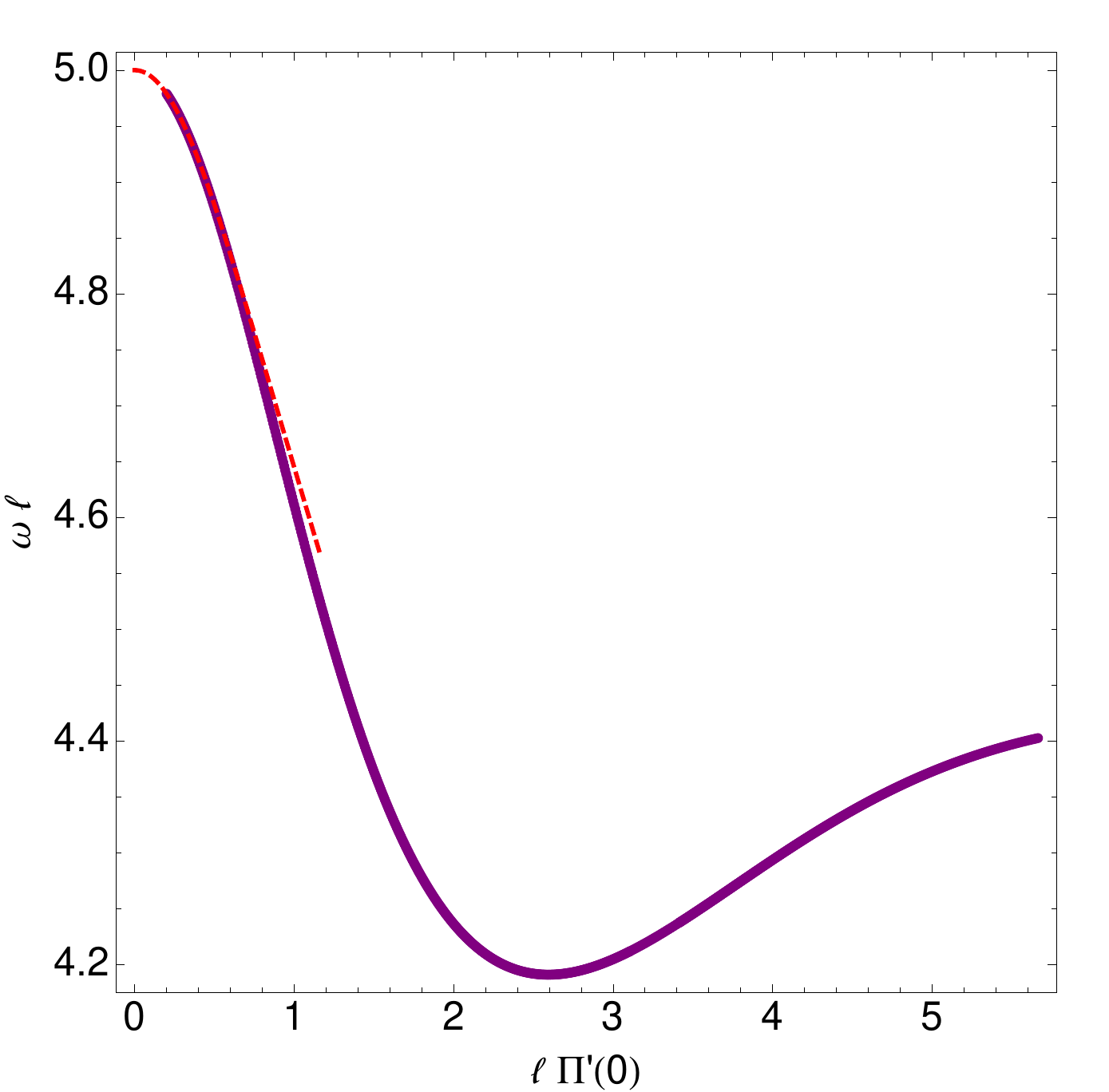}
\label{fig:plotssolitonsc}
}\hspace{1.5cm}
\subfigure[$\epsilon$ as a function of $\ell\, \Pi'(0)$.]{
\includegraphics[width = 0.4 \textwidth]{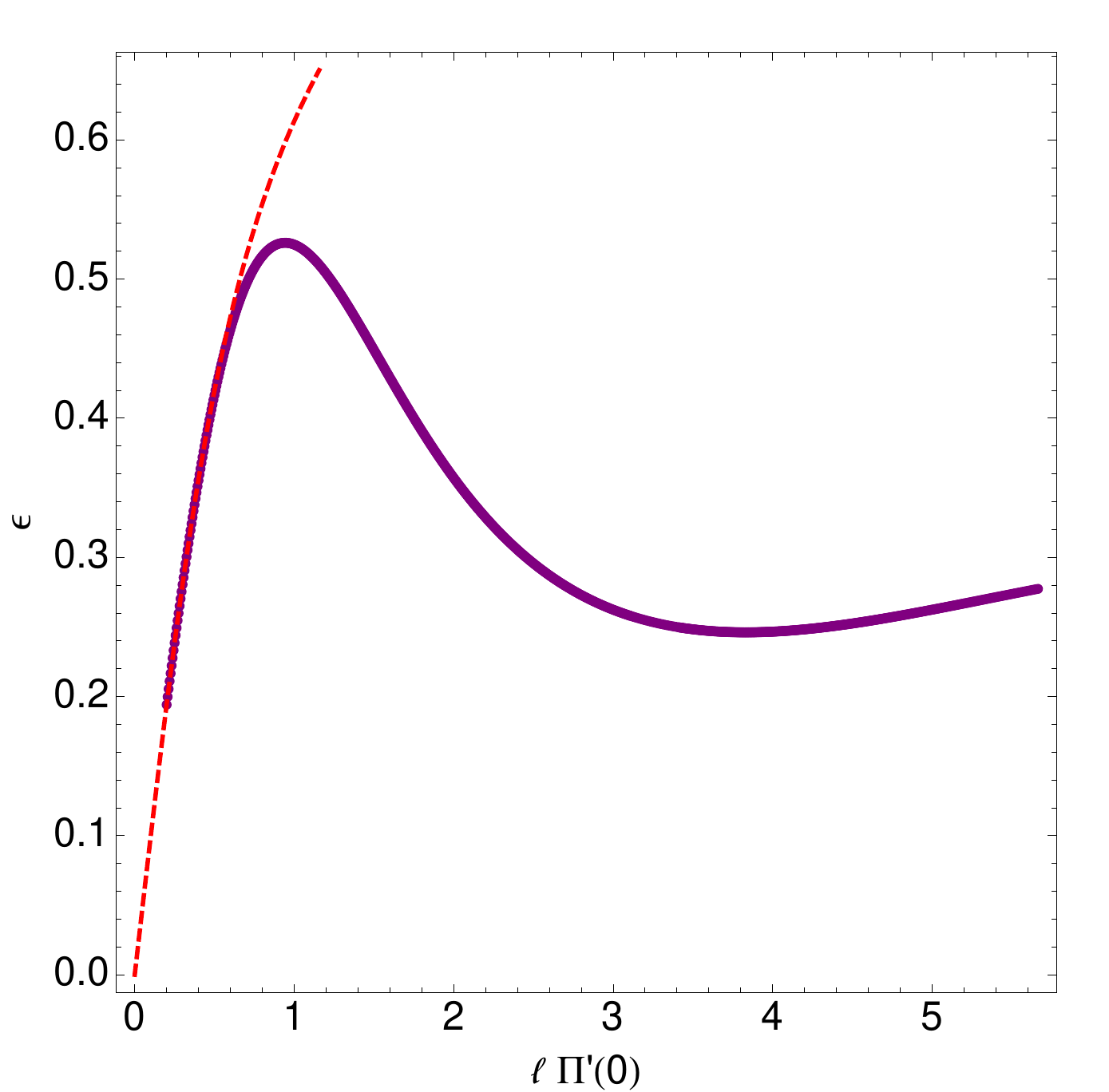}
\label{fig:plotssolitonsd}
}
\\
\subfigure[$\epsilon$ as a function of $\omega\,\ell$.]{
\includegraphics[width = 0.4 \textwidth]{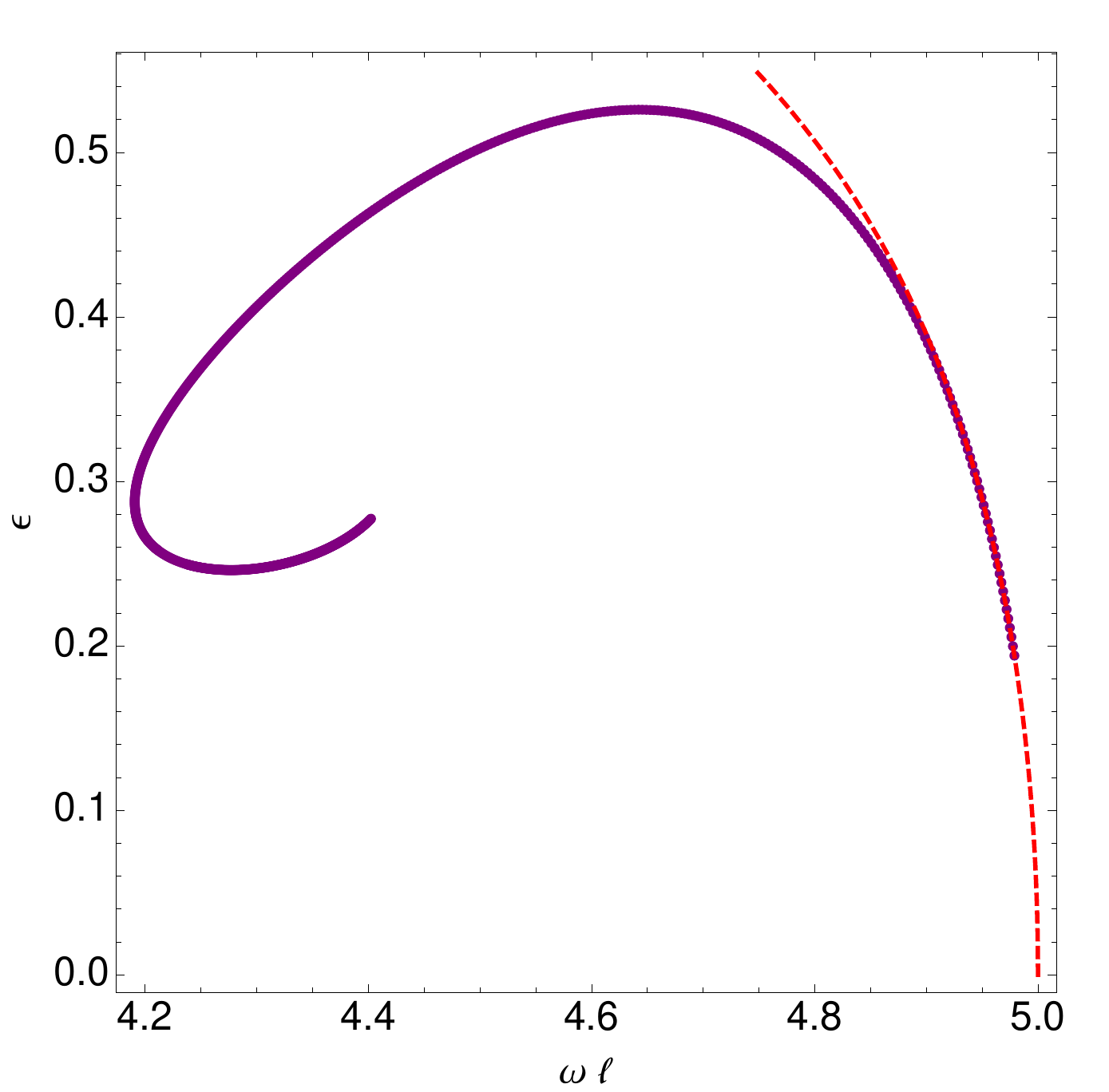}
\label{fig:plotssolitonse}
}\hspace{1.5cm}
\subfigure[$K_s$ as a function of $\omega\,\ell$.]{
\includegraphics[width = 0.4 \textwidth]{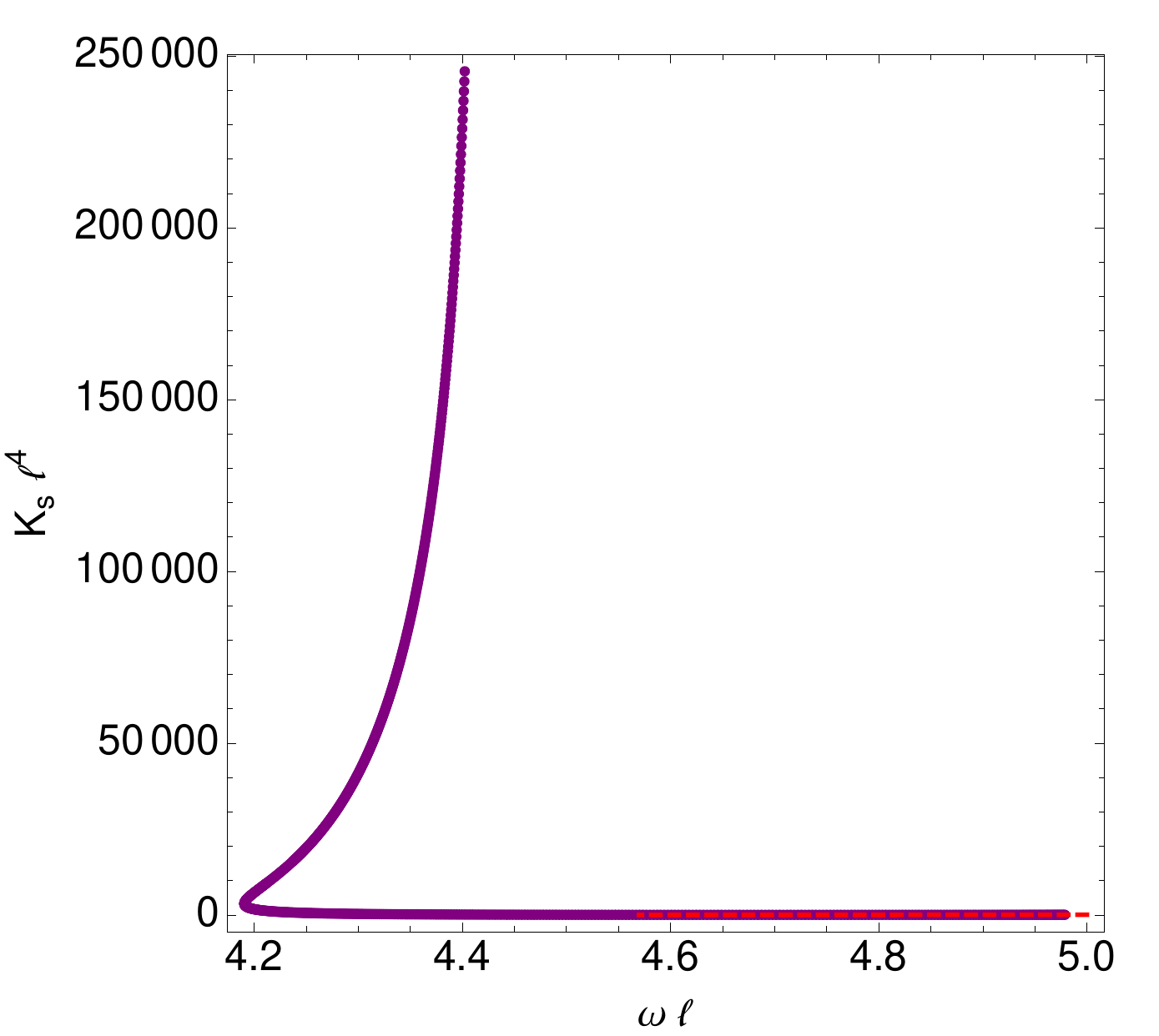}
\label{fig:plotssolitonsf}
}
\caption{\label{figs:plotssolitons}Some properties of rotating  boson stars. The solid curves represent the exact solution, whereas the dashed lines show the analytical estimates of section \ref{sec:PertBstar}.}
\end{figure}

In Figs.~\ref{fig:plotssolitonsa}-\ref{fig:plotssolitonsd} we show how the dimensionless energy, angular momentum, frequency and asymptotic scalar field amplitude change as we increase the energy density. The solid lines represent the exact numerical solution, whereas the dashed lines show the analytical estimates of section \ref{sec:PertBstar}. The physics involved here is in many respects similar to the flat space counterpart \cite{Hartmann:2010pm}. We see that there is a sign of a damped oscillatory behavior, similar to the charged solitonic solutions in $AdS_5\times S_5$ found in \cite{Bhattacharyya:2010yg}. This behavior occurs around a non-zero critical value of the energy, angular momentum, scalar frequency and scalar amplitude, respectively. We will label these values by $\{E_c,J_c,\omega_c,\epsilon_c\}$, respectively. The analytic calculations developed in section \ref{sec:PertBstar} are in very good agreement with our numerical data in their regime of validity, \emph{i.e.} for small enough $\Pi'(0)$. Note however that the perturbative analysis does not capture the interesting phenomenon of the damped oscillations.

There is a simple consequence of the damped oscillatory behavior. Consider for example a plot of the condensate strength $\epsilon$ as a function of the scalar field frequency $\omega$ (we could alternatively choose other quantities, but the general picture would be unchanged). Due to the damped oscillations of Figs.~\ref{fig:plotssolitonsc}-\ref{fig:plotssolitonsd}, we can view $\epsilon$ and $\omega$ as coordinates of a two dimensional damped harmonic oscillator. Typical motion in the  $\epsilon - \omega$ plane is a spiral and that is indeed what we find  in Fig.~\ref{fig:plotssolitonse}.

It is now clear that the line of boson stars ends at a specific energy and angular momentum, namely at $E_c$ and $J_c$, respectively. As we observe in the top two panels of Fig.~\ref{figs:plotssolitons}, we would need an infinite energy density or, equivalently, an infinite $\Pi'(0)$ to reach the critical values $E_c$ and $J_c$. A possible detection of this singular behavior can be readily observed by computing the Kretschmann scalar $K_s$ at the center of our geometry. This quantity, in terms of the  spectral functions defined in (\ref{eqs:spectralfs}), is simply given by
\begin{equation}
\ell^4 K_s \equiv \ell^4 R^{abcd}R_{abcd}{\bigl |}_{r = 0} = 40-32\,q_0^2+\frac{32}{3}\,q_0^4+96\,q_h'(0)^2.
\end{equation}
Fig.~\ref{fig:plotssolitonsf} shows how the Kretschmann scalar evolves with $\omega\ell$. In particular, it seems to diverge  at $\omega_c\ell$, in accordance with our expectations. 

We expect an infinite number of  damped oscillations in the plots of Figs.~\ref{fig:plotssolitonsa}-\ref{fig:plotssolitonsd}, as the system evolves along the curves towards the critical values  $\{E_c,J_c,\omega_c,\epsilon_c\}$. As a consequence, the associated spirals are predicted to continue \emph{ad aeternum} in Fig.~\ref{fig:plotssolitonse}. In particular, this means that for a line of constant $\omega = \omega_c$, we expect to find an infinite number of solutions with different values of $J$, say. The reason why we do not see the spirals continuing further inwards in our plots (\ie why we do not observe the other arms of the spiral) can be well understood by inspecting what is happening to the curvature invariant $K_s$. Indeed,  Fig.~\ref{fig:plotssolitonsf} clearly shows that the curvature is quickly diverging as we move inward along the spiral. Unfortunately, this makes it considerably harder to find numerically the solutions.

As a check on our numerics, we have verified that our solutions satisfy the first law  \eqref{eq:BS1stlaw}  with only  $0.005\%$ error. This provides a  measure of the error in determining these solutions.

\subsection{\label{subsec:rhbhj}Rotating hairy black holes}

The perturbative construction of section \ref{sec:PertBHs} necessarily explores only the small mass and angular momentum region of the parameter space of hairy black holes. In this section, we want to go beyond this perturbative construction, and construct  hairy rotating black hole solutions with a single Killing vector field whose energy and angular momentum are not restricted to be small.
The construction procedure will be very similar to the one followed in the previous subsection for the boson stars, except that the black hole solutions will now depend on an extra parameter, namely the black hole horizon size $r_+$.  It is tempting to parameterize our solutions by $r_+$ and $\Omega_H$. Indeed these two quantities uniquely label the MP AdS black holes. However, we have already seen that for the boson stars $\omega$ is not single valued, and we know that the hairy black holes must have $\Omega_H = \omega$. This strongly suggests that the hairy black holes will not be uniquely described by $r_+$ and $\Omega_H$, and this is indeed the case.

It turns out that there is a simple geometric way to uniquely label the hairy black holes. These five-dimensional black holes have squashed $S^3$ horizons, which can be viewed as $S^1$ bundles over $S^2$. A natural parametrization of our solutions is just given in terms of the size of each of these spheres, which we will label by $r_1$ and $r_2$, respectively. In terms of the metric ansatz (\ref{eq:metricansatz}), these are given by 
\begin{equation}
r_1 =  r_+\,\sqrt{h(r_+)}\,,\qquad r_2 = r_+\,.
\end{equation}
We will explore this two-dimensional parameter space by fixing various values of $r_2$ and slowly increasing $r_1$.

Numerically, this is difficult to implement directly. To input $r_1$, we need to work in a gauge where $\omega = 0$. However, we then find that ``half way" through our computations, the code develops a numerical instability. This can be cured by changing to a second (different) code. In this new code, we work in a gauge where $\Omega(\infty) = 0$ and input $\omega$ directly. This makes the construction of these black holes challenging, because each line of constant $r_2/\ell$ needs at least two independent codes to be generated. In addition to this, all the black holes we found are rather small, namely $r_+/\ell \lesssim 0.14$, which means that the curvature scales involved in the construction of these solutions are large. For these reasons, we find appropriate to use a rather dense Chebyshev grid with no less than two hundred points.

For numerical reasons, it is convenient to introduce the new functions $\{q_f,q_h,q_g, q_\Pi, q_\Omega\}$ which are adapted to our boundary conditions \eqref{eq:asympBC} and \eqref{eq:BHhorizonBC}:
\begin{gather}
f(r) =\left(\frac{r^2}{\ell^2}+1\right)\left(1-\frac{r_+^2}{r^2}\right)\,q_f(r)\,,\quad h(r) = q_h(r)\,,\quad g(r) = q_g(r)\,,   \quad \Pi(r) = \frac{r \ell^4\, q_{\Pi}(r)}{(r^2+\ell^2)^{5/2}}\,, \\
 \Omega^{(1)}(r) = \frac{1}{\ell}\left(1-\frac{r_+^2}{r^2}\right)\,q_ \Omega(r)\,,\:\: \hbox{if gauge $\omega=0$}\,; \qquad  \Omega^{(2)}(r) = \frac{\ell^3}{(r^2+\ell^2)^2}\,q_\Omega(r)\,,\:\:  \hbox{if gauge $\Omega(\infty) = 0$}.\nonumber 
\label{eqs:spectralfbh}
\end{gather}
We emphasize that $\Omega(r)$ will either be given by $\Omega^{(1)}(r)$ or $\Omega^{(2)}(r)$, whether we use a gauge where $\omega = 0$ or $\Omega(\infty) = 0$, respectively. Moreover, we use the radial variable $y = 1-r_+^2/r^2$, which takes values in the unit interval, the horizon being at $y = 0$ and asymptotic infinity at $y = 1$. For $y =1$, we demand the spacetime to be asymptotically AdS, and at $y = 0$ we require regularity. The system then obeys the boundary conditions \eqref{eq:asympBC} and \eqref{eq:BHhorizonBC}. These, in turn, demand certain non-trivial relations among the functions $\{q_f,q_h,q_g, q_\Pi, q_\Omega\}$ and their derivatives evaluated at $y = 1$ or $y =0$, four of which are similar in many respects to the boson star boundary condition relations (\ref{eqs:bcs}). However, for the black holes, these four boundary conditions cannot be expressed in such a compact form as (\ref{eqs:bcs}), and as such shall not be presented here. The remaining boundary condition imposes $\Omega(r_+) = \omega$ and ensures regularity at the black hole horizon. 

\begin{figure}
\centering
\subfigure[$E/\ell^2$ as a function of $r_1/\ell$.]{
\includegraphics[width = 0.4 \textwidth]{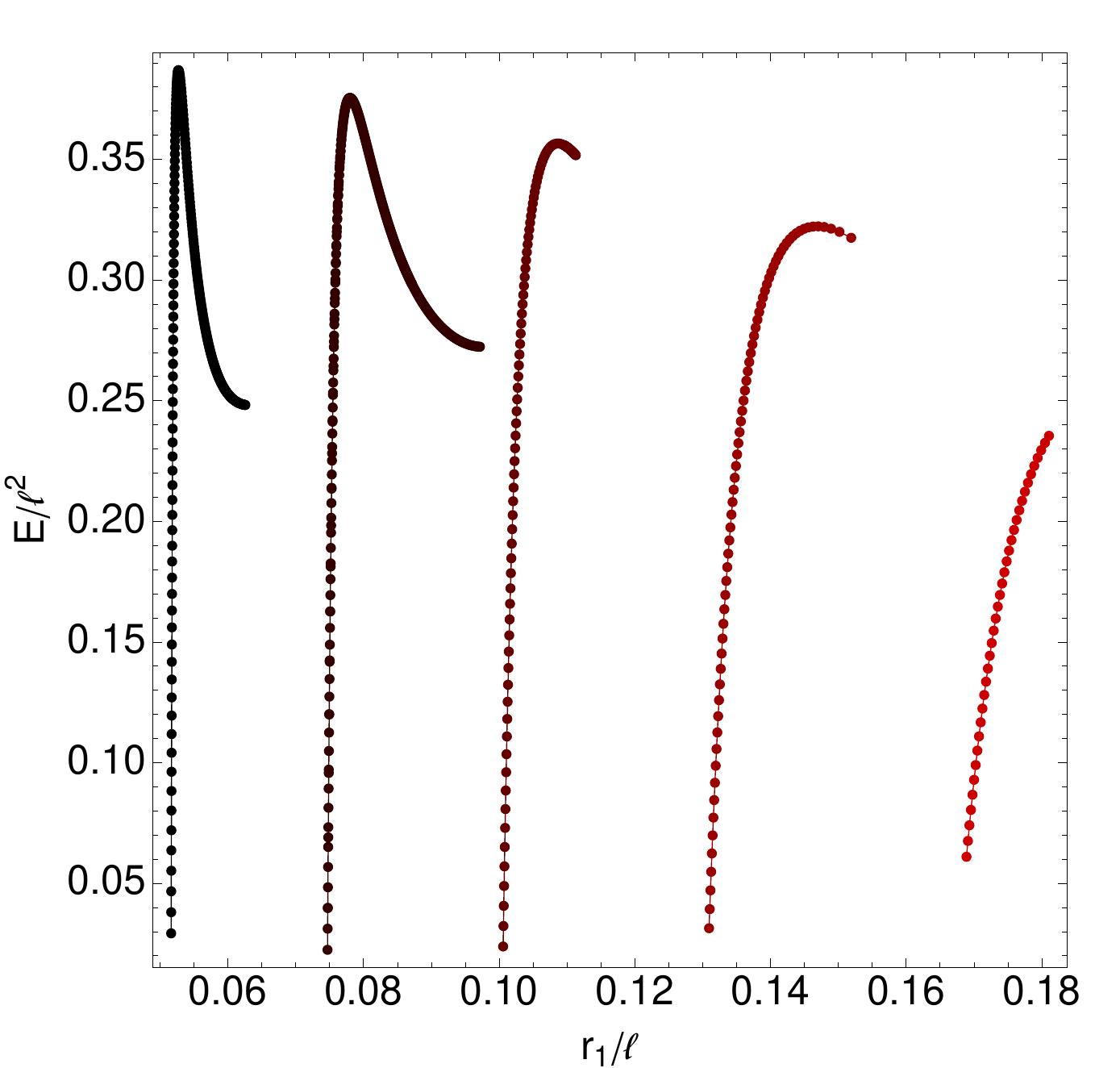}
\label{fig:ejotscvsq2a}
}\hspace{1.5cm}
\subfigure[$J/\ell^3$ as a function of $r_1/\ell$.]{
\includegraphics[width = 0.4 \textwidth]{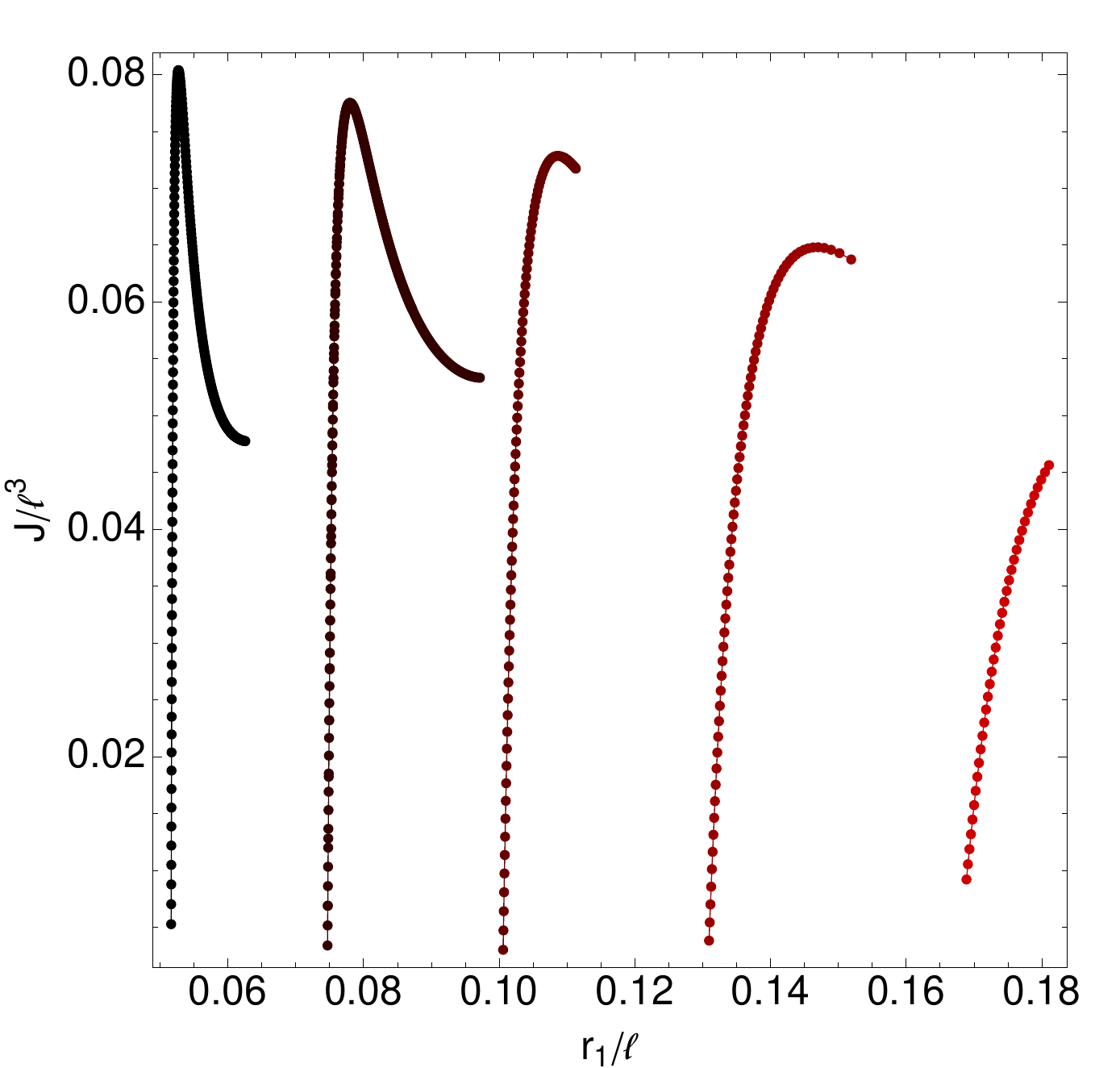}
\label{fig:ejotscvsq2b}
}
\\
\subfigure[$T_H\,\ell$ as a function of $r_1/\ell$.]{
\includegraphics[width = 0.4 \textwidth]{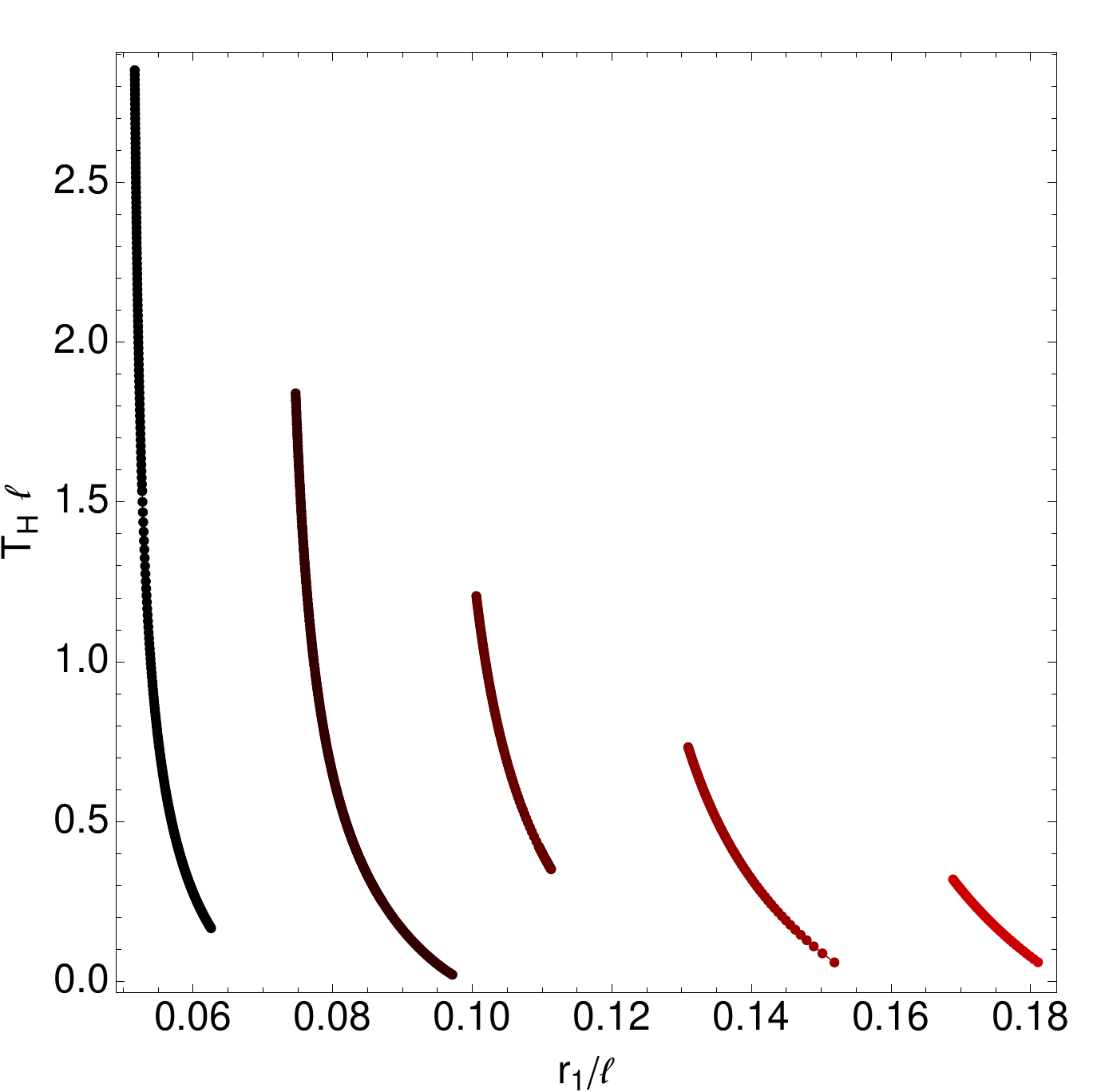}
\label{fig:ejotscvsq2d}
}\hspace{1.5cm}
\subfigure[$\Omega_H \,\ell$ as a function of $r_1/\ell$.]{
\includegraphics[width = 0.4 \textwidth]{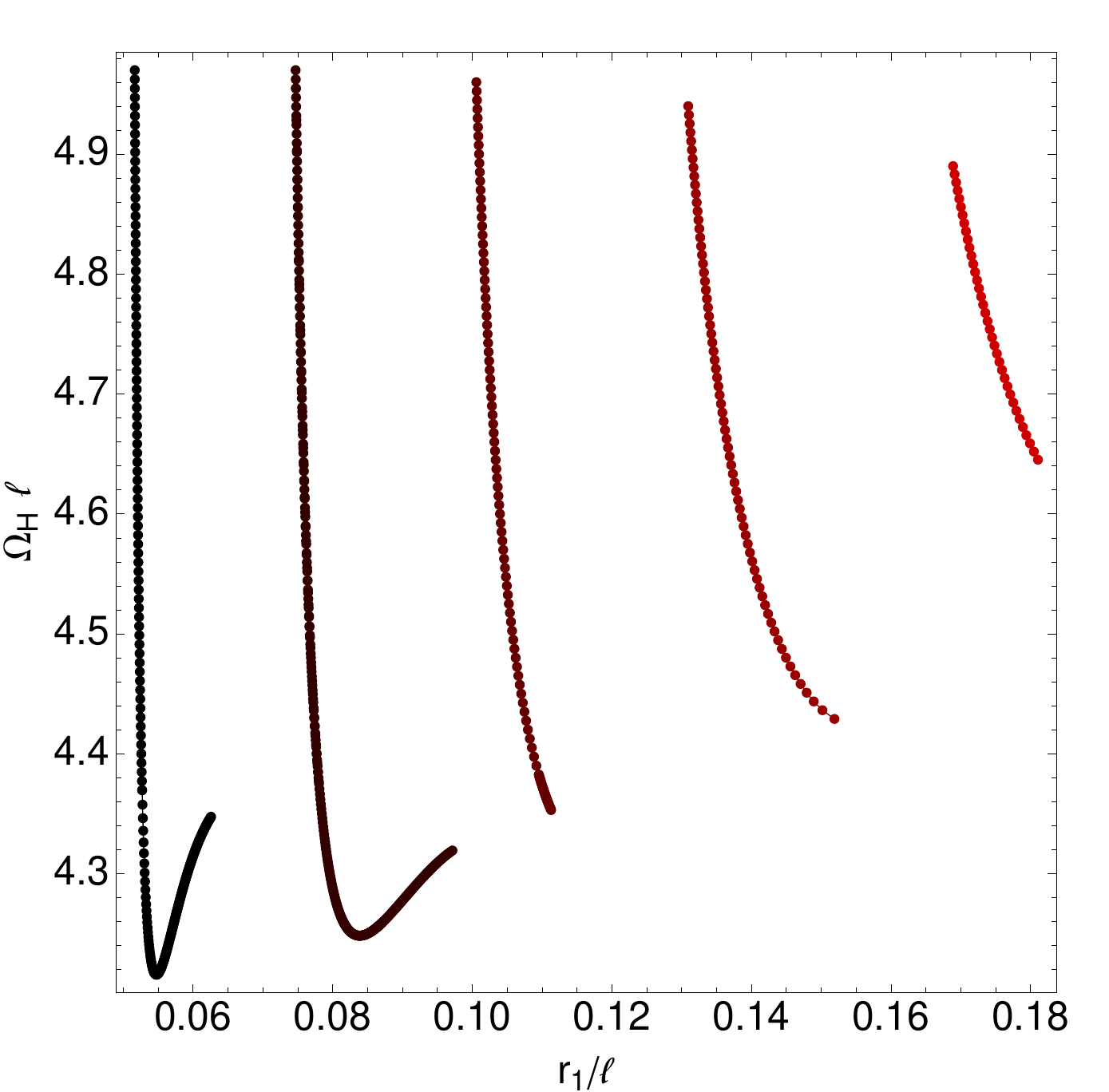}
\label{fig:ejotscvsq2c}
}
\\
\subfigure[$\ell^2T_{ab}\dot{X}^a\dot{X}^b|_{r=r_+}$ as a function of $r_1/\ell$.]{
\includegraphics[width = 0.4 \textwidth]{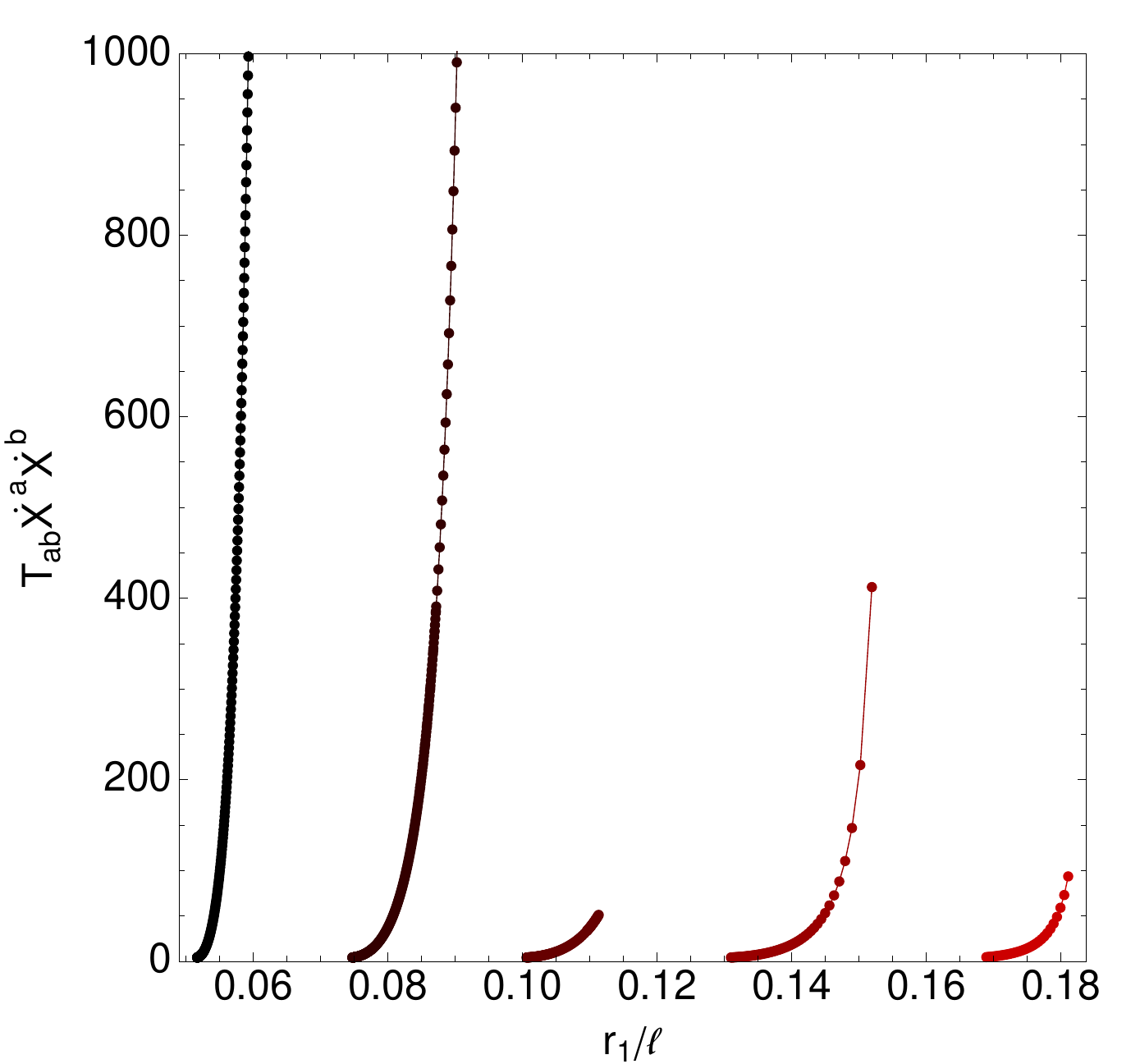}
\label{fig:ejotscvsq2e}
}\hspace{1.5cm}
\subfigure[$\epsilon$ as a function of $r_1/\ell$.]{
\includegraphics[width = 0.4 \textwidth]{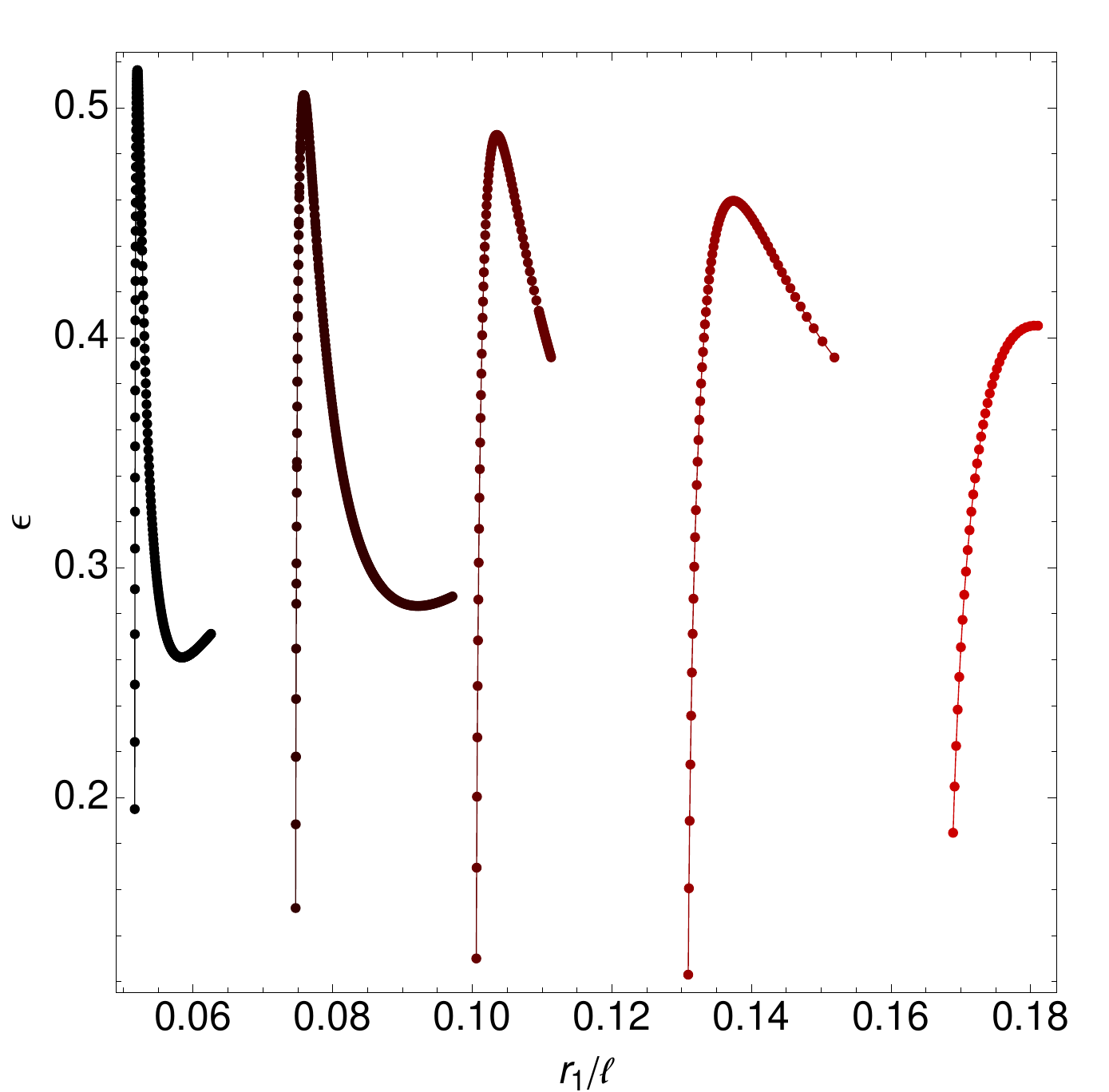}
\label{fig:ejotscvsq2j}
}
\caption{\label{figs:ejotscvsq2}Some properties of rotating hairy black holes. Each line corresponds to a different value of $r_2/\ell\equiv r_+/\ell$. From left to right in each panel, one has $r_2/\ell = 0.05,\,0.07,\,0.09,\,0.11,\,0.13$.}
\end{figure}

Our findings are compiled in Figs.~\ref{figs:ejotscvsq2}. Each panel contains a set of lines that label different values of $r_2/\ell$. From left to right in each panel, one has $r_2/\ell = 0.05,\,0.07,\,0.09,\,0.11,\,0.13$. Let us start with the two top panels, where we observe the variation of the energy $E$ and angular momentum $J$ with $r_1/\ell$. For small enough $r_2/\ell$, we observe an oscillatory behaviour that seems to stop abruptly when we reach extremality; see Fig.~\ref{fig:ejotscvsq2d}. Furthermore, as $r_2/\ell$ decreases to zero, the number of oscillations before reaching extremality seems to increase. This seems to make sense, since we expect an infinite number of oscillations for the boson stars, \emph{i.e.} as $r_2=r_+\to 0$. In Fig.~\ref{fig:ejotscvsq2c} we show how the  black hole angular frequency $\Omega_H=\omega$ varies with $r_1/\ell$ and, as anticipated, the frequency starts precisely at the $m=1$ zero-mode determined in the Merger subsection \ref{subsec:merger}; see the $m=1$ curve in Fig. \ref{fig:merger}. The relative error in determining the MP-AdS onset angular velocity using the linear results obtained in  subsection \ref{subsec:merger}, and the full non-linear results of this section (when the hairy black holes are close to the MP-AdS solution) is smaller than $10^{-4}\%$.  So, the onset of the $m=1$ superradiant instability signals, in a phase diagram, a merger or bifurcation line that connects the MP-AdS black holes with the $m=1$ rotating hairy black holes. Also quite interestingly, the black hole angular velocity is bounded below by the smallest frequency of the boson star: compare Fig.~\ref{fig:plotssolitonsc} with Fig.~\ref{fig:ejotscvsq2c}. In particular, we find no hairy black hole with $\Omega_H \ell \leq 1$.

We can also study the nature of the extremal hairy black hole solutions, in particular whether they are singular. In order to test this, we first studied the Kretschmann scalar evaluated at the horizon as a function of $r_1/\ell$ and $r_2/\ell$. However, we soon realised that the Kretschmann scalar remains finite as we approach extremality. This means that if the solution is singular, it cannot be detected by this curvature invariant. Fortunately, in \cite{Horowitz:1997uc}, the authors proposed a new mechanism to test whether a black hole is singular. In particular, they observed that some large static black holes have \emph{all} curvature invariants finite at the horizon, yet for such black holes, any object which falls in experiences enormous tidal forces outside the horizon. In order to check whether the hairy black holes are afflicted by the same problem, we need to calculate the geodesic motion of a timelike particle around our solutions. We label its geodesic tangent vector by $\dot{X}^a$. This turns out to be a rather simple exercise, because the form of the metric we are currently studying has enough symmetry to reduce this problem to a set of quadratures. In particular, we consider a radially infalling particle in the equatorial plane and initially at rest at asymptotic infinity, for which
\begin{equation}
\dot{X}^a = \left\{\frac{1}{2 f(r) g(r)},\frac{1}{2} \sqrt{\frac{1}{g(r)}-4 f(r)},\frac{\Omega (r)}{2 f(r) g(r)},0,0\right\}.
\label{eq:geodesic}
\end{equation}
In Fig.~\ref{fig:ejotscvsq2e}, we plot $\ell^2T_{ab}\dot{X}^a\dot{X}^b|_{r=r_+}$ as a function of $r_1/\ell$, and we see that it diverges precisely as we approach extremality, signalling a singular black hole solution, \emph{i.e.} rotating extremal hairy black holes are singular.

Finally, we point out that the hairy black holes solutions obey  the first law of thermodynamics \eqref{eq:BH1stlaw}  up to $0.005\%$ error. Like in the boson stars, we view this as a very good check on our numerics, and as providing a measure of the error in determining these solutions.

\subsection{\label{subsec:cpdj}Complete phase diagram}

In this section, we will sew all of the information gathered in the past two subsections, and construct the phase diagram for these rotating hairy solutions, \emph{i.e.} the phase diagram for MP-AdS black holes, $m = 1$ rotating hairy black holes and  rotating boson stars. As we shall see, the phase diagram is rather intricate, but the overall picture is quite simple.

\begin{figure}
\centering
\subfigure[Energy of the boson star (solid line) as a function of its angular momentum. The inset plot shows $(E_2-E_1)/\ell^2$ as a function of $J/\ell^3$. ($E_1,E_2$ are defined in the main text).]{
\includegraphics[width = 0.55 \textwidth]{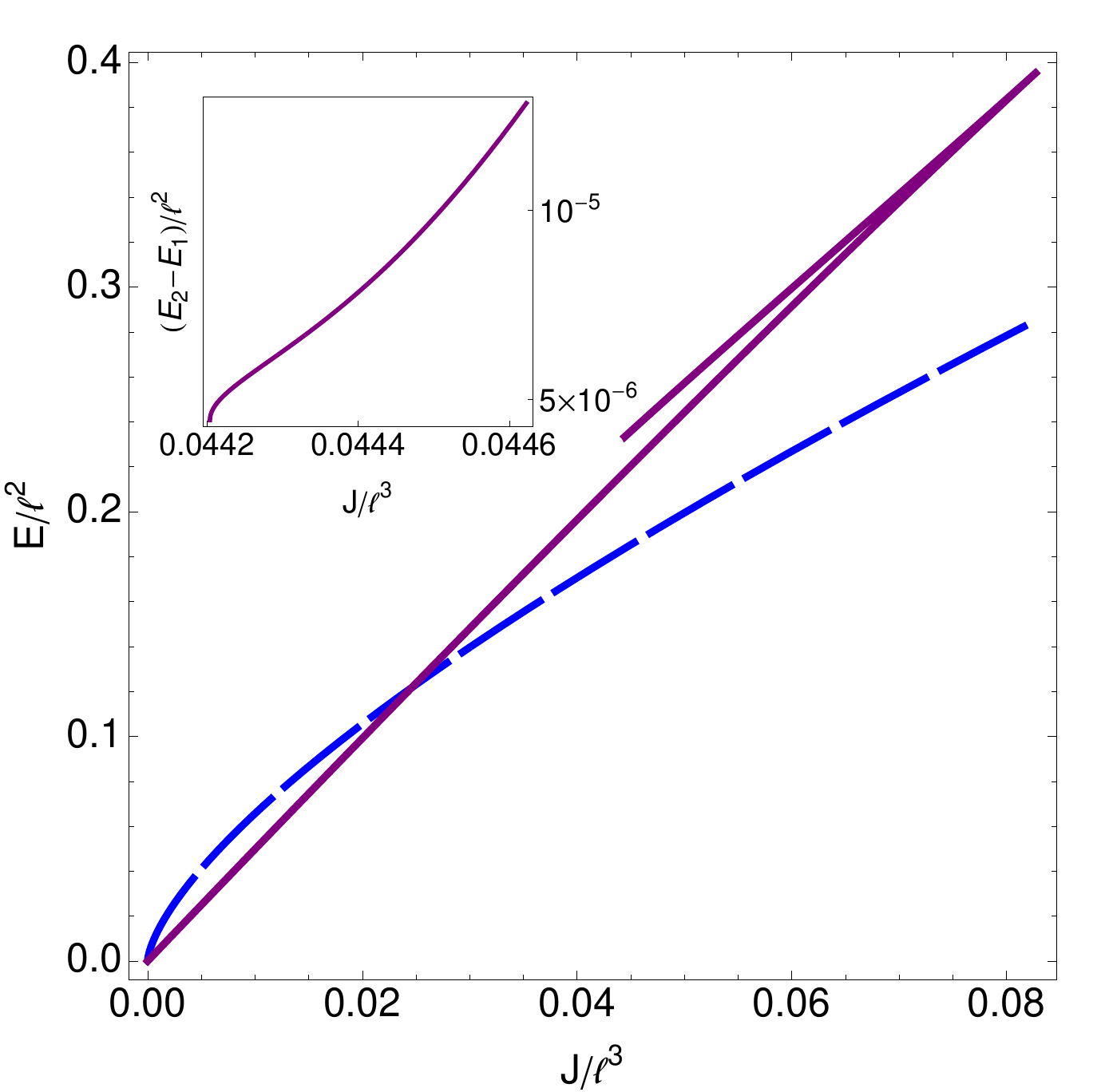}
\label{fig:phasediagramsol}
}
\subfigure[Complete phase diagram for rotating hairy solutions. $\Delta E$ is the energy relative to the lower branch of boson stars.]{
\includegraphics[width = 0.55 \textwidth]{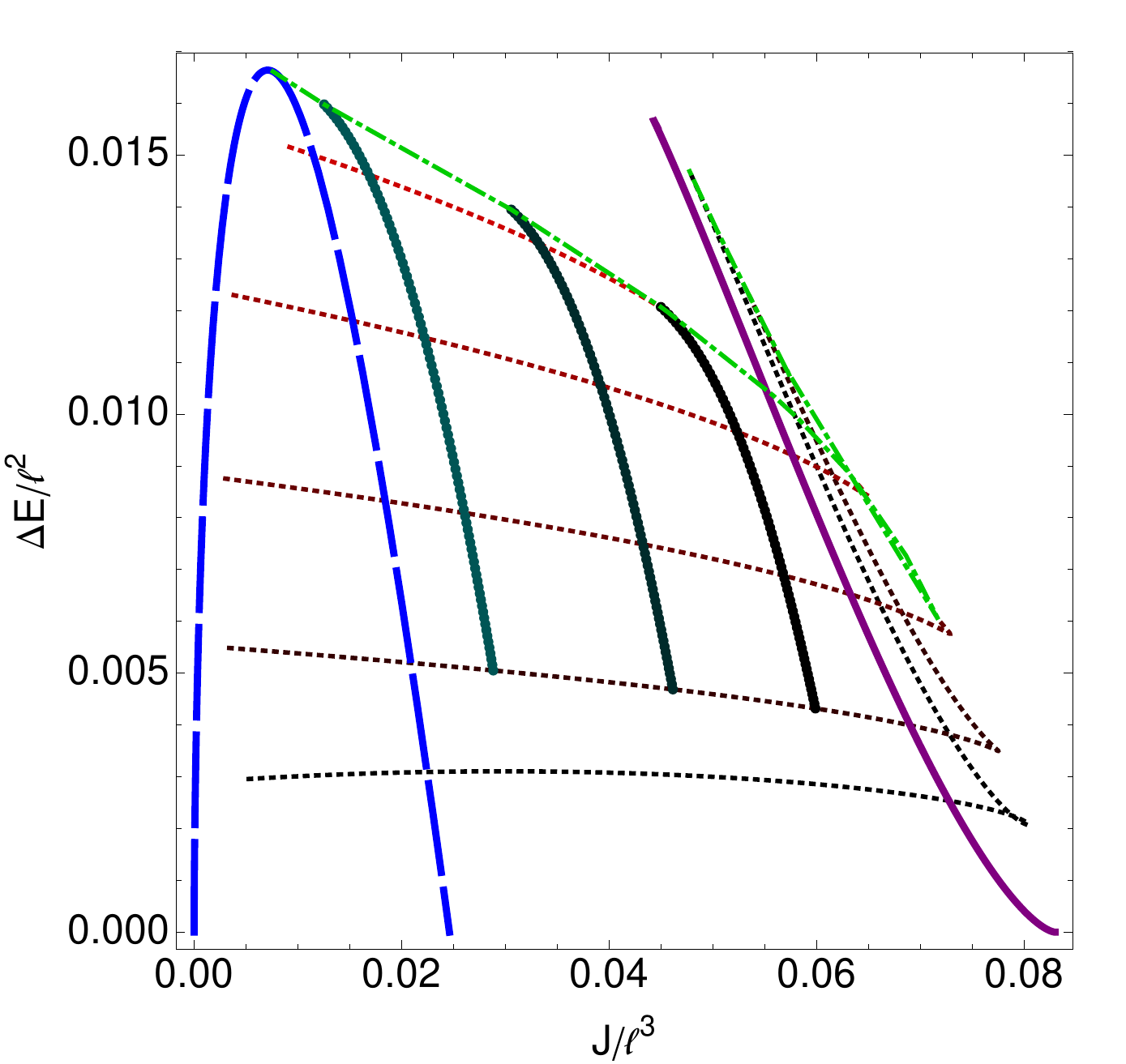}
\label{fig:phasediagram}
}
\caption{\label{figs:phasediagram}Phase diagrams of rotating hairy solutions. The dashed blue curve indicates the location of the extremal MP-AdS black holes, and regular MP-AdS black holes only exist above it.  The solid purple curve represents the boson stars, the dotted red curves represent lines of constant $r_2/\ell=r_+/\ell$ and the green ``vertical" solid curves are lines of constant $\Omega_H\ell$. Finally, the dotted-dashed green curve represents the best approximation to the singular extremal hairy black holes.}
\end{figure}

We will present the results in the microcanonical ensemble. In Figs.~\ref{figs:phasediagram}, the dashed blue line represents the MP-AdS extremal line, above which non-extremal regular MP-AdS exist. In both plots, the solid purple line represents boson stars.

In Fig.~\ref{fig:phasediagramsol} we show how the boson star energy varies with its angular momentum. One can understand the shape of this line as follows. As we have seen in section \ref{subsec:rbsj}, the energy  $E$ and angular momentum $J$ of the boson stars oscillate as a function of the energy density at the centre of the star, $\Pi'(0)$. However, unlike  Fig. \ref{fig:plotssolitonse}, the first law ensures that $E$ and $J$ have extrema at the same values of $\Pi'(0)$, which we label by $\Pi'(0)_i$. This means that instead of a spiral, the motion in the $E,J$ plane has cusps.   Let us define $E_i(J)$ to be the energy of the star as a function of $J$ for the branch where $\Pi'(0)\in(\Pi'(0)_i,\Pi'(0)_{i+1})$, where we identify $\Pi'(0)_0 \equiv 0$.   We can see up to two cusps (and thus three branches of boson stars), one of which is plotted as an inset plot. Furthermore, we find that some boson stars coexist with MP-AdS black holes. This happens when the solid purple line is above the dashed blue line.

The hairy black holes lie in a thin band above the boson star curve. To make this clearer, it is convenient to plot the energy relative to the lower line of the boson stars, \ie $\Delta E \equiv E(J)-E_0(J)$. This is shown in Fig.~\ref{fig:phasediagram}. The dashed blue line now returns to zero at the point where the boson star crosses the extremal MP AdS line in Fig.~\ref{fig:phasediagramsol}. The solid purple line on the far right is the second branch of boson stars. The green ``vertical" curves represent hairy black holes with constant angular frequency $\Omega_H \ell$, whereas the dotted red curves are lines of constant $r_2\equiv r_+$. Recall that $r_1$ is increasing along the lines of constant $r_2$. Thus, whenever these dotted curves cross we have hairy black holes with different horizon geometry but the same energy and angular momentum. In other words, we have non-uniqueness. Since there is also a MP-AdS solution, we have up to three coexisting black hole solutions.

Close to the merger, the entropy of the MP-AdS solutions is smaller than that of the hairy black holes, being equal exactly at the merger line. This means that this transition is a second order phase transition. However, for sufficiently large angular momentum, the MP-AdS black holes also coexist with the hairy solutions, their entropy being now larger than that of the hairy black holes. Moreover, the transition is now first order, because these solutions never merge for this range of angular momentum.  In sum, in a 3-dimensional plot of $\{S/\ell^3,\Delta E/\ell^2, J/\ell^3\}$ the hairy black hole family is a 2-dimensional surface bounded by the merger line and the boson star curve, if the angular momentum is smaller than the critical $J$ where the red and blue curves meet in Fig. \ref{figs:phasediagram}. This surface continues for larger values of $J$ where it is now bounded by the boson star line and the extremal hairy black hole curve. This  2-dimensional surface never intersects with itself and has a sequence of (regular) ``cusp lines".

We have provided evidence in section \ref{subsec:rhbhj} that the zero temperature limit of the hairy black holes is singular due to the appearence of large tidal forces. In order to estimate the shape of this singular curve in the phase diagram, we take the points of smallest temperature for each line of constant $r_2/\ell$ and $\Omega_H \ell$. This line is represented as a dotted-dashed green curve in Fig.~\ref{fig:phasediagram}. The entropy along this curve monotonically decreases as we move away from the merger line. We conjecture that this line should join to the endpoint of the boson star, which has zero entropy, perhaps spiraling an infinite number of times before reaching it.

\begin{figure}
\centerline{
\includegraphics[width=0.45\textwidth]{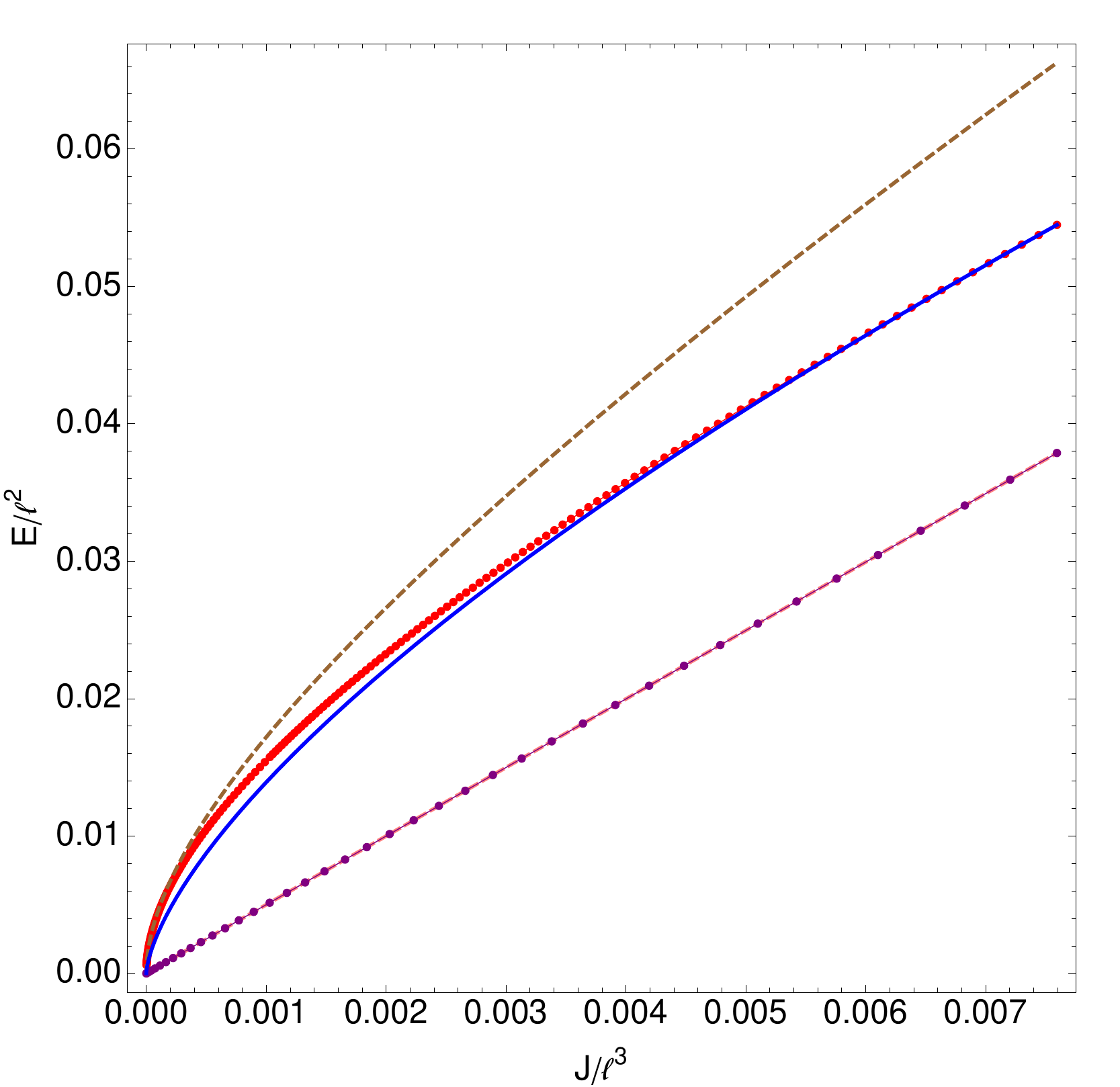}
}
\caption{Phase diagram for small values of the energy and angular momentum. From top to bottom, analytical estimate for the merger line, exact merger line, extremal line of the MP-AdS black holes, analytical estimate for the bosons stars and finally the exact data for the bosons stars.}
\label{fig:comparison}
\end{figure}

To end this section, we will compare the numerical results with the analytical estimates of section \ref{sec:PertBstar} and \ref{sec:PertBHs}. In Fig.~\ref{fig:comparison} we plot the analytical estimates and our exact numerical data. The agreement is excellent for small $E$ and $J$. We might have expected our perturbative analysis to be valid for a larger range of energy and angular momentum, because the black holes are always small. However, we point out that the curvatures involved here are rather large (\ie the coefficients of the expansion get quickly large as the order increases), and thus these expansions should have a small radius of convergence. This is the reason why we observe large discrepancies in the analytical estimate of the merger line for $J/\ell^3 \gtrsim 0.001$.

\section{Discussion}

By exploiting a clever ansatz for two (complex) scalar fields \cite{Hartmann:2010pm}, we have constructed hairy black holes with only a single Killing vector. The ansatz corresponds to an $m=1$ mode of the scalar fields. When the horizon area shrinks to zero, the solutions reduce to rotating boson stars. These solutions were constructed using two complementary approaches: a perturbative analytic approach for very small black holes, and numerical methods for larger ones.
The phase diagram of these solutions was given in Fig. 6 and it was found that they exhibit nonuniqueness. 

We now make a few comments about the stability of our solutions. First consider the boson stars. We know that AdS is linearly stable, and because boson stars with small $E$ and $J$  are just perturbations of AdS, we expect them to also be linearly stable. In fact, we expect the normal modes of these boson stars to be simple deformations of the AdS normal modes \eqref{eqP:BSlinearPi}. For larger $E$ and $J$, we expect the boson stars to eventually become unstable. For nonrotating boson stars this usually occurs at the maximum value of $E$, \ie, at the first cusp. However, in the rotating case, it is clear from Fig. 4 that the maximum of $E$ does not coincide with an extremum of the frequency $\omega$. Thus, if one looks for a zero mode to mark the transition between stability and instability, it will not occur at the maximum mass.

We now turn to black holes. If we consider $E$ and $J$  in the region where MP-AdS black holes are unstable, then our hairy black holes are almost certainly stable within our $m=1$ scalar ansatz. They represent the endpoint of the superradiant instability for this mode and have larger entropy than the MP-AdS solutions. Outside this region of $(E,J)$ the hairy black holes probably remain stable to $m=1$ perturbations for a while, but then become unstable (like the boson stars). However, in all cases,  since our hairy black holes have $\Omega_H \ell > 4$, it is clear from Fig. 1 that they are likely to be unstable to $m>1$ perturbations. It is natural to ask what is the endpoint of this instability in the full theory. This raises an interesting puzzle.

 If the black hole settles down,   the scalar fields at the horizon must satisfy $K^\mu \nabla_\mu \vec \Pi= 0$ where $K^\mu$ is the null generator of the horizon. This ensures that there is no flux of scalar field into the black hole. If the final black hole has $\Omega_H \ell > 1$ it will probably still be unstable to superradiance in a large $m$ mode, so we need $\Omega_H \ell \le 1$. Since evolution is likely to stop at the first stable configuration, we expect the final black hole to have $\Omega_H \ell = 1$. But since we can start with an arbitrarily small black hole, this would seem to require a boson star with $\Omega_H \ell = 1$, \ie invariant under $\partial_t + \partial_\psi$. Since this Killing vector does not have an ergoregion, such a boson star is unlikely to exist.

The only alternative would be that the black hole will never settle down.
It was shown in section (\ref{subsec:growth}) that  the time scale for the (linear) instability grows rapidly with  $m$. Thus, under evolution, the black hole  approaches the $m=1$ stable configuration discussed here. Then, the $m=2$ instability kicks in and the black hole evolves toward the $m=2$ stable configuration. This can repeat indefinitely, with the black hole developing structure on smaller and smaller scales over longer and longer times. However, since there is only a finite amount of energy that can be extracted from the black hole, one still expects the amplitude for the $m^{th}$ mode to go to zero, so it is not clear why it doesn't settle down.


We have checked that standard MP-AdS black holes always have lower free energy than our hairy black holes. This is true for both the Helmholtz free energy $F=E-TS$ (comparing black holes with the same temperature and angular momentum) and the Gibbs free energy $G = E - TS - \Omega_H J$ (comparing black holes with the same temperature and angular velocity). 

Our ansatz for the scalar fields has an immediate generalization to any odd dimensional spacetime. We expect that the results discussed here will extend to higher dimensions. We also expect that most of the properties that we have found for black holes coupled to massless scalar fields will carry over to black holes with no scalar fields. Vacuum black holes exhibit superradiance and with AdS boundary conditions we expect them to evolve toward black holes with gravitational waves propagating around them.  In this case, the metric itself (and not just the scalar hair) would be invariant under just a single Killing field. We cannot add a mass to our scalars and look for asymptotically flat five dimensional black holes with scalar hair. There is a superradiant instability for massive scalars only in four dimensions \cite{Cardoso:2005vk}.

We have considered only classical solutions, so it is natural to ask what happens if one includes Hawking radiation. Since our black holes are all small compared to the AdS radius, they have negative specific heat and one might expect them to evaporate completely. However this is incorrect since it would result in a decrease in entropy: A small spherical black hole with horizon radius $r_+$ has energy $E \sim r_+^2 /G$ and entropy $S_{BH} \sim  r_+^3/G$. The entropy of a gas with energy $E$ in $AdS_5$ is roughly $S_g \sim  (E\ell)^{4/5}$ where $\ell$ is the AdS radius. If all the energy in the black hole went into the gas, the entropy would increase only if $r_+/L < (\ell_p/\ell)^{3/7}$ where $\ell_p$ is the Planck length.  Our black holes can be much larger than this, so they  will start to evaporate but then quickly come into equilibrium with their  Hawking radiation. 

Given the recent work on the fluid/gravity correspondence (see, e.g., \cite{Rangamani:2009xk}), one might wonder if there is a dual interpretation of our solutions in terms of a rotating fluid. The answer appears to be no. The hydrodynamic approximation involves a restriction to long wavelength modes. Our black holes are all small compared to the AdS radius, so they do not describe a  state of a fluid. To describe  a thermal state of a fluid one needs a large black hole with $r_+\gg\ell$. Such a black hole can still have a superradiant instability, but only for modes with large $m$. This would again violate the long wavelength requirement of the hydrodynamic approximation. Even if there is no fluid interpretation, there must still be some dual interpretation of our hairy black holes in terms of gauge/gravity duality, and it would be interesting to understand what it is. 
\vskip .5cm
{\bf Note Added:} After this paper was submitted, we were informed of \cite{Ridgway:1995ac} which describes another type of black hole with only one symmetry. It is static and asymptotically flat, but has no spatial symmetries.

\vskip .5cm
\centerline{\bf Acknowledgements}
\vskip .2cm

It is a pleasure to thank B. Way for collaboration at an early stage of this work. This work was supported in part by the  National Science Foundation under Grant No.~PHY08-55415.
OJCD acknowledges financial support provided by the European Community through the Intra-European Marie Curie contract PIEF-GA-2008-220197.

\appendix

\section{\label{sec:AppExpansion}Perturbative fields for the  boson star and hairy black hole}

\subsection{\label{sec:AppExpanSoliton}Perturbative boson star}

In subsection \ref{sec:PertBstar} we explained how we can construct perturbatively the  boson stars. In this Appendix we explicitly write the expansion of the gravitational and doublet scalar fields up to order  $\mathcal{O}\left( \epsilon^4 \right)$: 

{\small
\begin{eqnarray} \label{BS:functions}
 &&\hspace{-1cm} f(r)=
\left(1+\frac{r^2}{\ell ^2} \right)-\epsilon^2\,\frac{r^2 \left(5 r^4 \ell ^2+20 r^2 \ell ^4+6 \ell ^6\right)}{9 \left(r^2+\ell ^2\right)^4}
-\epsilon^4\,\frac{r^2 \ell ^2}{1270080 \left(r^2+\ell ^2\right)^9} {\biggl (}514952 r^{14}+4631027 r^{12} \ell ^2\nonumber\\
&& \hspace{-0.5cm} +18512283 r^{10} \ell ^4+40913902 r^8 \ell ^6+51954798 r^6 \ell ^8+36154839 r^4 \ell ^{10}+11249595 r^2 \ell ^{12}+1315860 \ell ^{14} {\biggr )}
\!\!+\!\mathcal{O}\left( \epsilon^6\right),\nonumber \\
 &&\hspace{-1cm} g(r)=
1-\epsilon ^2\,\frac{2 \ell ^8 \left(4 r^2+\ell ^2\right)}{3 \left(r^2+\ell ^2\right)^5}-\epsilon ^4\,\frac{\ell ^4}{1270080 \left(r^2+\ell ^2\right)^{10}}  {\biggl (} 3541 r^{16}+35410 r^{14} \ell ^2+149055 r^{12} \ell ^4+3052440 r^{10} \ell ^6\nonumber\\
&& \hspace{0.0cm}+16099475 r^8 \ell ^8+34403186 r^6 \ell ^{10}+25939155 r^4 \ell ^{12}+7971520 r^2 \ell ^{14}+872290 \ell ^{16} {\biggr )}
+\mathcal{O}\left( \epsilon^6\right),\nonumber \\
 &&\hspace{-1cm} h(r)=1+\epsilon ^4\,\frac{r^2 \ell ^4 }{1270080 \left(r^2+\ell ^2\right)^9}  {\biggl (} 3541 r^{12}+31869 r^{10} \ell ^2+123066 r^8 \ell ^4+260694 r^6 \ell ^6+311661 r^4 \ell ^8\nonumber\\
&& \hspace{0.0cm}+183645 r^2 \ell ^{10}+22260 \ell ^{12}{\biggr )} +\mathcal{O}\left( \epsilon^6\right),\nonumber \\
 &&\hspace{-1cm} 
\Omega(r)=\epsilon ^2\,\frac{\ell ^3 \left(r^4+4 r^2 \ell ^2+6 \ell ^4\right)}{12 \left(r^2+\ell ^2\right)^4}+\epsilon^4\,\frac{\ell ^3}{2540160 \left(r^2+\ell ^2\right)^9} {\biggl (} 167242 r^{14}+1505178 r^{12} \ell ^2+6020712 r^{10} \ell ^4)\nonumber \\
 &&\hspace{0.0cm} +14139048 r^8 \ell ^6+20982192 r^6 \ell ^8+19004760 r^4 \ell ^{10}+8795055 r^2 \ell ^{12}+1598455 \ell ^{14}{\biggr )} )
+\mathcal{O}\left( \epsilon^6\right),\nonumber \\
 &&\hspace{-1cm} 
\Pi(r)=\epsilon\, \frac{r \ell ^4}{\left(r^2+\ell ^2\right)^{5/2}}+\epsilon^3\frac{r \ell ^6 \left(900 r^6+3935 r^4 \ell ^2+5548 r^2 \ell ^4+1540 \ell ^6\right)}{2016 \left(r^2+\ell ^2\right)^{13/2}}
+\epsilon^5\, \frac{r \ell ^6 }{853493760 \left(r^2+\ell ^2\right)^{25/2}}\nonumber \\
 &&\hspace{0.5cm}\times {\biggl (}428716940 r^{18}+4416801537 r^{16} \ell ^2+20395866890 r^{14} \ell ^4+55586393870 r^{12} \ell ^6+98320298706 r^{10} \ell ^8\nonumber \\
 &&\hspace{0.0cm}+115794392980 r^8 \ell ^{10}+88872056182 r^6 \ell ^{12}+41756607180 r^4 \ell ^{14}+10678880150 r^2 \ell ^{16}+1128452101 \ell ^{18}{\biggr )}
\nonumber\\ && \hspace{0.0cm}
+\mathcal{O}\left( \epsilon^7\right).
 \end{eqnarray}
}
The perturbative construction also generates the expansion for the frequency of the scalar field, which is explicitly written in equation \eqref{eqP:omegaBstar}. 

\subsection{\label{sec:AppExpanBH}Perturbative black hole: near-region and far-region field expansions}

In subsections \ref{sec:FRPertBHs}-\ref{sec:MatchPertBHs} we have explained how we can construct perturbatively the hairy black holes using a matched asymptotic expansion of the near and far region solutions. In this Appendix we explicitly write the expansion of the gravitational and doublet scalar fields in the near-region and far-region with the matching conditions already imposed. We do the expansion up to order $\mathcal{O}\left( \epsilon^n, (r_+/\ell)^p\right)$ with $(n+p)=4$. 

The far-region expansion of the gravitational and scalar fields is given by:
{\small
\begin{eqnarray} \label{FR:functions}
 &&\hspace{-0.5cm} f^{out}(r)=\left[\left(1+\frac{r^2}{\ell ^2}\right)-\frac{r_+^2}{r^2}-\frac{26}{r^2}\frac{r_+^4}{\ell ^2}+
 \mathcal{O}\left(\frac{r_+^6}{\ell^6}\right)\right] \nonumber \\
 &&\hspace{1cm} +\epsilon^2{\biggl [}-\frac{r^2 \left(5 r^4 \ell ^2+20 r^2 \ell ^4+6 \ell ^6\right)}{9 \left(r^2+\ell ^2\right)^4}+
\frac{\ell ^2 }{72 \left(r^2+\ell ^2\right)^5}{\biggl (}-191 r^8-535 r^6 \ell ^2-20 r^4 \ell ^4+84 r^2 \ell ^6+48 \ell ^8
 \nonumber \\
 && \hspace{1.7cm} +84 r^2 \left(r^2+\ell ^2\right) \left(5 r^4+20 r^2 \ell ^2+6 \ell ^4\right) {\rm log}\left(\frac{r^2}{r^2+\ell ^2}\right){\biggl)}\,\frac{r_+^2}{\ell ^2}
  + \mathcal{O}\left(\frac{r_+^4}{\ell^4}\right) {\biggl ]}    
\nonumber \\
 &&\hspace{1cm} -\epsilon^4{\biggl [}\frac{r^2 \ell ^2 }{1270080 \left(r^2+\ell ^2\right)^9}{\biggl (}514952 r^{14}+4631027 r^{12} \ell ^2 +18512283 r^{10} \ell ^4  +40913902 r^8 \ell ^6
 \nonumber \\
 &&\hspace{1.7cm} +51954798 r^6 \ell ^8 +36154839 r^4 \ell ^{10}+11249595 r^2 \ell ^{12}+1315860 \ell ^{14}{\biggr)}+ \mathcal{O}\left(\frac{r_+^2}{\ell^2}\right) {\biggl ]}\,, \nonumber \\
  &&\hspace{-0.5cm}
  g^{out}(r)=\left[  1 +\mathcal{O}\left(\frac{r_+^6}{\ell^6}\right)\right]
   +\epsilon^2{\biggl [}   -\frac{2 \ell ^8 \left(4 r^2+\ell ^2\right)}{3 \left(r^2+\ell ^2\right)^5}  
   +{\biggl (} \left( \frac{20}{3}-\frac{7 \ell ^8 \left(4 r^2+\ell ^2\right)}{\left(r^2+\ell ^2\right)^5}\right) {\rm log}\left(1+\frac{\ell ^2}{r^2}\right)\nonumber \\
 &&\hspace{1cm}
  -\frac{\ell ^2 \left(60 r^{10}+330 r^8 \ell ^2+740 r^6 \ell ^4+855 r^4 \ell ^6+378 r^2 \ell ^8+83 \ell ^{10}\right)} {9 \left(r^2+\ell ^2\right)^6}
  {\biggr )}\frac{r_+^2}{\ell ^2} + \mathcal{O}\left(\frac{r_+^4}{\ell^4}\right){\biggl ]}    \nonumber \\
 &&\hspace{1cm} -\epsilon^4{\biggl [}   \frac{\ell ^4 }{1270080 \left(r^2+\ell ^2\right)^{10}}{\biggl (}3541 r^{16}+35410 r^{14} \ell ^2+149055 r^{12} \ell ^4+3052440 r^{10} \ell ^6
\nonumber \\
&&\hspace{1.4cm}+16099475 r^8 \ell ^8 +34403186 r^6 \ell ^{10}+25939155 r^4 \ell ^{12}+7971520 r^2 \ell ^{14}+872290 \ell ^{16}{\biggl )}
 \!\! +\! \mathcal{O}\left(\frac{r_+^2}{\ell^2}\right) {\biggl ]}\,, \nonumber \\
  &&\hspace{-0.5cm} h^{out}(r)=\left[  1 +\mathcal{O}\left(\frac{r_+^6}{\ell^6}\right)\right] +\epsilon^2\left[ \mathcal{O}\left(\frac{r_+^4}{\ell^4}\right) \right] 
   +\epsilon^4{\biggl [}  \frac{r^2 \ell ^4 }{1270080 \left(r^2+\ell ^2\right)^9}{\biggl (}3541 r^{12}+31869 r^{10} \ell ^2 \nonumber \\
 &&\hspace{1cm}
+123066 r^8 \ell ^4+260694 r^6 \ell ^6+311661 r^4 \ell ^8+183645 r^2 \ell ^{10}+22260 \ell ^{12} {\biggr )}
  + \mathcal{O}\left(\frac{r_+^2}{\ell^2}\right) {\biggr ]}\,, \nonumber \\
  &&\hspace{-0.5cm}
 \ell\,\Omega^{out}(r)=\left[ \frac{5 r_+^4}{r^4} +\mathcal{O}\left(\frac{r_+^6}{\ell^6}\right)\right]
 +\epsilon^2 {\biggl [}  \frac{\ell ^4 \left(r^4+4 r^2 \ell ^2+6 \ell ^4\right)}{12 \left(r^2+\ell ^2\right)^4} \nonumber \\
&&\hspace{1.3cm}+ {\biggl (} \frac{129 \ell ^4}{r^4}+\frac{3 \ell ^2 \left(384 r^{12}+1728 r^{10} \ell ^2+2924 r^8 \ell ^4+2086 r^6 \ell ^6+238 r^4 \ell ^8-215 r^2 \ell ^{10}-43 \ell ^{12}\right)}
{r^4 \left(r^2+\ell ^2\right)^5} \nonumber \\
&&\hspace{1.7cm}
+36 \left(-32+\frac{7 \ell ^4 \left(r^4+4 r^2 \ell ^2+6 \ell ^4\right)}{\left(r^2+\ell ^2\right)^4}\right) {\rm log}\left(1+\frac{\ell ^2}{r^2}\right) {\biggr )}\,\frac{r_+^2}{288 \,\ell^2}
+ \mathcal{O}\left(\frac{r_+^4}{\ell^4}\right)  {\biggr ]}  \nonumber \\
 &&\hspace{1cm} +\epsilon^4{\biggl [}
 \frac{\ell ^4 }{2540160 \left(r^2+\ell ^2\right)^9}{\biggl (} 167242 r^{14}+1505178 r^{12} \ell ^2+6020712 r^{10} \ell ^4+14139048 r^8 \ell ^6 \nonumber \\
&&\hspace{1.7cm}
+20982192 r^6 \ell ^8+19004760 r^4 \ell ^{10} +8795055 r^2 \ell ^{12}+1598455 \ell ^{14}{\biggr)}  + \mathcal{O}\left(\frac{r_+^2}{\ell^2}\right) {\biggr ]}\,, \nonumber \\
  &&\hspace{-0.5cm}
 \Pi^{out}(r)=\epsilon\,\ell^4 {\biggl \{} 
 \frac{r}{\left(r^2+\ell ^2\right)^{5/2}}-\frac{ 11 r^2 \ell ^2+\ell ^4+21 r^2 \left(r^2+\ell ^2\right) {\rm log}\left[\frac{r^2}{r^2+\ell ^2}\right]}{4 r \left(r^2+\ell ^2\right)^{7/2}}\frac{r_+^2}{\ell ^2}
 \nonumber\\
&&\hspace{1.7cm}
+\frac{1}{64 \ell ^4 r^3\left(r^2+\ell ^2\right)^{9/2}}{\biggl [}
-24 r^4 \ell ^4 \left(r^2+\ell ^2\right)^2 \text{Li}_2\left(-\frac{r^2}{\ell ^2}\right)-2 r^{10} \ell ^2-\left(19+4 \pi ^2\right) r^8 \ell ^4 \nonumber\\
&&\hspace{1.7cm}   -2 \left(2163+4 \pi ^2\right) r^6 \ell ^6 -4 \left(1156+\pi ^2\right) r^4 \ell ^8-600 r^2 \ell ^{10}-5 \ell ^{12}+ 2 \left(r^2+\ell ^2\right) \nonumber\\
&&\hspace{1.7cm}
\times {\biggl (}-24 r^4 \ell ^4 \left(r^2+\ell ^2\right) {\rm log}^2\left(\frac{r}{\ell }\right)-2 \ell ^4 \left(r^2+\ell ^2\right) {\rm log}\left(\frac{r_+}{4 \ell }\right) \left(8 r^2 \ell ^2+\ell ^4+12 r^4 {\rm log}\left[1+\frac{\ell ^2}{r^2}\right]\right) \nonumber\\
&&\hspace{2.1cm} +{\rm log}\left(1+\frac{\ell ^2}{r^2}\right){\biggl (}r^{10}+9 r^8 \ell ^2+3745 r^6 \ell ^4+3267 r^4 \ell ^6-51 r^2 \ell ^8-\ell ^{10}\nonumber\\
&&\hspace{2.1cm}+441 r^4 \ell ^4 \left(r^2+\ell ^2\right) {\rm log}\left(1+\frac{\ell ^2}{r^2}\right){\biggr )} {\biggr )}
{\biggr ]}\frac{r_+^4}{\ell^4}
+\mathcal{O}\left(\frac{r_+^6}{\ell^6}\right){\biggr \}} \nonumber \\
&& +\epsilon^3\,\ell^4 {\biggl \{} \frac{r \ell ^2 \left(900 r^6+3935 r^4 \ell ^2+5548 r^2 \ell ^4+1540 \ell ^6\right)}{2016 \left(r^2+\ell ^2\right)^{13/2}} 
 + \frac{1}{48384 \,\ell ^4 r \left(r^2+\ell ^2\right)^{15/2}}  {\biggl [}
-14196 r^{14} \ell ^2  \nonumber\\
&&\hspace{1.7cm}   -7098 \left(25+4 \pi ^2\right) r^{12} \ell ^4+14 \left(6037-10140 \pi ^2\right) r^{10} \ell ^6+1183 \left(1667-240 \pi ^2\right) r^8 \ell ^8\nonumber\\
&&\hspace{1.7cm}  +65 \left(55327-4368 \pi ^2\right) r^6 \ell ^{10}+5 \left(450949-28392 \pi ^2\right) r^4 \ell ^{12}+3 \left(180149-9464 \pi ^2\right) r^2 \ell ^{14} \nonumber\\
&&\hspace{1.7cm} -9240 \ell ^{16} +6 \left(r^2+\ell ^2\right) {\biggl (} 2366 r^{12}+28392 r^{10} \ell ^2+34096 r^8 \ell ^4-43516 r^6 \ell ^6-8889 r^4 \ell^8\nonumber\\
&&\hspace{1.7cm} 
+145204 r^2 \ell ^{10}+41208 \ell ^{12} {\biggr )} {\rm log}\left(1+\frac{\ell ^2}{r^2}\right)
 -170352 r^2 \ell ^4 \left(r^2+\ell ^2\right)^5 \text{Li}_2\left(-\frac{r^2}{\ell ^2}\right)\nonumber\\
&&\hspace{1.7cm}+ 85176 r^2 \ell ^4 \left(r^2+\ell ^2\right)^5 \text{log}\left(1+\frac{r^2}{\ell ^2}\right) \text{log}\left(\frac{\ell ^2 \left(r^2+\ell ^2\right)}{r^4}\right)
 {\biggr ]}\frac{r_+^2}{\ell^2}
+\mathcal{O}\left(\frac{r_+^4}{\ell^4}\right)    {\biggl \}}  \nonumber \\
 &&\hspace{1cm} +\epsilon^5\,\ell^4 {\biggl [} 
\frac{r \ell^2}{853493760 \left(r^2+\ell^2\right)^{25/2}} {\biggl (} 428716940 r^{18}+4416801537 r^{16} \ell^2+20395866890 r^{14}  \ell^4 \nonumber\\
&&\hspace{2.1cm} +55586393870 r^{12}  \ell^6
 +98320298706 r^{10} \ell^8+115794392980 r^8 \ell^{10}+88872056182 r^6  \ell^{12}\nonumber\\
&&\hspace{2.1cm} +41756607180 r^4  \ell^{14}+10678880150 r^2 \ell^{16}+1128452101 \ell^{18} {\biggl )} 
+\mathcal{O}\left(\frac{r_+^2}{\ell^2}\right)    {\biggl ]} \,,
 \end{eqnarray}
}
where  $\text{Li}_n(x)=\sum _{k=1}^{\infty } x^k/k^n$ is the Polylogarithm function.

The near-region expansion of the gravitational and scalar fields is given by:
{\small
\begin{eqnarray} \label{NR:functions}
 && f^{in}(z)= \left[\left(1-\frac{\ell^2}{z^2}\right)+\frac{25\ell^4-26 z^2\ell^2+z^6/\ell^2}{z^4}\,\frac{r_+^2}{\ell^2}
-\frac{570 \left(\ell^4-z^2\ell^2\right)}{z^4}\,\frac{r_+^4}{\ell^4}+
 \mathcal{O}\left(\frac{r_+^6}{\ell^6}\right)\right] \nonumber \\
 &&\hspace{1cm} +\epsilon^2 {\biggl [} \frac{r_+^2}{\ell^2} \,\frac{\ell^2}{z^2}
 \int_1^{z/\ell} \frac{dx}{672 x^3 \left(1-x^2\right)}{\biggl (}-8480 \left(1-x^2\right)\nonumber \\
 &&
 \hspace{1.9cm} +21 \pi ^2 x^2 {\biggl (}\left(3-16 x^2+16 x^4\right) \left(P_{1/2}\left(2 x^2-1\right)\right){}^2
   +3 \left(P_{3/2}\left(2 x^2-1\right)\right){}^2\nonumber \\
 &&\hspace{1.9cm} +6 \left(1-2 x^2\right) P_{1/2}\left(2 x^2-1\right) P_{3/2}\left(2 x^2-1\right) {\biggr )} {\biggr )}
 +\mathcal{O}\left(\frac{r_+^4}{\ell^4}\right){\biggr ]}
 +\epsilon ^4 \mathcal{O}\left(\frac{r_+^2}{\ell^2}\right) \,, \nonumber\\
 &&g^{in}(z)=\left[1-\frac{25 \ell^4}{z^4}\frac{r_+^2}{\ell ^2}+\frac{5\left(125\ell^8-114 z^4\ell^4\right)}{z^8}\frac{r_+^4}{\ell^4}+
 \mathcal{O}\left(\frac{r_+^6}{\ell^6}\right)\right] \nonumber \\
  &&\hspace{1cm} +\epsilon^2 {\biggl [}
 -\frac{2}{3}+\frac{r_+{}^2}{\ell^2}{\biggl (}
 \left(\frac{635}{504}+\frac{7}{8}{\rm log}(2)+\frac{2}{3}{\rm log}\left(\frac{r_+}{\ell}\right)\right)\nonumber \\
 &&\hspace{1.6cm}
 +\!\int_1^{z/\ell}\!dx\,\frac{-870 \left(1-x^2\right)^2+7 x^4 \left(\left(1-2 x^2\right) E\left(1-x^2\right)
  +x^2 K\left(1-x^2\right)\right)^2}{21 x^5 \left(1-x^2\right)^2} {\biggr )}
  +\mathcal{O}\left(\frac{r_+^4}{\ell^4}\right){\biggr ]}\nonumber \\
  && \hspace{1cm} +\epsilon^4 {\biggl [}-\frac{87229}{127008}+\mathcal{O}\left(\frac{r_+^2}{\ell^2}\right){\biggr ]} \,,\nonumber\\
 &&h^{in}(z)=\left[1+\frac{25\ell^4}{z^4}\frac{r_+^2}{\ell ^2}+\frac{570\ell^4}{z^4}\frac{r_+^4}{\ell ^4}+
 \mathcal{O}\left(\frac{r_+^6}{\ell^6}\right) \right]+\epsilon^2\left[ \frac{265 \ell^4}{42 z^4}\frac{r_+^2}{\ell^2}
 +\mathcal{O}\left(\frac{r_+^4}{\ell^4}\right) \right]+\epsilon ^4 \mathcal{O}\left(\frac{r_+^2}{\ell^2}\right) \,, \nonumber\\
 &&\ell\,\Omega^{in}(z)={\biggl [} \frac{5\ell^4}{z^4}+\frac{-125\ell^8+122 z^4\ell^4}{z^8}\frac{r_+^2}{\ell^2}
  \nonumber \\
  &&\hspace{2cm} +\frac{50000\ell^{12}-94400 z^4\ell^8+z^8\ell^4\left(43441+24 {\rm log}\left(\frac{r_+}{4\ell}\right)\right)}{16 z^{12}}\frac{r_+^4}{\ell^4}
 +\mathcal{O}\left(\frac{r_+^6}{\ell^6}\right){\biggr ]} \nonumber \\
  &&\hspace{1cm} +\epsilon^2{\biggl [}
  \left(-\frac{15}{28}+\frac{29}{28}\left(1-\frac{\ell^4}{z^4}\right)\right)+\frac{r_+^2}{\ell^2}
  {\biggl (}-\frac{4469}{840}+\int_1^{z/\ell}\frac{dx}{x^5}{\biggl (}\frac{152095763}{1720320}-\frac{1565 {\rm log}(2)}{1024 }\nonumber \\
  &&\hspace{2cm} +\frac{45 {\rm log}^2(2)}{128}
  -10 {\rm log}\left(\frac{r_+}{\ell}\right)+\int_1^x\frac{5\, d\eta}{168\eta^5 \left(\eta^2-1\right)^2}{\biggl (}-6080\left(\eta^2-1\right)^2\nonumber \\
  &&\hspace{2cm}
  +21 \pi^2 \eta^4 {\biggl (}\left(-3-2 \eta^2\left(\eta^2-1\right)\left(\eta^4+5\right)\right) \left(P_{1/2}\left(2 \eta ^2-1\right)\right){}^2\nonumber \\
  &&\hspace{2cm}
  +6 \left(2 \eta ^2-1\right) P_{1/2}\left(2 \eta ^2-1\right)P_{3/2}\left(2 \eta^2-1\right)
  -3 \left(P_{3/2}\left(2 \eta^2-1\right)\right){}^2{\biggr )}{\biggr )} {\biggr )}{\biggr )} +\mathcal{O}\left(\frac{r_+^4}{\ell^4}\right)
   {\biggr]}\nonumber \\
  && \hspace{1cm} +\epsilon^4 {\biggl [} -\frac{22456447}{35562240}+\frac{14944939}{11854080}\left(1-\frac{\ell^4}{ z^4}\right)
  +\mathcal{O}\left(\frac{r_+^2}{\ell^2}\right){\biggr ]} \,,  \\
 &&\Pi^{in}(z)=\epsilon {\biggl [}\frac{\pi }{4} P_{1/2}\left(2 \frac{z^2}{\ell^2}-1\right)\frac{r_+}{\ell }
+\frac{r_+^3}{\ell^3}\, {\biggl (}  Q_{1/2}\left(2 \frac{z^2}{\ell^2}-1\right)\int _1^{z/\ell} dx \, s(x) P_{1/2}(1+2 x)\nonumber \\
  &&\hspace{2cm} 
 - P_{1/2}\left(2  \frac{z^2}{\ell^2}-1\right)\int _1^{z/\ell} dx \,s(x) Q_{1/2}(1+2 x){\biggr )} +\mathcal{O}\left(\frac{r_+^5}{\ell^5}\right){\biggr ]}+ \epsilon ^3 \mathcal{O}\left(\frac{r_+^3}{\ell^3}\right)
+ \epsilon^5 \mathcal{O}\left(\frac{r_+}{\ell}\right)\,,\nonumber
 \end{eqnarray}
}
where $z=r \,\ell/r_+$ and we defined 
{\small
\begin{eqnarray} \label{NR:functionsAux}
 && s(x)=\frac{1}{6 (1+x)^{3/2}}{\biggr [} 2\sqrt{1+x}\,{\bigr [}-11+x (-79+x (43+11 x)){\bigr ]}{\bigr [}8(1+2 x)E(-x)-(5+8 x)K(-x){\bigr ]}\nonumber \\
  && \hspace{3cm} +\pi  \sqrt{-x} {\bigr [}122+x (122-x (97+22 x)){\bigr ]} P_{3/2}(1+2 x){\biggr ]}\, \\
  && 
  \hbox{and}\qquad  E(x)=\int _0^1\left(1-\eta ^2\right)^{-1/2}\left(1-x \eta ^2\right)^{1/2}d\eta\,,\qquad  K(x)=\int_0^1 \left[\left(1-\eta ^2\right)\left(1-x \eta ^2\right)\right]^{-1/2} \, d\eta  \nonumber
 \end{eqnarray}
}
are, respectively, the complete elliptic integral and the complete elliptic integral of the first kind.

The matched asymptotic expansion construction also generates the expansion for the frequency of the scalar field, that is equal to the angular velocity of the hairy black hole, and explicitly written in equation \eqref{eqP:omegaBH}. 



\end{document}